\documentclass{ws-ijmpd}
\usepackage[super,compress]{cite}
\usepackage[breaklinks]{hyperref}
\usepackage{breakurl}
\usepackage{footmisc}
\usepackage{longtable,lscape}
\usepackage{rotating}

\def\bib{B\kern-.05em{I}\kern-.025em{B}\kern-.08em}
\def\btex{B\kern-.05em{I}\kern-.025em{B}\kern-.08em\TeX}

\def\gr{$\gamma$-ray}
\def\eiso{$E_{\rm \gamma,iso}$}
\def\ee{$\epsilon_{\rm e}$}
\def\eb{$\epsilon_{\rm B}$}

\begin{document}

\markboth{Lara Nava}
{High-energy emission from GRBs}

%
\catchline{}{}{}{}{}
%

\title{HIGH-ENERGY EMISSION FROM GAMMA-RAY BURSTS}

\author{LARA NAVA}

\address{
INAF - Osservatorio Astronomico di Brera\\ via E. Bianchi 46, I-23807, Merate, Italy\\
lara.nava@brera.inaf.it}

\address{
INAF - Osservatorio Astronomico di Trieste\\ via G.~B. Tiepolo 11, I-34143, Trieste, Italy}

\address{
INFN - Istituto Nazionale di Fisica Nucleare, Sezione di Trieste\\ via Valerio 2, I-34127, Trieste, Italy}

\maketitle

\begin{history}
\received{Day Month Year}
\revised{Day Month Year}
\end{history}

\begin{abstract}
The number of Gamma-Ray Bursts (GRBs) detected at high energies ($\sim\,0.1-100$\,GeV) has seen a rapid increase over the last decade, thanks to observations from the 
{\it Fermi}-Large Area Telescope.
The improved statistics and quality of data resulted in a better characterisation of the high-energy emission properties and in stronger constraints on theoretical models. 
In spite of the many achievements and progresses, several observational properties still represent a challenge for theoretical models, revealing how our understanding is far from being complete.
This paper reviews the main spectral and temporal properties of $\sim\,0.1-100$\,GeV emission from GRBs and summarises the most promising theoretical models proposed to interpret the observations.
Since a boost for the understanding of GeV radiation might come from observations at even higher energies, the present status and future prospects for observations at very-high energies (above $\sim$\,100\,GeV) are also discussed.
The improved sensitivity of upcoming facilities, coupled to theoretical predictions, supports the concrete possibility for future ground GRB detections in the high/very-high energy domain.

\end{abstract}


\ccode{PACS numbers:}

\tableofcontents

\section{Introduction}
The origin of the radiation detected from Gamma-Ray Bursts (GRBs) is still enigmatic in many aspects. 
The nature of the radiative processes and energy dissipation mechanisms at work have not been clearly identified yet. 
The limited understanding of the mechanisms responsible for radiation strongly affects the possibility to constrain the properties of the emitting region, and ultimately, disclose the composition of the jet itself and the nature of the progenitor.

Currently, GRBs are detected mostly by the space missions {\it Swift} and {\it Fermi}.
{\it Swift}, launched in 2004, finds and localises GRBs with the Burst Alert Telescope (BAT, 15-150\,keV) and follows their afterglow with the X-ray Telescope (XRT, 0.3-10\,keV) and the UV/Optical Telescope (UVOT, 170 - 600\,nm).
{\it Fermi}, launched in 2008, is dedicated to hard X-ray and $\gamma$-ray observations, covering the 8\,keV-300\,GeV energy range thanks to its two instruments: the Gamma-ray Burst Monitor (GBM, 8\,keV - 30\,MeV) and the Large Area Telescope (LAT, 20\,MeV-300\,GeV).
Considering both satellites, the current GRB detection rate is approximately one event per day.
Follow-up observations in the radio, optical, and soft X-ray bands complete the observational picture by providing information on the afterglow emission.

Exciting non-electromagnetic channels for the investigation of GRBs include gravitational waves\cite{abbott17}, cosmic-rays\cite{globus15,baerwald15}, and neutrinos\cite{aartsen16,albert17}.
However, a relatively new window on the study of GRBs has recently opened also within the electromagnetic domain, thanks to GRB detections above several tens of MeV. 
Since August 2008, the LAT has detected on average $\sim$\,14~GRBs per year\footnote{\label{fn:lat_table}http://fermi.gsfc.nasa.gov/ssc/observations/types/grbs/lat\_grbs/table.php.}, with photon energies up to $\sim$\,100\,GeV\cite{ackermann14}. 
Observations at higher energies, performed by all major Imaging Atmospheric Cherenkov Telescopes (IACTs) and Extensive Air Shower (EAS) arrays, have so far resulted only in upper limits\cite{hoischen17,acciari11,aliu14,berti17,alfaro17}. The comparisons between such flux limits and theoretical models suggest that an improvement in the instrument sensitivity, such as the one warranted by the next generation of Cherenkov telescopes, should be sufficient for GRB ground detections at 50-100\,GeV, and possibly even beyond.

The least understood phase of the GRB phenomenon probably remains the prompt emission, lacking a basic understanding of the nature of the dominant radiation mechanism. 
Spectra between $\sim$\,10\,keV and $\sim$\,1\,MeV are generally well described by an empirical function (called Band function\cite{band93}) composed by two smoothly connected power-laws (PLs). The low-energy PL has photon index $\alpha$ ($N_\nu\propto \nu^{\alpha}$) with distribution peaked around the value $\langle\alpha\rangle\sim-1$, in contrast with expectations from the synchrotron model ($\alpha^{\rm syn}=-1.5$). 
The high-energy PL  has photon index $\beta$ ($N_\nu\propto \nu^{\beta}$) smaller than -2, with typical values around -2.2 -- -2.3, consistent with a non-thermal distribution of relativistic electrons with spectral index $p\simeq 2.4 - 2.6$ (in the notation $N(\gamma)\propto\gamma^{-p}$).
The spectral peak energy $E_{\rm peak}$ has a wide distribution centred around $\sim$\,200\,keV and ranging from a few keV to several MeV\cite{nava11,goldstein13,gruber14}.

Due to the inconsistency between the low-energy part of the observed spectra and the synchrotron spectrum, a large number of alternative models have been proposed and investigated\cite{thompson94,ghisellini99,ryde04,eichler00,rees05,lazzati10,toma11}. 
Within the synchrotron scenario, a reconciliation with the observed spectra is possible if one considers a synchrotron spectrum produced by electrons in moderately fast cooling regime and/or modified by inverse Compton scatterings in Klein-Nishina (KN) regime\cite{daigne11}.
In spite of all the efforts, the origin of the prompt radiation still remains one of the most puzzling open issues in GRB physics.

Recent progresses in the characterisation of prompt spectra have outlined the presence (at least in a fraction of GRBs) of a third PL behaviour at low energies (below a {\it break energy} that assumes values in the range 2-100\,keV)\cite{oganesyan17a,oganesyan17b,ravasio17}. When accounting for this additional PL, the spectral fits are consistent with $N_\nu\propto \nu^{-2/3}$ and $N_\nu\propto \nu^{-3/2}$ below and above the break energy, respectively, as expected in synchrotron radiation spectra. These recent findings call now for a re-evaluation of the consistency between prompt spectra and synchrotron radiation.

As for the temporal characteristics, the duration of the prompt phase provided the first hint for the existence of two different classes of GRBs.
The prompt duration ($T_{90}$) has indeed a bimodal distribution\cite{kouveliotou93}, based on which GRBs are classified as short or long, depending whether their $T_{90}$ is smaller or larger than 2\,s. The fact that we are in presence of two different populations (reflecting the existence of two different types of progenitors) is supported and reinforced by other observational differences among the two classes (e.g., spectral hardness, properties of the host galaxies, association with supernovae) and by the recent detection of a gravitational wave signal consistent with a binary neutron star merger and associated to a short GRB\cite{abbott17}. 
Both long and short GRBs have been detected in the high-energy ($0.1-100$\,GeV) domain, with similar behaviours.

The interpretation of afterglow radiation is less problematic as compared to the prompt, since observations are in general agreement with the standard model (i.e., the synchrotron external shock scenario\cite{sari98}). Nevertheless, features such as plateaus, bumps, and rebrightenings, commonly present in afterglow lightcurves, still represent an open issue.

A recent review on the current status of GRB observations, theoretical understanding, and open issues can be found in Ref.~\refcite{kumar15}.
The present work focuses instead on one specific aspect, that is the high-energy emission detected in a fraction of GRBs during their prompt and/or afterglow phases. 
Throughout the whole paper, the term {\it high-energy} (HE) will be used with reference to emission above a few tens of MeV.\\

The paper is organised as follows: I first present a brief history of high-energy observations through the history of the main detectors that allowed us to reveal and probe  the presence of HE emission from GRBs (section~\S\ref{sec:missions}). 
In \S\ref{sec:observations}, I summarise the observational properties of this radiation. 
I review the most promising theoretical models in section \S\ref{sec:origin}. 
Since relevant information may be derived also from the lack of detections, HE flux upper limits and their implications for physical models are discussed in section \S\ref{sec:lack}. 
Finally, in section \S\ref{sec:vhe}, the present status and future prospects for GRB observations at very-high energies (VHE, $\gtrsim 100$\,GeV) are summarised.
Conclusions are presented in section \S\ref{sec:conclusions}.


\section{Detectors for observations of GRBs at high-energies}\label{sec:missions}
The very first hints for GRBs emitting radiation above 0.1\,GeV date back to more than thirty years ago, when the Gamma-Ray Spectrometer onboard the Solar Maximum Mission detected $\sim$0.1\,GeV photons from a very bright event occurred on 1984 August 5 \cite{share86}. Since then, much has been learned on the temporal and spectral properties of the radiation detected at these energies, and photons have been detected at even higher energies, the record holder being a 95\,GeV photon (corresponding to 128\,GeV in the frame at rest with the progenitor) from GRB~130427A \cite{ackermann14}.

In this section, I review the (past and current) major detectors that mostly contributed to the study of HE radiation from GRBs, and I highlight their main discoveries and accomplishments.

\subsection{{\it CGRO}-EGRET}
\begin{figure}[th]
\centering
\includegraphics[scale=0.1]{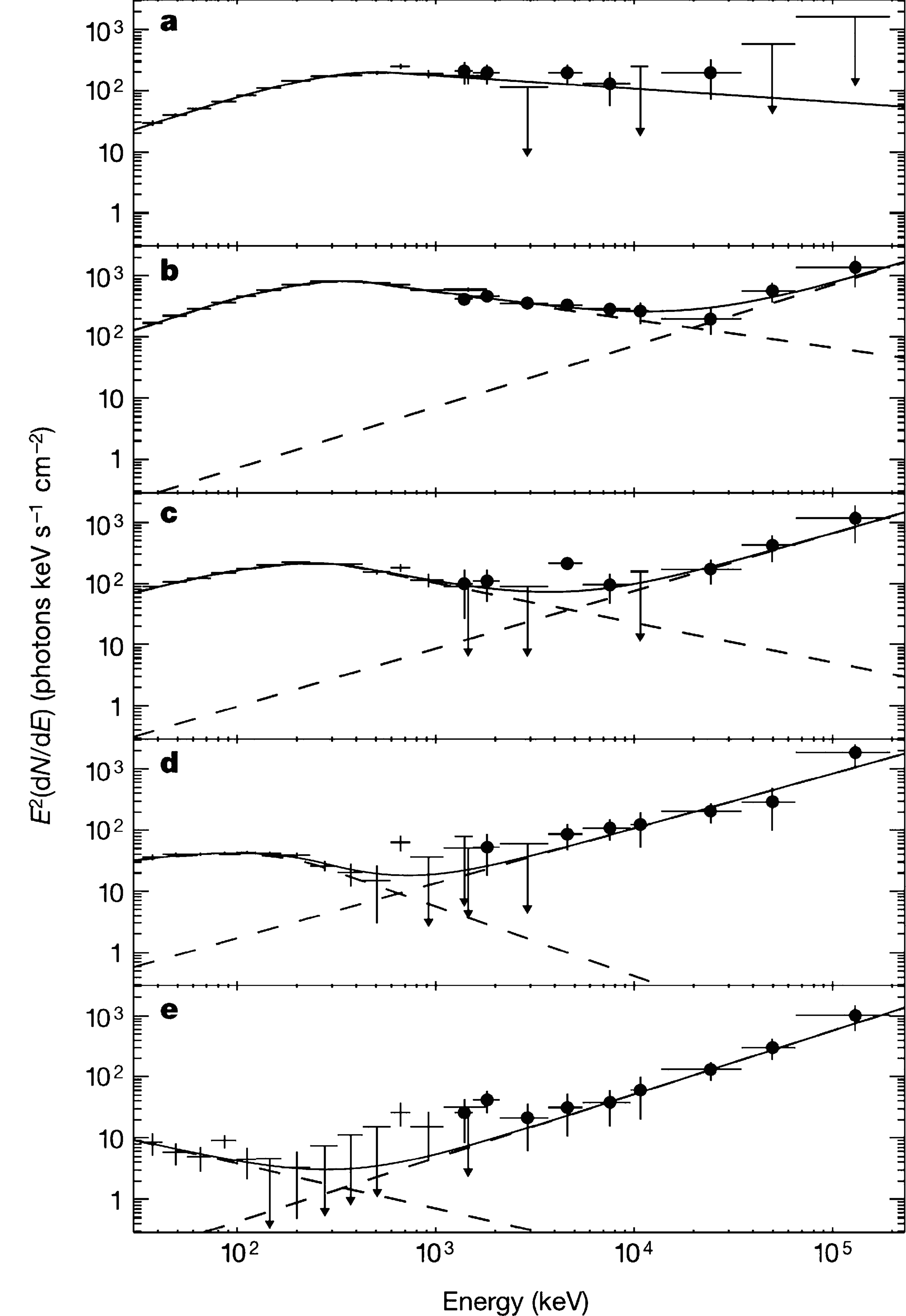}
\caption{Combined spectrum of GRB~941017, observed by the {\it CGRO}, in five different time intervals (panels {\it a} to {\it e}, referring to intervals -18--14\,s, 14--47\,s, 47--80\,s, 80--133\,s, and 133--211\,s, respectively). 
Crosses denote BATSE-LAD data, filled circles denote TASC data. 
Two different spectral components (a Band function at low energy and a simple power-law at high energy) are present in the time intervals {\it b} to {\it e}. 
From Ref.~\protect\refcite{gonzalez03}.
\label{fig:941017}}
\end{figure}
The Compton Gamma-Ray Observatory ({\it CGRO}, 1991-2000) with its instruments devoted to the study of the energetic transient Universe, has marked the beginning of a new era for GRB observations. 
The Burst And Transient Source Experiment (BATSE), sensitive in the energy range $\sim$\,20-2000\,keV, has detected 
prompt emission from about 2700~GRBs, allowing us to infer spectral and temporal properties for a large sample of events and over a wide range of energies.
At even higher energies, GRBs could be studied thanks to EGRET, the Energetic Gamma-Ray Experiment Telescope, sensitive in the energy range $\sim$\,20\,MeV-30\,GeV.

EGRET, featuring a field of view of $\sim0.5$\,sr and a dead time of $\sim100$\,ms, has detected five GRBs with its spark chamber (GRB~910503, GRB~910601, GRB~930131, GRB~940217, and GRB~940301\cite{merck95}), plus one (GRB~941017) that triggered the calorimeter beneath the spark chamber (EGRET-TASC, Total Absorption Scintillator Calorimeter).
 
These first few detections immediately revealed the diversity in GRB spectral and temporal properties at high energies. 
The radiation detected by EGRET from GRB~930131 \cite{sommer94} was consistent with being the high-energy continuation of the keV-MeV spectrum observed by BATSE, suggesting a common origin of the radiation from keV to GeV energies.
On the contrary, the HE emission from GRB~940217 pointed to the presence of an additional emission component with a distinct origin as compared to the prompt:
the HE emission lasted indeed much longer than the prompt phase, being detected up to 90 minutes after the burst, with an 18\,GeV photon arrived after 75 minutes\cite{hurley94}. 
A clear evidence for the presence of an additional spectral component was found also in GRB~941017\cite{gonzalez03}. 
In this GRB, the combined use of BATSE and TASC data revealed the presence of an additional and hard power-law spectral component, observed up to 200\,MeV (Fig.~\ref{fig:941017}, panels $b$ to $e$).

\subsection{{\it AGILE}-GRID}
The {\it AGILE} satellite\cite{tavani09}, launched in 2007, features onboard a pair-tracking telescope operating in the 30\,MeV-50\,GeV energy range, called the Gamma Ray Imaging Detector (GRID\cite{prest03}). 
With its larger field of view ($\sim$\,2.5\,sr) and reduced dead time ($\sim$\,200\,$\mu$s) it represents a substantial improvement over the capabilities of EGRET, and is particularly suited for observations of GRBs.
{\it AGILE} can also observe GRBs at hard X-rays (18-60\,keV), thanks to SuperAGILE\cite{feroci07}.
Moreover, the AGILE Mini-Calorimeter (MCAL), besides being part of the GRID, can autonomously detect and study GRBs in the 0.35-100\,MeV range with excellent timing.
\begin{figure}[b]
{\includegraphics[scale=0.337]{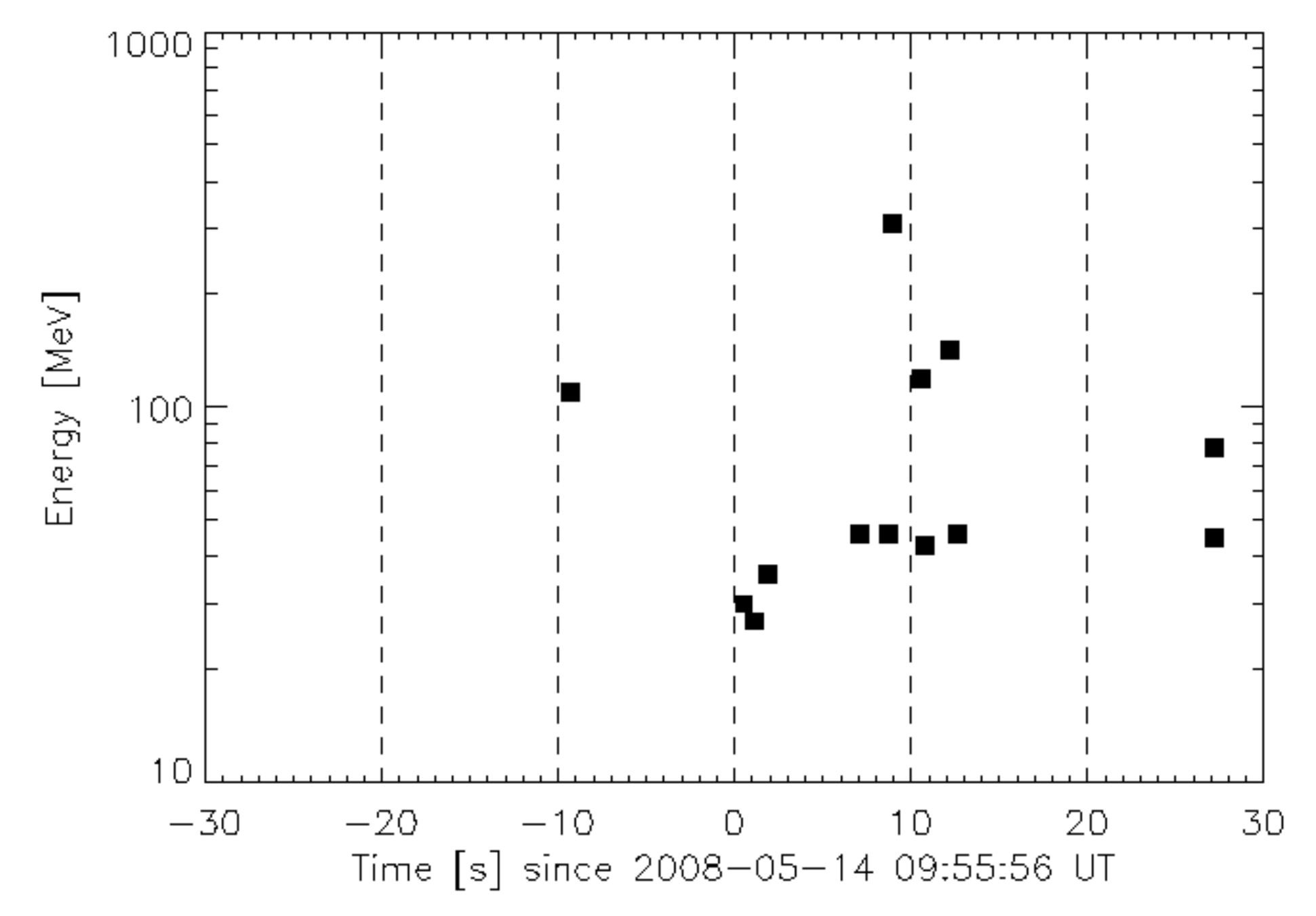}
\includegraphics[scale=0.29]{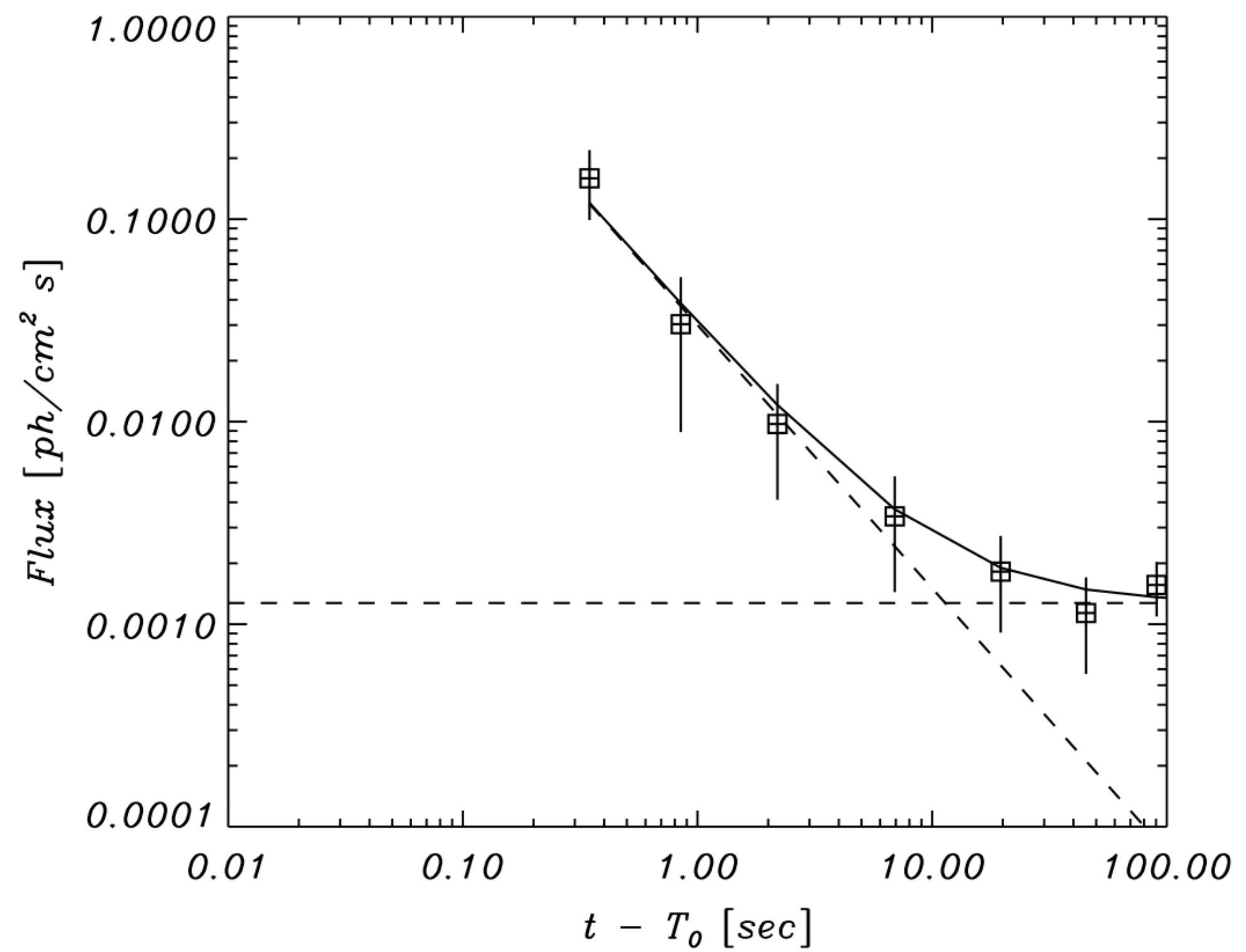}}
\caption{Left: energy of photons detected by the {\it AGILE}-GRID from GRB~080514B as a function of their arrival time (from Ref.~\protect\refcite{giuliani08}). The prompt emission detected in the 17-50\,keV range lasted about 7\,s.
Right: {\it AGILE}-GRID lightcurve of GRB~090510. 
The two dashed lines show a constant background flux and a power-law with index $-1.3$.  The solid line is the sum of the two components (from Ref.~\protect\refcite{giuliani10}).
\label{fig:agile}}
\end{figure}

After the EGRET era, {\it AGILE} has been the first mission to detect radiation above several tens of MeV from GRBs (GRB~080514B\cite{giuliani08}). 
The energy of photons detected by the GRID from GRB~080514B versus their arrival time is shown in Fig.~\ref{fig:agile} (left-hand panel). The duration of this burst at keV energies was $\sim$\,7\,s.

{\it AGILE} detected so far 9 GRBs\footnote{http://agile.rm.iasf.cnr.it/gnc/gcn.html} with emission above 30\,MeV \cite{giuliani10,delmonte11,giuliani14}.
The GRID lightcurve of the short GRB~090510 is shown in Fig.~\ref{fig:agile} (right-hand panel): significant emission above 30\,MeV is visible well after the prompt phase, that ended at $t\sim$\,1\,s. The decay of the flux as a function of time can be modelled by a PL function ($F\propto t^{-1.3}$), as shown by the dashed line.

\subsection{{\it Fermi}-LAT}
The {\it Fermi} satellite was launched in June 2008 and began operations in early August 2008. 
It has two instruments onboard: the Gamma-ray Burst Monitor (GBM) \cite{meegan09}, and the Large Area Telescope (LAT) \cite{atwood09}.
The GBM is a full sky monitor including 12 sodium iodide (NaI) detectors and two bismuth germanate (BGO) detectors, sensitive respectively in the 8\,keV-1\,MeV and 150\,keV-30\,MeV energy range. 
It localises GRBs and determines their position with a $\sim\,3^\circ$-4$^\circ$ accuracy (but see Ref.~\refcite{connaughton15}).
The GBM is detecting GRBs at an average rate of 240 per year\footnote{\label{fn:gbm_cat}https://heasarc.gsfc.nasa.gov/W3Browse/fermi/fermigbrst.html}, which is about the expected rate. 
Around the 16\% are short bursts. 

The LAT is a pair conversion telescope covering the energy range from $\sim$\,20\,MeV to $\sim$\,300\,GeV.
The LAT features a large field of view (2.4\,sr at 1\,GeV), a broad energy range, a low dead time per event $<50\,\mu$s), and a large effective area at all energies.
As compared to the EGRET and GRID experiments, these features 
resulted in a much larger number of detections and an improved capability of inferring the temporal and spectral properties of the HE emission.
The LAT is detecting GRBs at an average rate of 14 per year. 
The full list of GRBs detected by the LAT until December 2017 (141 events\footref{fn:lat_table}) is reported in Table~\ref{tab:table}.
Given that about half of the GBM-detected GRBs fall inside the LAT field of view, the detection rate of LAT corresponds to $\sim$\,12\% of GBM bursts. 

While in the first 2 years of observations the analysis of GRB data was restricted to photons with energy larger than 100\,MeV, in 2010 a new technique was introduced to reconstruct the signal between $\sim$\,30 and 100\,MeV\cite{pelassa10}. 
The introduction of the LAT low-energy (LLE) event class allowed to fill the gap between the GBM and the emission above 100\,MeV. Around 15\% of the LAT GRBs have been detected only below 100\,MeV (LLE-only events).

Most of our present knowledge on HE emission from GRBs is based on {\it Fermi}-LAT observations and will be presented in the next section.

\section{Observational properties of GRBs with high-energy emission}\label{sec:observations}
The full list of GRBs detected by the LAT until December 2017 and their main properties can be found in Table~\ref{tab:table}.
The different scenarios for the interpretation of these properties will be discussed in section~\ref{sec:origin}.

\subsection{Detection rate and redshift distribution}
Simulations performed before the launch of {\it Fermi} predicted an LAT detection rate of 10-12~GRBs/year above 100\,MeV, including 6-8~GRBs/year above 1\,GeV\cite{band09}. These pre-launch estimates were based on simple PL spectral extrapolations of the prompt keV-MeV component into the LAT energy range.
For these calculations, the spectrum of the prompt component has been described as a Band function with parameters distributed according to the properties of BATSE bursts.
Estimates based on this purely phenomenological approach are subject to two main sources of uncertainty, affecting the predicted rate in opposite directions. 
A rate based on PL extrapolations of the Band prompt component might lead to overestimate the actual rate in case high-energy spectral cutoffs are present at $\lesssim$\,GeV energies. Cutoffs around these energies are expected as a result of $\gamma$-$\gamma$ absorption and they have been indeed observed in a few cases (see section~\ref{subsec:spectral_properties}). 
On the other hand, a rate based on keV-MeV prompt spectra might lead to underestimate the actual rate in case an additional spectral component is present in the LAT energy range (e.g., an inverse Compton component of internal or external origin). Additional spectral components have been observed in several LAT GRBs, as it will be discussed in section~\ref{subsec:spectral_properties}.
\begin{figure}
{\includegraphics[scale=0.295]{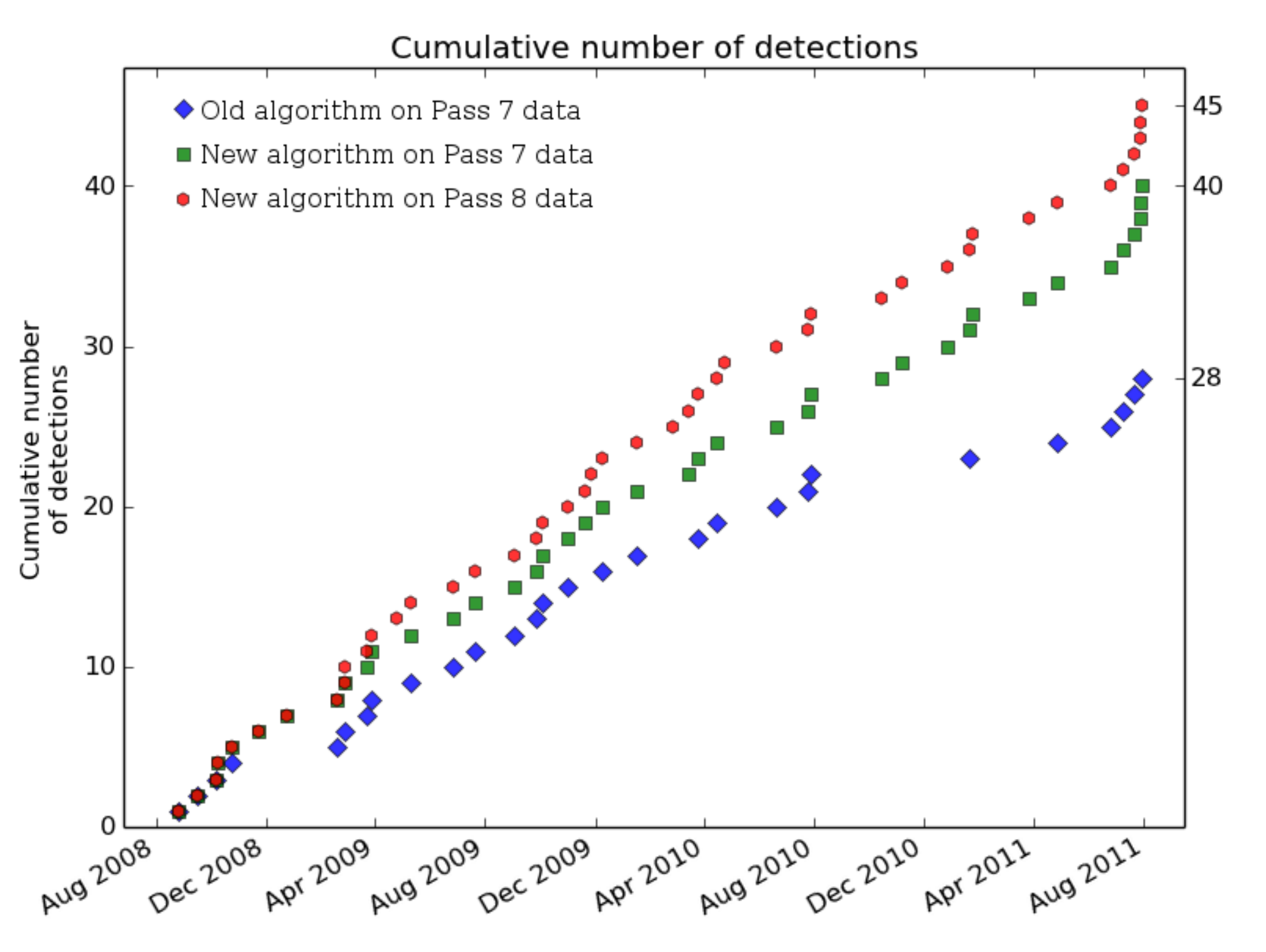}
\includegraphics[scale=0.237]{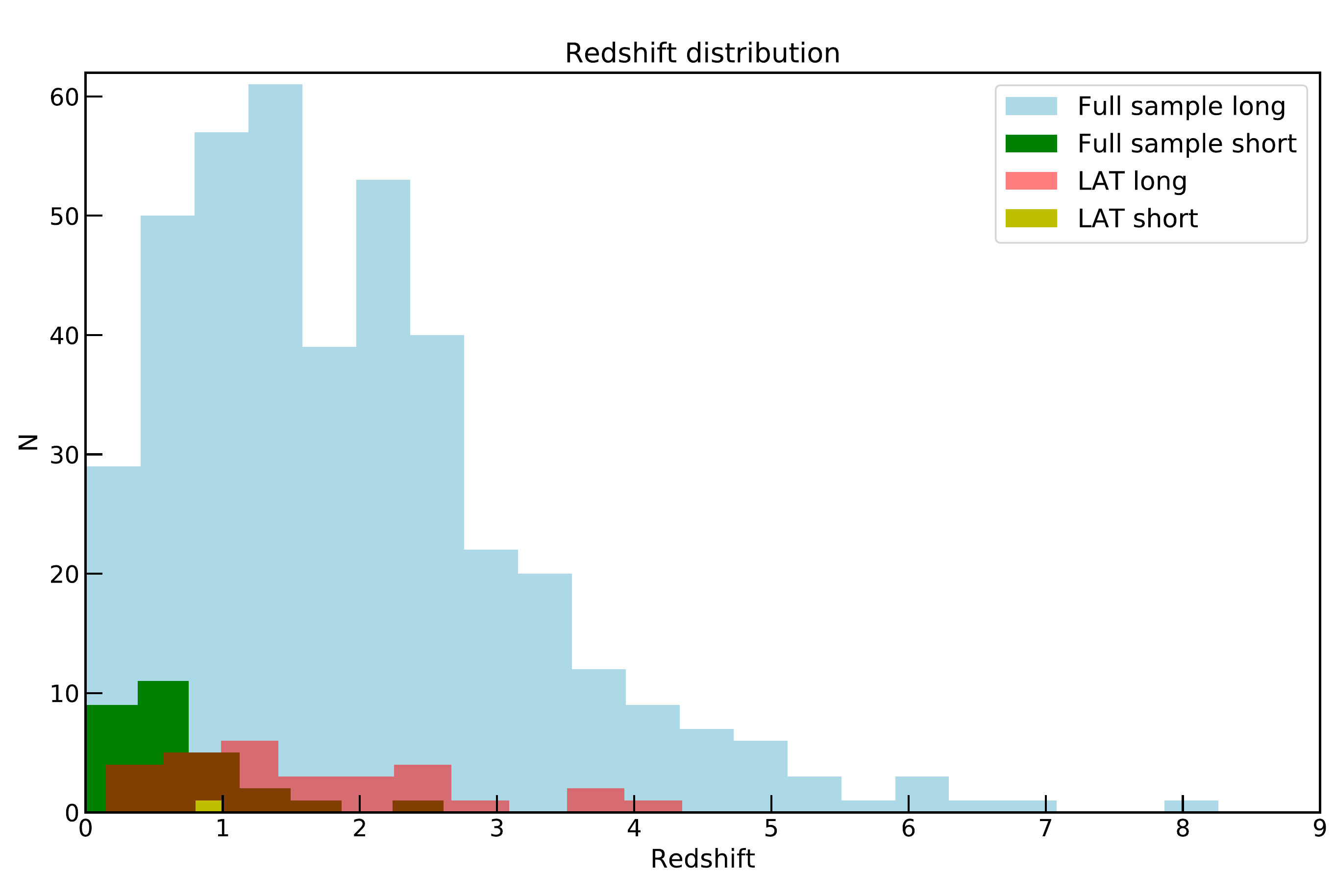}}
\caption{Left: cumulative number of GRB detections by the LAT during the first three years of observations (the time covered by the first LAT GRB catalog\cite{LATcatalog13}). Only GRBs with photons above 100\,MeV are included. Diamonds (blue) denote the 28 events with photon energies in excess of 100\,MeV included in the first LAT catalog. Squares (green) show the effect of the new detection algorithm, and circles (red) show the combined effect of the improved detection algorithm\protect\cite{vianello15} and of the improvement in the event analysis from Pass~7 to Pass~8 (from Ref.~\protect\refcite{vianello15}). 
Right (this work): the red and yellow histograms show the redshift distribution of LAT long and short GRBs up to December 2017 (see Table~\ref{tab:table}), compared to the full sample\protect\footref{fn:greiner_website}
 (also updated to December 2017, blue and green histogram for long and short GRBs, respectively).
\label{fig:detection_redshift}}
\end{figure}

During the first 3~years of the {\it Fermi} mission, the LAT detection rate were slightly below pre-launch expectations\cite{LATcatalog13} (28 detections above 100\,MeV, see the cumulative number in Fig.~\ref{fig:detection_redshift}, left-hand panel, blue diamonds). 
This was interpreted as the evidence for the frequent presence of cutoffs/breaks in the high-energy part of prompt spectra.
The development of new event analyses for the LAT and the introduction of a new search algorithm accounting for the large GBM position uncertainty \cite{vianello15} have however increased the LAT detection rate to 14~GRBs/year ($\sim$\,12~GRBs/year with photons above 100\,MeV), consistent with pre-launch estimates. 
The effects of the 'Pass 8' event analysis and of the new detection algorithm on the LAT detection rate until August 2011 (corresponding to the time covered by the first LAT GRB catalog\cite{LATcatalog13}) are shown in Fig.~\ref{fig:detection_redshift} (left-hand panel).
The good agreement between the observed detection rate and the predicted one must not be considered as a demonstration that LAT GRB detections are always caused by the HE part of the prompt emission spectrum, that is the scenario on which the pre-launch estimates were based. 
More likely, the present rate is the result of cases where the Band spectrum continues up to high-energies, cases with MeV cutoffs resulting in no HE detections, and cases where the HE detection is caused by an additional spectral component.

Among the 141\,GRBs detected by the LAT in about nine and a half years of operations (until December 2017), 30 have known redshift (see Table~\ref{tab:table}), including one short event (GRB~090510). 
Their redshift distribution is shown in Fig.~\ref{fig:detection_redshift} (right-hand panel, red and yellow histograms for long GRBs and for the short one, respectively), where they are compared with the total distribution of spectroscopic redshifts\footnote{http://www.mpe.mpg.de/~jcg/grbgen.html\label{fn:greiner_website}}, also updated to December 2017.
GRBs with detected HE emission span a large range in redshift, from 0.145 to 4.35.

\begin{figure}
\centering
\includegraphics[scale=0.32]{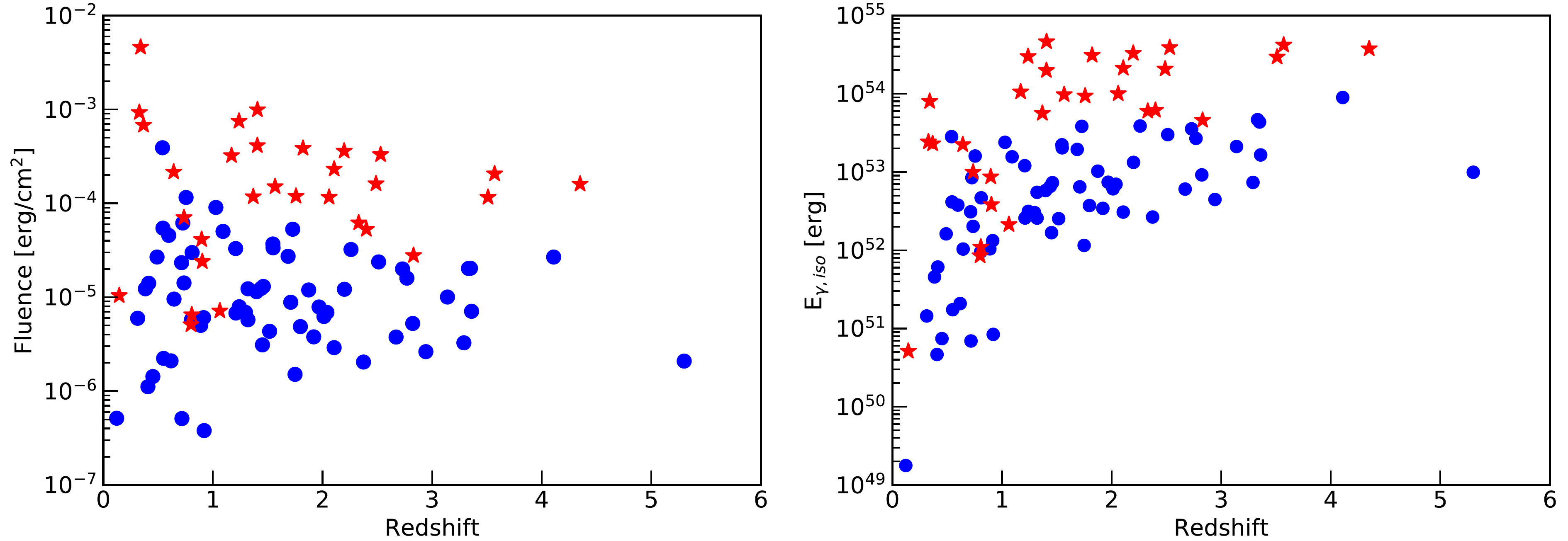}
\caption{Prompt fluence (left-hand panel) and energy \eiso\ (right-hand panel) integrated between 1\,keV and 10$^4$\,keV versus redshift, for GRBs detected by the GBM up to December 2017. Stars denote GRBs detected also by the LAT.
\label{fig:fluence_eiso_redshift}}
\end{figure}

A comparison between LAT, BAT and GBM+BAT samples showed that their respective redshift distributions are all consistent with each other\cite{racusin11}. There are indications, however, that for a given redshift, LAT-detected GRBs are generally brighter than the average\cite{cenko11,racusin11}.
Fig.~\ref{fig:fluence_eiso_redshift} shows the prompt fluence and prompt \eiso\ (both integrated in the energy range 1-10$^4$\,keV) versus redshift $z$ (left-hand and right-hand panels, respectively) for all GBM events with measured redshift. Stars denote the subsample of GRBs detected also by the LAT. Their fluence and \eiso\ are also reported in Table~\ref{tab:table}.


\begin{landscape}
\scriptsize[\refcite{LATcatalog13}]
\begin{longtable}{llcrcccr|ccrl} 
\caption{List of GRBs detected by LAT until December 2017}\label{tab:table}\\[-1.5ex]
\hline\\[-2.2ex]
Name    &    $z$    & Class& $T_{90}$\footnote{\tiny{Between 50-300\,keV}}&   $E_{\rm peak}$      &   Fluence\footnote{\tiny{Integrated from the extrapolated spectrum in the range 1-10$^4$\,keV if $E_{\rm peak}$ is known, from the measured spectrum between 10-1000\,keV otherwise.}}      & $E_{\rm\gamma,iso}$\footnote{\tiny{Integrated from the extrapolated spectrum in the range 1-10$^4$\,keV (rest frame) if $E_{\rm peak}$ is known, from the measured spectrum between 10-1000\,keV otherwise}} & Ref.\footnote{\tiny{References for the prompt (sub-MeV) emission. Five-digit numbers refer to the GCN number, GBMcat refers to the online GBM catalog\footref{fn:gbm_cat}.}}    & EE\footnote{\tiny{Presence of long-lasting HE emission. Possible cases: the duration of the HE emission is (i) $>3T_{90}$ (Y), (ii) $\lesssim T_{90}$ (N), (iii) intermediate or no information available (-).}}  &  HE photon\footnote{\tiny{Energy of the most energetic photon detected by LAT}} & Arr. time\footnote{\tiny{Arrival time of the most energetic photon detected by LAT}}& Ref.\footnote{\tiny{Reference for the LAT emission. Five-digit numbers refer to the number of the GCN, numbers between square brackets refer to published papers.}} \\
             &          &          &  [s]~~           &    [keV]       	  &     [erg cm$^{-2}$]	 &      [erg]~		       &               &     &  [GeV]	&   [s]~~~~~  &       \\[0.4ex]
\hline\\[-1.9ex]
171212B &        &  L  &      32  &     74$\pm$9          &  $(7.58\pm0.40)10^{-6}$  &                       & 22249      &  Y  &   0.93   &     544     & 22247 \\ 
171210A &	 &  L  &     143  &    145$\pm$3	  &  $(9.76\pm0.11)10^{-5}$  &		       	       & 22233      &  Y  &	12	&    1400     & 22228 \\
171124A &	 &  L  &      26  &   1120$\pm$280	  &  $(2.2\pm0.6)10^{-5}$	   &		       	       & 22164      &  -  &	 3.5	&       4     &	22156 \\
171120A &	 &  L  &      44  &    141$\pm$12	  &  $(2.68\pm0.07)10^{-5}$  &		       	       & 22138      &  Y  &	 3.4	&    4800     &	22136 \\
171102A &	 &  L  &      35  &    171$\pm$6	  &  $(3.52\pm0.06)10^{-5}$  & 		       	       & 22083      &  Y  &   0.5  	&     328     & 22081 \\
171022A &	 &  L  &      13  &    151$\pm$5	  &  $(8.3\pm0.2)10^{-6}$    & 		       	       & 22046      &  Y  &     -    &       -     & 22045,22071 \\
171010A &  0.3285 &  L  &     104  &    154$\pm$1	  &  $(9.29\pm0.02)10^{-4}$  &$(2.45\pm0.01)10^{53}$& 21992     &  Y  &	 7   	&     768     & 21985 \\
170906A &	 &  L  &      79  &    288$\pm$12	  &  $(1.98\pm0.02)10^{-4}$  & 		   	       & 21839      &  -  &   3.6  	&     200     & 21827 \\
170810A &	 &  L  &     137  &    267$\pm$34	  &  $(1.0\pm0.6)10^{-5}$    & 		   	       & 21453      &  -  &   2.3  	&      34     & 21452 \\
170808B &	 &  L  &      18  &    262$\pm$5	  &  $(1.71\pm0.01)10^{-4}$  & 		   	       & 21429      &  Y  &   1.6  	&     162     & 21430 \\
170522A &	 &  L  &     7.4  &    341$\pm$9	  &  $(2.41\pm0.04)10^{-5}$  & 		   	       & GBMcat   &  -  &   3.7  	&	8     & 21127 \\
170510A &	 &  L  &     128  &    366$\pm$24	  &  $(8.45\pm0.12)10^{-5}$  & 		   	       & GBMcat   &  Y  &   1.8  	&     410     & 21080 \\
170409A &	 &  L  &      68  &    954$\pm$16	  &  $(5.79\pm0.02)10^{-4}$  & 		   	       & GBMcat   &  Y  &    10   	&     442     & 21004 \\
170405A &  3.51  &  L  &      79  &    267$\pm$9	  &  $(1.16\pm0.01)10^{-4}$  &$(2.4\pm0.1)10^{54}$ & GBMcat   &  Y  &	 0.9  	&     445     & 20987 \\
170329A &	 &  L  &      34  &    915$\pm$84	  &  $(2.23\pm0.05)10^{-5}$  & 		   	       & GBMcat   &  Y  &   0.8  	&	4     & 20942 \\
170306B &	 &  L  &      19  &    289$\pm$10    	  &  $(3.19\pm0.06)10^{-5}$  & 		   	       & 20827      &  Y  &   0.5  	&      46     & 20826 \\
170228A &	 &  L  &      61  &    752$\pm$65	  &  $(3.62\pm0.08)10^{-5}$  & 		   	       & GBMcat   &  -  &     5    &     130     & 20785 \\
170214A &  2.53  &  L  &     123  &    402$\pm$27	  &  $(3.30\pm0.02)10^{-4}$  &$(3.89\pm0.02)10^{54}$& GBMcat  &  Y  &	 7.8  	&     105     & 20676,[\refcite{tang17}]\\
170115B &	 &  L  &      44  &   1310$\pm$59   	  &  $(1.45\pm0.01)10^{-4}$  & 		   	       & GBMcat   &  -  &	   -    &	-     & \\
161202A &	 &  L  &     181  &   222$^{+201}_{-63}$  &  $3.50^{+0.48}_{-0.88}10^{-5}$  & 		       & 20238      &  Y  &     4    &    2000     & 20229 \\
161109A &	 &  L  &      24  &    251$\pm$21	  &  $(3.58\pm0.08)10^{-5}$  & 		   	       & GBMcat   &  Y  &   3.5  	&     600     & 20155 \\
160910A &	 &  L  &      24  &    322$\pm$11   	  &  $(1.42\pm0.01)10^{-4}$  & 		   	       & GBMcat   &  Y  &   0.5  	&    9000     & 19902 \\
160905A &	 &  L  &      34  &    751$\pm$72	  &  $(1.89\pm0.02)10^{-4}$  & 		   	       & GBMcat   &  Y  &     8    &     350     & 19890 \\
160829A &	 &  S  &    0.51  &    584$\pm$150   	  &  $(5.54\pm0.60)10^{-7}$  & 		   	       & GBMcat   &  -  &   9.5    &       2     & 19879 \\
160821A &	 &  L  &      43  &    941$\pm$16   	  &  $(1.07\pm0.01)10^{-3}$  & 		   	       & GBMcat   &  -  &   4.7  	&     212     & 19836 \\

\multicolumn{7}{r}{{\tablename} \thetable{} -- {Continued}}\\[1.ex]
\hline\\[-1.4ex]
160816A &	 &  L  &      11  &    240$\pm$4   	  &  $(3.51\pm0.04)10^{-5}$  & 		   	       & GBMcat   &  Y  &   9.5  	&    1100     & 19802 \\
160709A &	 &  S  &     5.6  &    2211$\pm$131 	  &  $(1.21\pm0.03)10^{-5}$  & 		   	       & 19676      &  -  &     1    &	2     & 19675 \\
160625B &  1.406 &  L  &     455  &    649$\pm$9	  &  $(9.93\pm0.03)10^{-4}$  & $(4.63\pm0.01)10^{54}$& GBMcat &  Y  &    15   	&     345     & 19586 \\
160623A &  0.367 &  L  &     225  &    562$\pm$23   	  &  $(6.8\pm0.1)10^{-4}$    & $(2.3\pm0.1)10^{53}$  & 19554    &  Y  &	 $>1$  	&	-     & 19553 \\
160521B &	 &  L  &     2.8  &    149$\pm$5	  &  $(1.71\pm0.03)10^{-5}$  & 		   	       & GBMcat   &  Y  &    12    &     420     & 19444 \\
160509A &  1.17  &  L  &    370   &    355$\pm$10	  &  $(3.23\pm0.02)10^{-4}$  & $(1.1\pm0.01)10^{54}$& GBMcat  &  Y  &    52    &      77     & 19413 \\
160503A &	 &  L  &     59   &    138$\pm$23	  &  $(1.97\pm0.21)10^{-6}$  & 		   	       & GBMcat   &  Y  &	$<1$    &       -     & 19379 \\
160422A &	 &  L  &     12   &    228$\pm$5	  &  $(1.29\pm0.08)10^{-4}$  & 		   	       & GBMcat   &  Y  &    12    &     770     & 19329 \\
160325A &	 &  L  &     43   &    238$\pm$15	  &  $(2.87\pm0.05)10^{-5}$  & 		   	       & GBMcat   &  Y  &     3  	&     100     & 19227\\
160314B &	 &  L  &     98   &    717$\pm$141	  &  $(1.47\pm0.07)10^{-5}$  & 		   	       & GBMcat   &  Y  &   0.9  	&     630     & 19197 \\
160310A &	 &  L  &     18   &    124$\pm$8	  &  $(5.50\pm0.21)10^{-6}$  & 		   	       & GBMcat   &  Y  &    30  	&    5800     & 19158 \\
160101B &	 &  L  &     22   &    843$\pm$410	  &  $(4.4\pm0.3)10^{-6}$    & 		    	       & GBMcat   &  N  &	$<0.1$  &       -     & 18799 \\
151006A &	 &  L  &     93   &    227$\pm$53	  &  $(5.3\pm0.1)10^{-5}$    & 		    	       & GBMcat   &  N  &	$<0.1$ 	&       -     & 18406 \\
150902A &	 &  L  &     14   &    347$\pm$7	  &  $(1.40\pm0.01)10^{-4}$  & 		    	       & GBMcat   &  Y  &    11    &     100     & 18228 \\
150724B &	 &  L  &     38   &    732$\pm$39	  &  $(4.86\pm0.06)10^{-5}$  & 		    	       & GBMcat   &  Y  &	   2    &      50     & 18065 \\
150702A &	 &  L  &     46   &   1273$\pm$176	  &  $(3.73\pm0.10)10^{-5}$  & 		    	       & GBMcat   &  Y  &	 0.8  	&    1200     & 17989 \\
150627A &	 &  L  &     65   &    226$\pm$5	  &  $(2.85\pm0.01)10^{-4}$  & 		    	       & GBMcat   &  Y  &	 8.1  	&     259     & 17971 \\
150523A &	 &  L  &     82   &    553$\pm$20	  &  $(4.34\pm0.06)10^{-5}$  & 		    	       & GBMcat   &  -  &	   7    &     118     & 17864 \\
150514A &  0.807 &  L  &     11   &	78$\pm$4	  &  $(5.6\pm0.2)10^{-6}$  & $(9.7\pm0.3)10^{51}$  & GBMcat   &  Y  &	 6.5  	&     400     & 17816 \\
150513A &	 &  L  &    159   &     99$\pm$16	  &  $(1.65\pm0.06)10^{-5}$  & 		    	       & GBMcat   &  -  &	 2.1  	&      60     & 17812 \\
150510A &	 &  L  &     52   &   1574$\pm$62	  &  $(2.11\pm0.01)10^{-4}$  & 		    	       & GBMcat   &  -  &	 1.5    &     170     & 17806 \\
150416A &	 &  L  &     33   &    940$\pm$160	  &  $(3.15\pm0.09)10^{-5}$  & 		    	       & GBMcat   &  -  &	   -   	&       -     & \\
150403A &  2.06  &  L  &     22   &    429$\pm$17	  &  $(1.16\pm0.02)10^{-4}$  &$(1.00\pm0.02)10^{54}$& GBMcat  &  Y  &	   5  	&     630     & 17667 \\
150314A &  1.758 &  L  &     11   &    347$\pm$7	  &  $(1.19\pm0.01)10^{-4}$  &$(9.37\pm0.06)10^{53}$& GBMcat  &  Y  &	0.67  	&      78     & 17576 \\
150210A &	 &  L  &     31   &   2943$\pm$221	  &  $(9.78\pm0.08)10^{-5}$  & 		    	       & GBMcat   &  N  &	   1  	&       2     & 17438 \\
150202B &	 &  L  &    167   &    234$\pm$14	  &  $(5.49\pm0.09)10^{-5}$  & 		    	       & GBMcat   &  N  & $<$0.1   &       -     & 17385 \\
150127A &	 &  L  &     53   &   1480$\pm$194	  &  $(4.50\pm0.09)10^{-5}$  & 		    	       & GBMcat   &  N  & $<$0.1   &       -     & 17356 \\
150118B &	 &  L  &     40   &    771$\pm$19	  &  $(2.08\pm0.01)10^{-4}$  & 		    	       & GBMcat   &  -  &	    2   &      50     & 17307 \\
141222A &	 &  S  &    2.8   &   3292$\pm$1080	  &  $(2.15\pm0.04)10^{-5}$  & 		    	       & GBMcat   &  N  &     20  	&     0.1     & 17218 \\
141207A &	 &  L  &     21   &    776$\pm$110	  &  $(5.87\pm0.09)10^{-5}$  &			       & GBMcat   &  Y  &	  5.5   &     734     & [\refcite{arimoto16}] \\
141102A &	 &  S  &    2.6   &    638$\pm$106	  &  $(2.4\pm0.1)10^{-6}$  & 		   	       & GBMcat   &  Y  &	    -   &        -    & 17022 \\
141028A &  2.33  &  L  &     31   &    294$\pm$16	  &  $(6.2\pm0.1)10^{-5}$  & $(6.0\pm0.1)10^{53}$  & GBMcat   &  Y  &	  3.9   &     160     & 16969 \\
140928A &	 &  L  &     18   &    304$^{+80}_{-66}$  &  $(9.1\pm0.5)10^{-5}$  & 	    	    	       & GBMcat   &  Y  &     35   &    3100     & 16847 \\
140810A &	 &  L  &     82   &    289$\pm$9	  &  $(1.76\pm0.02)10^{-4}$  & 		    	       & GBMcat   &  Y  &     16   &    1500     & 16678 \\
140729A &	 &  L  &     56   &    787$\pm$92	  &  $(1.63\pm0.05)10^{-5}$  & 		    	       & GBMcat   &  Y  &	  1.3  	&      44     & 16633 \\
140723A &	 &  L  &     56   &   1205$\pm$258	  &  $(2.43\pm0.06)10^{-5}$  & 		    	       & GBMcat   &  Y  &	  1.8  	&     163     & 16623 \\
140619B &	 &  S  &    2.8   &   1478$\pm$250	  &  $(4.06\pm0.22)10^{-6}$  & 		    	       & GBMcat   &  N  &     24  	&     0.61    & 16420 \\
140523A &	 &  L  &     19   &    280$\pm$6	  &  $(5.67\pm0.05)10^{-5}$  & 		    	       & GBMcat   &  Y  &	  6.5   &      43     & 16322 \\
140402A &	 &  S  &   0.32   &   1054$\pm$271	  &  $(1.3\pm0.1)10^{-6}$    & 		    	       & GBMcat   &  Y  &	  3.5   &     6.8     & 16069 \\

\multicolumn{7}{r}{{\tablename} \thetable{} -- {Continued}}\\[1.5ex]
\hline\\[-1.4ex]
140329A &	 &  L  &     22   &    220$\pm$7	  &  $(1.24\pm0.01)10^{-4}$  & 		    	       & GBMcat   &  Y  &	    -   &        -    & 16047 \\
140323A &	 &  L  &    111   &    143$\pm$5	  &  $(3.25\pm0.06)10^{-5}$  & 		    	       & GBMcat   &  Y  &	  2.5   &     220     & 16034 \\
140219A &	 &  L  &     77   &   2777$\pm$70  	  &  $(1.15\pm0.02)10^{-3}$  & 		    	       & GBMcat   &  Y  &	  1.6  	&    1350     & 15867 \\
140206B &	 &  L  &    147   &    538$\pm$163	  &  $(2.31\pm0.02)10^{-4}$  & 		    	       & GBMcat   &  Y  &	  3.5   &     460     & 15791 \\
140110A &	 &  L  &    9.5   &   1769$\pm$306	  &  $(1.97\pm0.06)10^{-5}$  & 		    	       & GBMcat   &  Y  &	 0.74  	&     160     & 15714 \\
140104B &	 &  L  &    188   &    726$\pm$102	  &  $(5.74\pm0.12)10^{-5}$  & 		    	       & GBMcat   &  Y  &	  1.8   &     800     & 15684 \\
140102A &	 &  L  &    3.6   &    207$\pm$5	  &  $(1.92\pm0.03)10^{-5}$  & 		    	       & GBMcat   &  Y  &	    8   &     520     & 15659 \\
131231A &  0.642 &  L  &     31   &    178$\pm$4	  &  $(2.16\pm0.01)10^{-4}$  &$(2.24\pm0.01)10^{53}$& GBMcat  &  Y  &	  9.7   &     700     & 15640\\ 131216A &	 &  L  &     19   &    542$\pm$448	  &  $(1.62\pm0.04)10^{-5}$  & 		    	       & GBMcat   &  N  &	$<0.1$  &       -     & 15598 \\
131209A &	 &  L  &     14   &    301$\pm$13	  &  $(1.53\pm0.04)10^{-5}$  & 		    	       & GBMcat   &  -  &	    -   &       -     &  \\
131108A &  2.40  &  L  &     18   &    318$\pm$38	  &  $(5.29\pm0.07)10^{-5}$  &$(6.16\pm0.08)10^{53}$ & GBMcat &  Y  &	   1.5  &      66     & 15472 \\
131029A &	 &  L  &    104   &    273$\pm$19	  &  $(2.87\pm0.10)10^{-5}$  & 		    	       & GBMcat   &  -  &	   1.3  &      70     & 15399 \\
131018B &	 &  L  &     40   &	77$\pm$7	  &  $(2.2\pm0.2)10^{-6}$    & 		    	       & GBMcat   &  Y  &      13  &     250     & 15357 \\
131014A &	 &  L  &    3.2   &    326$\pm$3	  &  $(2.57\pm0.01)10^{-4}$  & 		    	       & GBMcat   &  -  &	   1.8  &      15     & 15333 \\
130907A &  1.238 &  L  &    250   &    394$\pm$11	  &  $(7.5\pm0.1)10^{-4}$    & $(3.0\pm0.1)10^{54}$        & GBMcat   &  Y  &    55    &   17200     & [\refcite{tang14}] \\
130828A &	 &  L  &    136   &    233$\pm$13	  &  $(8.18\pm0.13)10^{-5}$  & 		               & GBMcat   &  -  &	   1.5  &     225     & 15128 \\
130821A &	 &  L  &     87   &    305$\pm$15	  &  $(1.19\pm0.01)10^{-4}$  & 		    	       & GBMcat   &  Y  &	     6  &     219     & 15115 \\
130702A &  0.145 &  L  &     59   &		   	  &  $(1.04\pm0.04)10^{-5}$    & $(5.13\pm0.13)10^{50} $    & 14972      &  Y  &	   1.5  &     260     & 14971 \\
130606B &	 &  L  &     52   &    537$\pm$16	  &  $(3.81\pm0.02)10^{-4}$  & 		               & GBMcat   &  -  &	    -   &       -     & 14795 \\
130518A &  2.488 &  L  &     49   &    398$\pm$16	  &  $(1.62\pm0.02)10^{-4}$  &$(2.07\pm0.02)10^{54}$ & GBMcat &  N  &	    -   &       -     & 14675 \\
130504C &	 &  L  &     73   &    538$\pm$25	  &  $(2.41\pm0.02)10^{-4}$  & 		    	       & GBMcat   &  Y  &	     5  &     251     & 14574 \\
130502B &	 &  L  &     24   &    299$\pm$5	  &  $(1.55\pm0.01)10^{-4}$  & 		    	       & GBMcat   &  -  &    30    &     220     & 14532 \\
130427A &  0.34  &  L  &    276   &   1028$\pm$20	  &  $(4.6\pm0.1)10^{-3}$  & $(8.0\pm0.2)10^{53}$  & GBMcat   &  Y  &    95    &     243     & [\refcite{maselli14,ackermann14}]\\
130327B &	 &  L  &     31   &    342$\pm$6	  &  $(5.51\pm0.06)10^{-5}$  & 		               & GBMcat   &  -  &	$>0.1$  &       -     & 14347 \\
130325A &	 &  L  &    6.9   &    207$\pm$20	  &  $(1.41\pm0.03)10^{-5}$  & 		    	       & GBMcat   &  Y  &	     -  &       -     & 14332 \\
130310A &	 &  S  &     16   &   2071$\pm$837	  &  $(3.67\pm0.06)10^{-5}$  & 		    	       & GBMcat   &  Y  &	     3  &  200-800    & 14282 \\
130305A &	 &  L  &     26   &    632$\pm$29	  &  $(9.83\pm0.14)10^{-5}$  & 		               & GBMcat   &  N  &  $<0.1$  &        -    & 14260 \\
130228A &	 &  L  &    111   &    326$\pm$35	  &  $(1.17\pm0.08)10^{-5}$  & 		    	       & GBMcat   &  -  &	  0.69  &      133    & 14258 \\
130206A &	 &  L  &     92   &    133$\pm$34  	  &  $(3.6\pm0.1)10^{-6}$  & 		    	       & GBMcat   &  N  & $<0.075$ &        -    & 14190 \\
121225B &	 &  L  &     58   &    281$\pm$40	  &  $(1.18\pm0.01)10^{-4}$  & 		    	       & GBMcat   &  -  & $<0.075$ &        -    & 14101 \\ 
121011A &	 &  L  &     66   &   1160$\pm$410	  &  $(1.7\pm0.1)10^{-5}$  & 		    	       & GBMcat   &  N  &  $<0.1$  &        -    & 13859 \\ 
120916A &	 &  L  &     53   &    344$\pm$26	  &  $(1.96\pm0.06)10^{-5}$  & 		               & GBMcat   &  -  &  $>0.1$  &        -    & 13777 \\
120911B &	 &  L  &     69   &   1200$\pm$55	  &  $(3.9\pm0.1)10^{-4}$  & 		    	       & 13757      &  -  &	$>0.1$  &        -    & 13756 \\
120830A &	 &  S  &    0.9   &   1157$\pm$127	  &  $(6.88\pm0.21)10^{-6}$  & 		               & GBMcat   &  N  &	   0.5  &      0.8    & 13704 \\
120729A &  0.80  &  L  &     25   &			  &  $(5.1\pm0.3)10^{-6}$  & $(8.54\pm0.43)10^{51}$ & GBMcat   &  -  &	     -  &        -    & \\
120711A &  1.405 &  L  &     44   &   1319$\pm$46	  &  $(4.12\pm0.02)10^{-4}$  &$(1.84\pm0.01)10^{54}$ & GBMcat &  Y  &	     2  &  800-7000   & 13444 \\
120709A &	 &  L  &     27   &    491$\pm$43	  &  $(1.76\pm0.04)10^{-5}$  & 		    	       & GBMcat   &  N  &	   3.1 &       2.0   & 13427 \\
120624B &  2.197 &  L  &    271   &    592$\pm$62	  &  $(3.60\pm0.03)10^{-4}$  &$(3.29\pm0.03)10^{54}$ & GBMcat &  -  &	 $>0.1$ &        -    & 13379 \\ 
120328B &	 &  L  &     30   &    164$\pm$5	  &  $(1.35\pm0.01)10^{-4}$  & 		               & GBMcat   &  -  &	    -   &        -    & \\

\multicolumn{7}{r}{{\tablename} \thetable{} -- {Continued}}\\[1.5ex]
\hline\\[-1.4ex]
120316A &	 &  L  &     27   &    708$\pm$45	  &  $(3.38\pm0.05)10^{-5}$  & 		   	       & GBMcat   &  -  &    1.98  &       27    & 13070 \\
120226A &	 &  L  &     53   &    297$\pm$14	  &  $(9.1\pm0.1)10^{-5}$  & 		    	       & GBMcat   &  -  &	    -   &        -    & \\
120107A &	 &  L  &     23   &    229$\pm$22	  &  $(7.15\pm0.35)10^{-6}$  & 		    	       & GBMcat   &  -  &	   1.8  &      1.18   & 12822,[\refcite{zheng12a}]\\
110731A &  2.83  &  L  &    7.5   &    287$\pm$46	  &  $(2.78\pm0.05)10^{-5}$  &$(4.59\pm0.08)10^{53}$ & GBMcat &  Y  &	   3.4  &       436   & [\refcite{LATcatalog13}] \\
110721A &	 &  L  &     22   &   1000$^{+300}_{-500}$&  $(2.5\pm0.1)10^{-5}$  & 		               & GBMcat   &  N  &	   1.7  &       0.7   & [\refcite{LATcatalog13}]\\
110709A &	 &  L  &     43   &    489$\pm$21	  &  $(5.06\pm0.07)10^{-5}$  & 		    	       & GBMcat   &  N  &	  0.42  &      41.8   & [\refcite{LATcatalog13}] \\
110625A &	 &  L  &     27   &    166$\pm$4	  &  $(9.70\pm0.08)10^{-5}$  & 		    	       & GBMcat   &  Y  &	   2.4  &     272.4   & [\refcite{LATcatalog13}] \\
110529A &	 &  S  &    0.5   &    997$\pm$221	  &  $(5.83\pm0.23)10^{-6}$  & 		    	       & GBMcat   &  -  &	    -   &        -    & [\refcite{LATcatalog13}] \\
110428A &	 &  L  &    5.6   &    164$\pm$11	  &  $(2.57\pm0.04)10^{-5}$  & 		    	       & GBMcat   &  Y  &	   2.6  &      14.8   & [\refcite{LATcatalog13}] \\ 
110328B &	 &  L  &    141   & 932$^{+248}_{-170}$   &  $(4.3\pm0.1)10^{-5}$  & 		    	       & GBMcat   &  N  &	    -   &        -    & [\refcite{LATcatalog13}] \\
110120A &	 &  L  &     28   &    856$\pm$74	  &  $(3.52\pm0.07)10^{-5}$  & 		               & GBMcat   &  Y  &	   1.8  &      72.5   & [\refcite{LATcatalog13}] \\
101123A &	 &  L  &    104   &    485$\pm$17	  &  $(2.22\pm0.02)10^{-4}$  & 		    	       & GBMcat   &  N  &	 $<0.1$ &        -    & [\refcite{LATcatalog13}] \\
101014A &	 &  L  &    449   &    202$\pm$6	  &  $(2.98\pm0.02)10^{-4}$  & 		    	       & GBMcat   &  N  &	 $<0.1$ &        -    & [\refcite{LATcatalog13}] \\
100826A &	 &  L  &     85   &    323$\pm$12 	  &  $(2.8\pm0.1)10^{-4}$  & 		    	       & GBMcat   &  N  &	    -	&        -    & [\refcite{LATcatalog13}] \\
100728A &  1.567 &  L  &    165   &    290$\pm$8	  &  $(1.50\pm0.1)10^{-4}$  & $(9.8\pm0.1)10^{53}$ & GBMcat   &  Y  &    13.5  &      5461   & [\refcite{LATcatalog13}] \\
100724B &	 &  L  &    115   &    358$\pm$8	  &  $(4.2\pm0.1)10^{-4}$  & 		    	       & GBMcat   &  N  &	  0.22  &      61.8   & [\refcite{LATcatalog13}] \\
100620A &	 &  L  &     52   & 324$^{+101}_{-69}$    &  $(7.3\pm0.1)10^{-6}$  & 		    	       & GBMcat   &  -  &	  0.27  &       3.8   & [\refcite{LATcatalog13}] \\
100414A &  1.368 &  L  &     26   &    668$\pm$15	  &  $(1.18\pm0.01)10^{-4}$& $(5.63\pm0.04)10^{53}$& GBMcat   &  Y  &	   4.7  &     288.3   & [\refcite{LATcatalog13}] \\
100325A &	 &  L  &    7.1   &    163$\pm$13	  &  $(3.08\pm0.17)10^{-6}$  & 		    	       & GBMcat   &  N  &	  0.84  &      0.35   & [\refcite{LATcatalog13}] \\
100225A &	 &  L  &     13   &    432$\pm$51	  &  $(7.40\pm0.38)10^{-6}$  & 		    	       & GBMcat   &  N  &	    -   &        -    & \\ 
100116A &	 &  L  &    103   &   1083$\pm$80	  &  $(6.26\pm0.07)10^{-5}$  & 		    	       & GBMcat   &  Y  &    13.1  &     296.4   & [\refcite{LATcatalog13}] \\
091208B &  1.063 &  L  &     12   &    127$\pm$13	  &  $(7.15\pm0.33)10^{-6}$& $(2.13\pm0.10)10^{52}$& GBMcat   &  -  &	   1.2  &       3.4   & [\refcite{LATcatalog13}] \\
091031A &	 &  L  &     34   &    548$\pm$43	  &  $(2.50\pm0.06)10^{-5}$  & 		    	       & GBMcat   &  Y  &	   1.2  &      79.8   & [\refcite{LATcatalog13}] \\
091003A &  0.897 &  L  &     20   &    426$\pm$21	  &  $(4.12\pm0.06)10^{-5}$& $(8.69\pm0.14)10^{52}$& GBMcat   &  Y  &	   2.8  &       6.5   & [\refcite{LATcatalog13}] \\
090926A &  2.106 &  L  &     14   &    304$\pm$13	  &  $(2.32\pm0.01)10^{-4}$& $(2.13\pm0.01)10^{54}$& GBMcat   &  Y  &    19.6  &      24.8   & [\refcite{LATcatalog13}] \\
090902B &  1.822 &  L  &     19   &    788$\pm$58	  &  $(3.86\pm0.02)10^{-4}$& $(3.11\pm0.01)10^{54}$& GBMcat   &  Y  &    33.4  &      81.8   & [\refcite{LATcatalog13}] \\
090720B &	 &  L  &     11   &   1481$\pm$236	  &  $(2.09\pm0.05)10^{-5}$  & 		               & GBMcat   &  N  &	  1.45  &      0.22   & [\refcite{LATcatalog13}] \\
090626A &	 &  L  &     49   &    176$\pm$10	  &  $(1.27\pm0.01)10^{-4}$  & 		    	       & GBMcat   &  Y  &	   2.1  &       112   & [\refcite{LATcatalog13}] \\
090531B &	 &  S  &    0.77  &   1692$\pm$410	  &  $(3.78\pm0.22)10^{-6}$  & 		    	       & GBMcat   &  N  &	   -    &        -    & \\
090510A &  0.903 &  S  &   0.96	  &   4772$\pm$353	  &  $(2.40\pm0.06)10^{-5}$& $(3.83\pm0.09)10^{52}$& GBMcat   &  Y  &    31.3  &       0.8   & [\refcite{LATcatalog13}] \\
090328A &  0.736 &  L  &     62   &    704$\pm$41	  &  $(7.05\pm0.08)10^{-5}$& $(9.97\pm0.11)10^{52}$& GBMcat   &  Y  &	   5.3  &       698   & [\refcite{LATcatalog13}] \\
090323A &  3.57  &  L  &    135   &    471$\pm$74	  &  $(2.06\pm0.02)10^{-4}$& $(4.20\pm0.04)10^{54}$& GBMcat   &  Y  &	   7.5  &     195.4   & [\refcite{LATcatalog13}] \\
090227B &	 &  S  &    0.3   &   1718$\pm$202	  &  $(2.92\pm0.06)10^{-5}$  & 		    	       & GBMcat   &  N  &	  0.24  &      0.48   & [\refcite{LATcatalog13}] \\
090217A &	 &  L  &     33   &    677$\pm$39	  &  $(4.16\pm0.06)10^{-5}$  & 		    	       & GBMcat   &  -  &	   1.2  &     179.1   & [\refcite{LATcatalog13}] \\
081024B &	 &  S  &    0.64  & 1400$^{+1280}_{-620}$   &  $(1.4\pm0.1)10^{-6}$   & 		    	       & GBMcat   &  Y  &	   3.1  &       0.5   & [\refcite{LATcatalog13}] \\
081006A &	 &  L  &    6.4   &  817$^{+827}_{-340}$ 	  &  $(1.96\pm0.20)10^{-6}$  & 		    	       & GBMcat   &  Y  &	   0.8  &      26.5   & [\refcite{LATcatalog13}],[\refcite{zheng12b}] \\
080916C &  4.35  &  L  &     63   &    670$\pm$157	  &  $(1.60\pm0.02)10^{-4}$& $(3.77\pm0.05)10^{54}$& GBMcat   &  Y  &    13.2  &      16.5   & [\refcite{LATcatalog13}] \\
080825C &	 &  L  &     21   &    174$\pm$6	  &  $(5.20\pm0.08)10^{-5}$  & 		    	       & GBMcat   &  -  &	  0.57  &      28.3   & [\refcite{LATcatalog13}] \\[0.6ex] \hline
\end{longtable}
\normalsize
\end{landscape}


\subsection{Lightcurves}
HE lightcurves of different GRBs share some recurrent properties.
The emission above 100\,MeV usually starts with a small delay (of the order of seconds) as compared to the onset of the keV-MeV signal and,
in most cases, lasts much longer than the keV-MeV prompt component.
The presence/absence of HE temporally extended emission (EE) is reported in Table~\ref{tab:table}. Tentatively, we consider the LAT emission temporally extended when it lasts at least 3 times longer than the prompt (sub-MeV) emission. The duration of the prompt emission is identified by its $T_{90}$ (fourth column in Table~\ref{tab:table}). When no information about the possible presence of HE extended emission is reported in the table, either the information is not available or the HE emission lasts longer than the prompt but less than $3\times T_{90}$.

When a long lasting component is present, at late time (i.e., approximately after the end of the prompt phase) its flux decays smoothly, according to a PL behaviour in time\cite{omodei09,ghisellini10}. 
During the prompt emission the temporal behaviour is instead more erratic and with characteristics resembling the prompt emission. While the paucity of events above 1\,GeV makes difficult to assess the presence of flux variability and cross-correlate the LAT above 1\,GeV with GBM lightcurves, hints of LAT variability below 1\,GeV have been found. 
It is likely that at these energies two different components are contributing to the emission: one related to the prompt (and then of internal origin) and the other one lasting much longer, and then probably related to external shocks.
The temporal properties of the HE emission both during and after the prompt phase are summarised in the following sections.

\subsubsection*{Temporal delay}
One of the common features characterising most of the HE detected GRBs is the temporal delay between the onset of the prompt keV-MeV and the detection of the first high-energy photons\cite{omodei09,LATcatalog13,castignani14}. 
Measured delays vary from half a second (for the short GRB~090510) to a few tens of seconds, with typical values between 1 and 10\,s\cite{LATcatalog13}.
This temporal lag behaves in an opposite way as compared to the lag usually observed at lower energies, where prompt pulses peak earlier at higher energies.
While in faint LAT GRBs the initial lack of events can be simply caused by the small number of photons, 
the observation of these delays also in bright LAT events, where the statistics is high, suggests that it has indeed a physical origin, and is not merely an instrumental effect. This behaviour is then considered an intrinsic feature that any physical model aimed at interpreting the GeV emission should be able to explain\cite{castignani14}.
Proposed interpretations will be discussed in section~\ref{sec:origin}.

\subsubsection*{Variability and temporal correlation with keV-MeV prompt emission}
Temporal variability in the HE emission has been measured with a certain statistical significance only in a few GRBs, at early times, when the prompt emission in the sub-MeV energy range is still ongoing.
In these cases, the variability seems to be correlated with the 
temporal variation observed in the prompt keV-MeV emission.

\begin{figure}
{\includegraphics[scale=0.335]{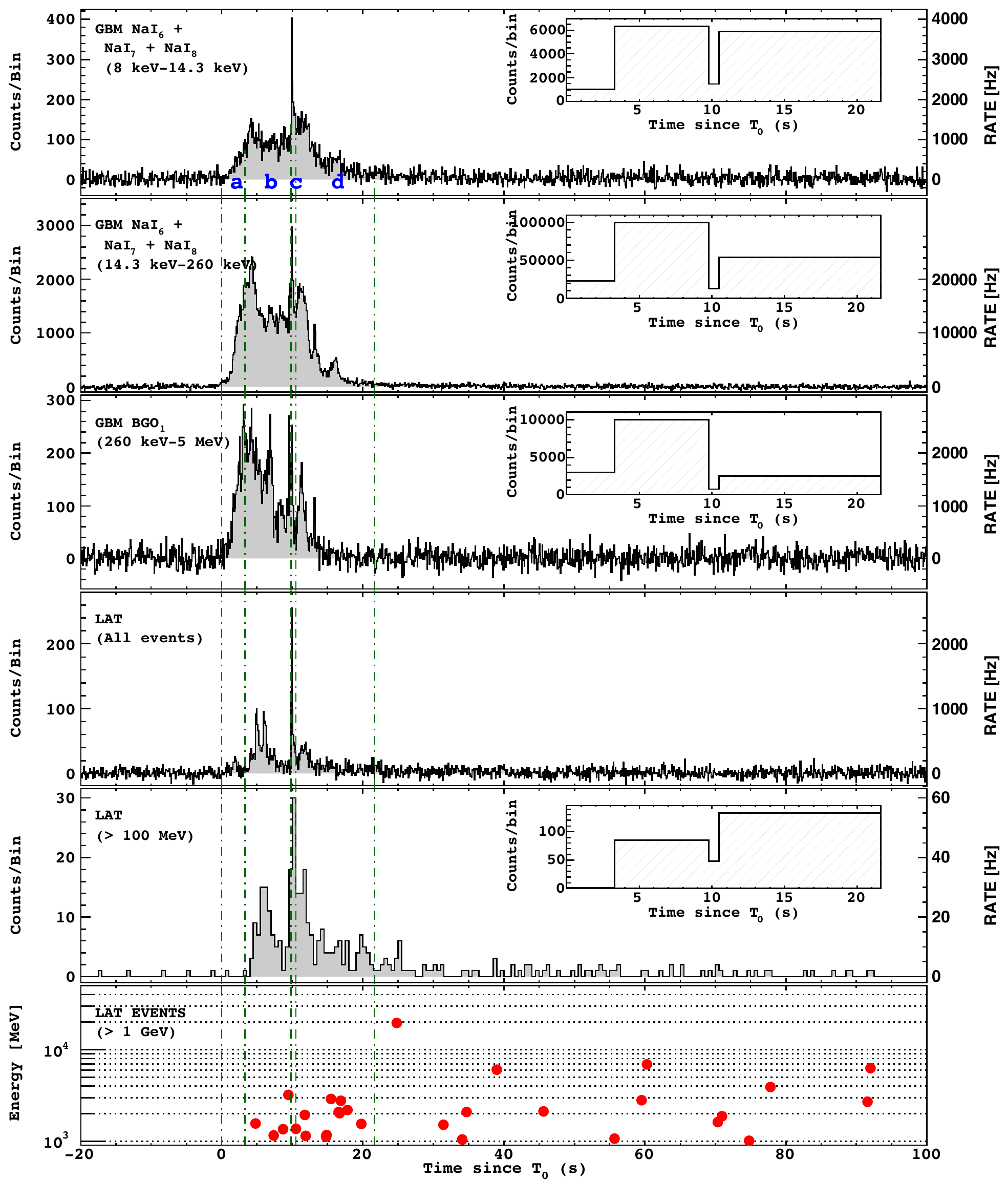}
\includegraphics[scale=0.32]{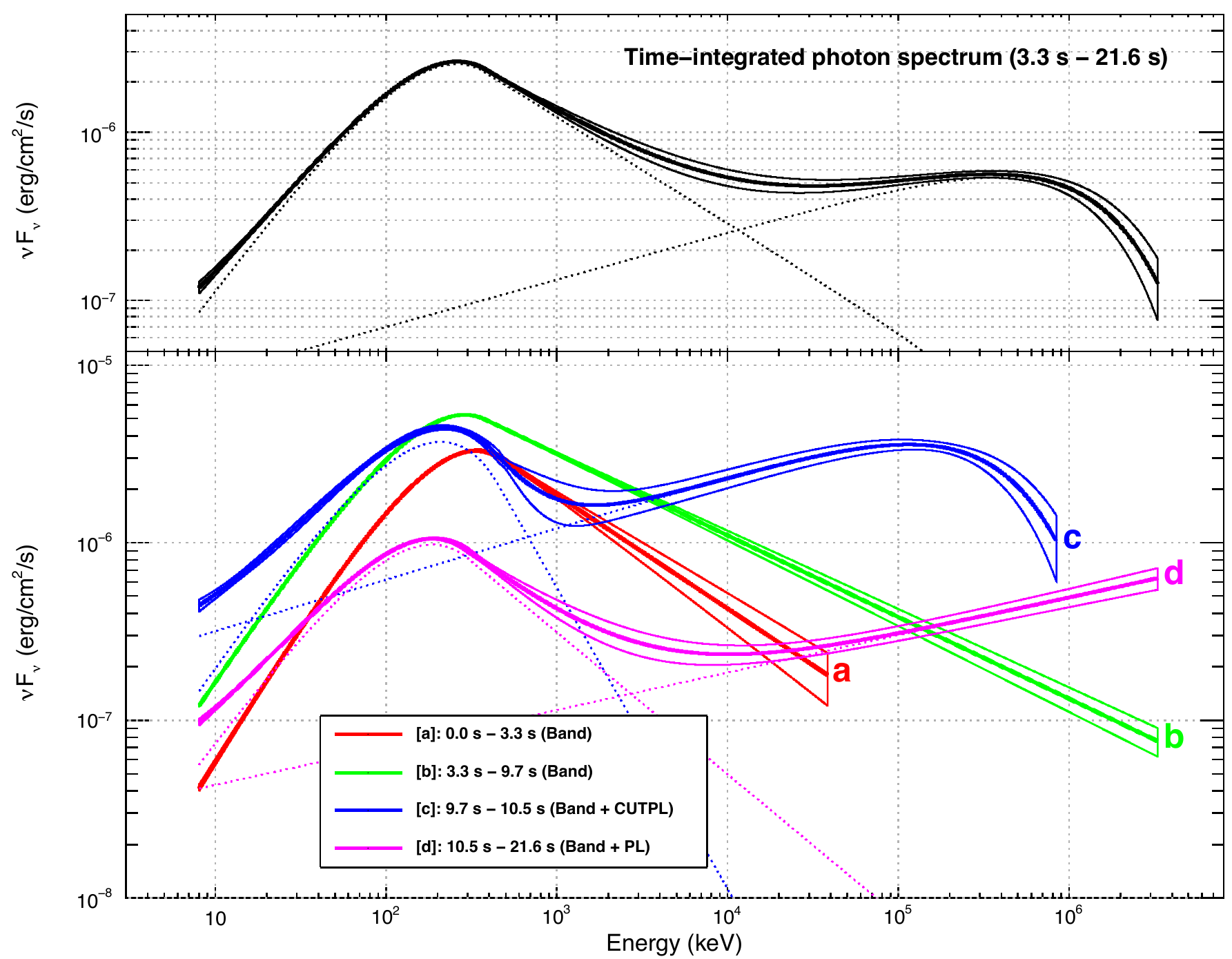}}
\caption{Left: lightcurve of GRB~090926A in different energy bands.
Right: best fit models obtained from the joint GBM-LAT time-integrated (upper panel) and time-resolved (lower panel) spectral analysis of GRB~090926A. Both figures are taken from Ref.~\protect\refcite{ackermann11}.  
\label{fig:090926A}}
\end{figure}

The most noticeable example is represented by GRB~090926A\cite{ackermann11,yassine17}. 
The lightcurve in six different energy ranges is reported in Fig.~\ref{fig:090926A} (left-hand panel).
A simultaneous, short spike in the lightcurve around $T=10$\,s is present across all energies, from several keV to at least 1\,GeV. 
A time resolved spectral analysis (right-hand panel in Fig.~\ref{fig:090926A}) revealed that during the spike (time interval {\it c}) an additional spectral component peaking between 0.1 and 1\,GeV is present. 
Above 1\,GeV the paucity of photons makes difficult to assess the presence of variability and temporal correlation with the emission at lower energies.

A recent analysis on a sample of 5 bright GRBs (080916C, 090510, 090902B, 090926A, and 170214A) is reported in Ref.~\refcite{tang17}. 
In their study, the authors search for a temporal correlation between the HE emission and the keV-MeV emission, selecting GRBs with good photon statistics. 
Moderate evidence of variability and temporal correlation with the prompt component is found in all five cases.
However, this result might be driven by photons below 1\,GeV, where a contamination from the prompt component is likely.

\subsubsection*{Late time component}
The flux of the long-lived LAT emission decays following a PL behaviour in time, which is reminiscent of the smooth temporal decay of the X-ray and optical afterglow fluxes, rather than the variable temporal structure typical of the prompt keV-MeV flux. 
Examples of temporally extended LAT lightcurves above 0.1\,GeV can be found in Fig.~\ref{fig:090510} for GRB~090510 and in Fig.~\ref{fig:clustering} (left-hand panel) for a sample of 10~GRBs with measured redshift.
In all these examples, the LAT emission lasts much longer than the prompt phase, being detected up to $10^2-10^3$\,s after the trigger (up to $\sim3\times10^4$\,s for the very bright event GRB~130427A).

In an early work\cite{ghisellini10} focused on the characterisation of this long lasting component, the authors showed that lightcurves are on average steeper (i.e. $F_{\rm LAT}\propto t^{-1.4}$) than what usually measured at later times in soft X-rays ($F_{\rm X}\propto t^{-1.2}$), and suggested (within the context of afterglow emission) that the blast-wave is initially in a radiative regime, where the dissipated energy is efficiently radiated. 
In the first Fermi LAT GRB catalog\cite{LATcatalog13}, only two GRBs are found to have a steep decay rate, while all the others have more standard decays, with temporal indices between -1 and -1.2. 
Given the paucity of data points and their large errors, the inferred temporal indices strongly depend on the chosen starting time, and different results are found, depending whether the fit is performed starting from the peak of the flux (that usually occurs during the prompt phase) or later, after the end of the prompt, and depending whether the fit is performed on the energy flux or on the photon flux lightcurve.

In three cases (GRB~090510, GRB~090902B, and GRB~090926A), a break from a steeper to a shallower temporal decay provides a better fit to the data\cite{LATcatalog13}, and might indicate a transition from prompt to afterglow dominated emission.
Recently, the lightcurves of 24~LAT GRBs have been analysed in Ref.~\refcite{panaitescu17}, where a possible correlation between the LAT lightcurve brightness and its decay rate (with brighter lightcurves decaying faster than dimmer ones) has been suggested.

\subsection{Spectral properties}\label{subsec:spectral_properties}
A large variety of behaviours is found from the investigation of the spectral properties of the $>$\,30\,MeV emission.
Due to the differences in the temporal behaviours during and after the prompt emission, it is more appropriate to discuss the spectral properties separately for the two temporal domains (i.e., the prompt and the afterglow phases).
Note that during the prompt, joint GBM-LAT spectral fits can be performed, providing spectral coverage over 7 orders of magnitude.

In what follows, I first discuss the spectral properties of the HE emission during the prompt phase, and then I focus on the spectral behaviour after the end of the prompt.
In both temporal domains, different cases are found: single and multiple spectral components, extending at high-energies as a simple PL or displaying high-energy cutoffs.

\subsubsection*{During the prompt phase: continuation of the Band spectrum}
In several cases, the joint GBM-LAT spectral fits performed during the prompt phase revealed that the LAT detection is caused by photons belonging to the high-energy continuation of the Band spectrum. With reference to the first LAT GRB catalog\cite{LATcatalog13}, examples of GRBs where the prompt LAT emission lies on the spectral PL continuation of the keV-MeV prompt emission are GRB~080825C, 090217, 090323, 090720B, 091003, 100116A, 100414A, 110709A, 110721A. These cases are interesting due to the possibility to infer a lower limit on the bulk Lorentz factor.

In a small fraction of cases in which the HE radiation is the spectral continuation of the sub-MeV component, a clear cutoff in the LAT energy range has been identified.
Ref.~\refcite{vianello17} reported the existence of two remarkable cases (GRB~100724B and GRB~160509A) where the combined GBM-LAT data show that the prompt keV-MeV spectrum has an evident softening (located at 20-60\,MeV and 80-150\,MeV, respectively), well modelled by an exponential cutoff. 
This feature in the high-energy part of the prompt spectrum is expected as the result of opacity to pair production within the source.
An alternate explanation invokes the maximum synchrotron photon energy that the radiative mechanism is able to achieve (reflecting the maximum electron energy achieved during particle acceleration). 
However, this limit is expected to be at higher energies (see a discussion and derivation in sections~\ref{sec:HEphotons} and \ref{sec:synchro}) and an interpretation of the cutoff as caused by pair production is more likely.
Under this last hypothesis, the cutoffs measured in GRB~100724B and GRB~160509A lead to estimated bulk Lorentz factors in the range $\Gamma=100-300$ for both GRBs\cite{vianello17}.

Further cases of HE cutoffs in the prompt spectra have been identified with a more indirect method, namely the non-detection of emission by the LAT. LAT flux upper limits compared to extrapolations of the Band component revealed the necessity for a spectral break/cutoff between the GBM and the LAT at least in six cases, out of a sample of 288 GBM GRBs\cite{LATcatalogUL12}. This use of flux upper limits will be discussed in detail in section~\ref{sec:lack}.

\subsubsection*{During the prompt phase: presence of an additional component}
In some cases, such as 
GRB~080916C\cite{LATcatalog13}, 
the short burst GRB 090510\cite{ackermann10}, 
GRB 090902B\cite{abdo09}, 
GRB~090926A\cite{ackermann11,yassine17}, 
GRB~110731A\cite{ackermann13b}, and
GRB~130427A\cite{ackermann14}
a second, harder spectral component (in addition to the common Band component) has been clearly identified during the prompt phase.
The quality of the data usually does not allow a detailed study of its spectral shape, and a simple PL provides an acceptable fit. 
An example is provided by the spectrum labelled as $d$, in the time-resolved analysis of GRB~090926A (see Fig.~\ref{fig:090926A}, right-hand panel).
Typically, the PL photon index of this component is  $\gtrsim-2$, that is harder than the high-energy part of the prompt keV-MeV spectrum.
In one case (GRB~090926A) the additional component is modeled by a PL with an exponential cutoff\cite{ackermann11}, both in the time-integrated analysis and in the time resolved spectrum accumulated over 0.8\,s around the peak of the emission (see the right-hand panel in Fig.~\ref{fig:090926A}). The cutoff energy is located at 1.4\,GeV in the time-integrated spectrum and at 0.4\,GeV at the peak of the emission.

In general, the additional component is energetically sub-dominant (from 10 to 50\% of the total prompt energy, see section~\ref{sec:energetics}).
Similarly to the main Band component, it evolves with time.
In at least two cases (GRB~090902B\cite{abdo09} and GRB~090510\cite{ackermann10}), the additional component seems to be correlated with an excess at low energy, as the extra PL exceeds the Band function not only in the LAT range but also below $\sim10-50$\,keV. 
While in some GRBs the presence of an additional component during the prompt is firmly established, in other cases its presence is identified only in the time-integrated analysis, while the time-resolved shows no strong evidence for the existence of this component.  
 
\subsubsection*{After the end of the prompt}
After the end of the keV-MeV prompt emission, the LAT photon spectrum is well modelled by a simple PL component with spectral index around -2 and no noticeable spectral evolution in time. 
In Ref.~\refcite{tam13} the simultaneous presence of two spectral components during the afterglow emission has been claimed. The harder one, dominating the emission in the high-energy part of the LAT energy range, has been interpreted as synchrotron-self Compton (SSC) radiation.
However, in Ref.~\refcite{ackermann14} the authors argue that there is no statistically significant evidence for the two-component fit, and the observed concave spectral shape might be the result of spectral evolution.
The presence of an SSC component would explain the detection of photons with energies in excess of the maximum energy achievable by synchrotron photons (see figure~\ref{fig:HEph_arrivaltime} and the next paragraph), as it will be discussed in section~\ref{sec:ssc}.

\subsection{High-energy photons}\label{sec:HEphotons}
\begin{figure}
\centering
\includegraphics[scale=0.3]{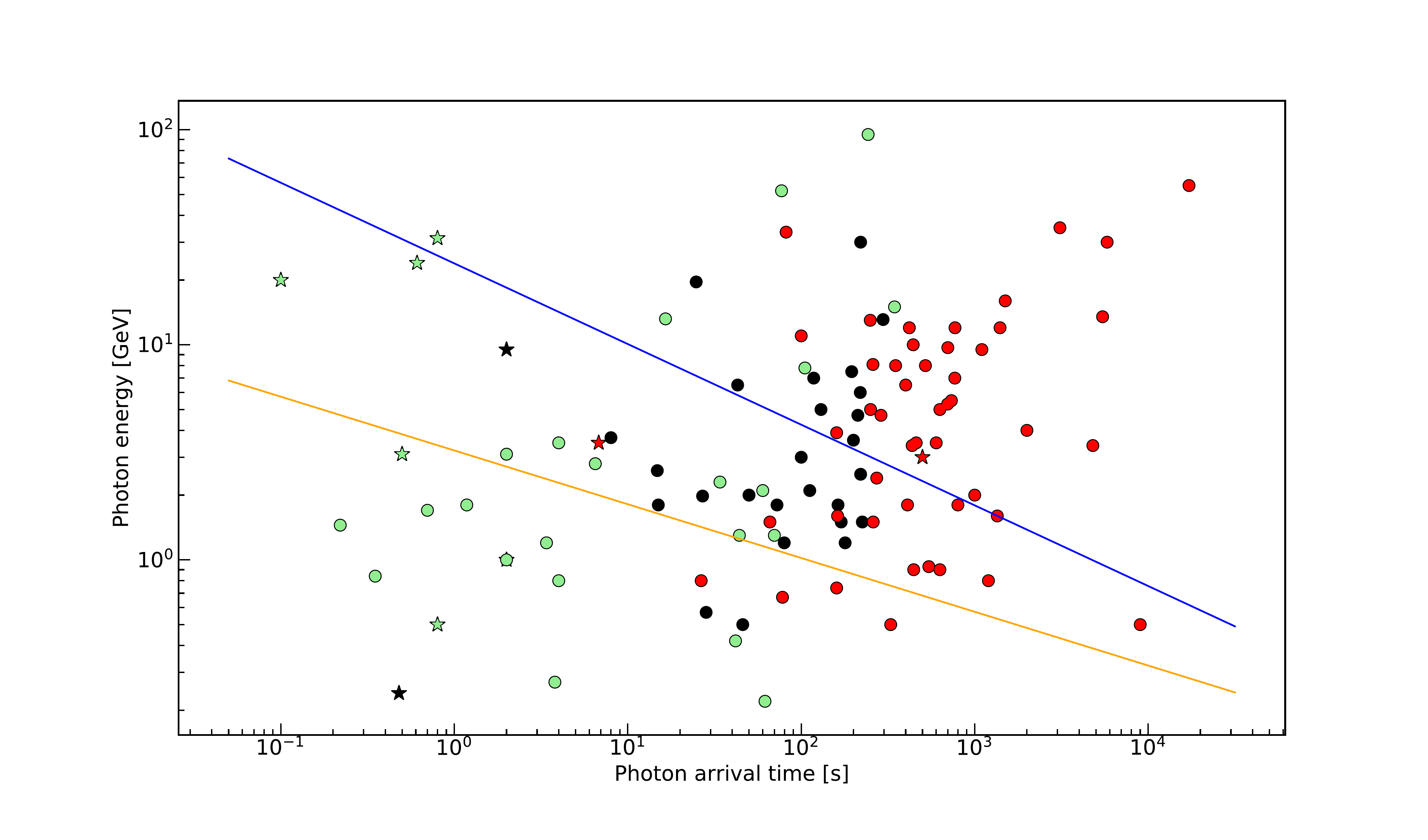}
\caption{For each GRB detected by LAT, the energy of the most energetic photon is plotted versus its arrival time. Energies and times are in the observer frame. Red symbols denote photons arriving well after the end of the prompt phase. Green symbols denote photons arrived during the prompt phase. Black symbols mark intermediate cases where the highest energy photon arrives after the end of the prompt emission, but on a comparable timescale. Short GRBs are marked with stars. The blue and orange lines show the limit obtained from $E^{\rm obs}_{\rm syn,max}=50$\,MeV$\times \Gamma/(1+z)$, where $\Gamma$ has been estimated assuming $\eta_\gamma=0.2$, fluence $F=4\times10^{-6}$\,erg\,s$^{-1}$, $z=1$, for a homogeneous medium with density $n=1$\,cm$^{-3}$ (blue line) and for a wind-like medium with density $n=3\times10^{35} R^{-2}$cm$^{-1}$ (orange line).
\label{fig:HEph_arrivaltime}}
\end{figure}
The LAT has observed several photons with energies well above 10\,GeV (up to 95\,GeV) coming from bright GRBs, which in the case of large redshift translate into photons exceeding 100\,GeV in the rest frame of the progenitor. 
This result poses a big challenge for their interpretation as synchrotron photons from electrons accelerated at the external shock. 
The emitting electrons, indeed, can be accelerated up to a maximum energy above which the timescale for radiative synchrotron losses becomes shorter than the acceleration timescale. 
Under simple assumptions, the estimated maximum energy for synchrotron photons is around $\sim50\,$MeV in the comoving frame, corresponding to $E^{\rm obs}_{\rm syn,max}\simeq50\,$MeV$\times \Gamma /(1+z)$ in the observer frame (a derivation and discussion of the underlying assumptions is provided in section~\ref{sec:synchro}). 
When the jet starts decelerating, the Lorentz factor $\Gamma$ decreases with time, implying that also $E^{\rm obs}_{\rm syn,max}$ decreases. The detection of photons with energy in excess of $\sim$\,1-10\,GeV at late time (i.e., well after the end of the prompt and the beginning of the deceleration) represents then a challenge, either for particle acceleration models or for the interpretation of these photons as synchrotron radiation.

Fig.~\ref{fig:HEph_arrivaltime} shows, for each GRB, the energy of the most energetic photon detected by the LAT versus its arrival time (estimated from the trigger time). 
Both quantities are in the observer frame and are listed in Table~\ref{tab:table}. 
Green symbols denote photons detected when the prompt emission was still ongoing (i.e., within the $T_{90}$), red symbols refer to photons detected well after the end of the prompt phase (i.e., with arrival time at least 3 times larger than $T_{90}$), black symbols refer to intermediate cases (arrival time between $T_{90}$ and $3\times T_{90}$).
Short GRBs are marked with star symbols, and long GRBs with circles.
Assuming that the Lorentz factor during the afterglow phase decreases according to the equation derived by Blandford \& McKee\cite{blandford76}, the two solid lines show the limiting value of the photon energy $E^{\rm obs}_{\rm syn,max}$ for redshift $z=1$, bolometric fluence $F=4\times10^{-6}$\,erg\,cm$^{-2}$, prompt efficiency $\eta_\gamma=0.2$, and for two different assumptions on the radial profile of the external medium: constant, with $n=1$\,cm$^{-3}$ (blue line) and wind-like, with $n=3\times10^{35}\,R^{-2}$\,cm$^{-1}$ (orange line).
These curves are valid during the deceleration phase. Before the deceleration, when the Lorentz factor is constant ($\Gamma=\Gamma_0$), the maximum energy is also constant in time, and is simply given by $E^{\rm obs}_{\rm syn,max}\simeq50\,$MeV$\times\Gamma_0 /(1+z)$. 

A large fraction of photons lie above the limiting curves. Even though the disagreement can be reduced by using a different choice of the parameters for the estimate of the limiting curves (see section~\ref{sec:synchro}), very extreme and contrived values are required to account for most of the photons in a simple shock acceleration/synchrotron radiation scenario.

The tension between observations and synchrotron model can be eased by considering a more complicated scenario for particle acceleration, for example where the magnetic field decays downstream on a length scale smaller than the region occupied by shocked particles\cite{kumar12}. 
Another possibility is to abandon the idea that these photons are produced by synchrotron radiation and invoke a separated component (e.g., inverse Compton) dominating the LAT emission (or, at least, the high-energy part of the LAT energy range).
The interpretation of the origin of these photons will be discussed in section~\ref{sec:origin}.

\subsection{Short bursts}
Among the 141 GRBs detected by the LAT in the first nine and a half years of operations (till December 2017), 13 are tentatively classified as short GRBs, based on their $T_{90}$ and spectral hardness.
This corresponds to a detection rate of 9\%, smaller than the overall GBM short GRB detection rate (16\%).
The redshift has been measured only in one case (GRB~090510, $z$=0.903, Fig.~\ref{fig:detection_redshift}, right-hand panel).

\begin{figure}
\centering
\includegraphics[scale=0.55]{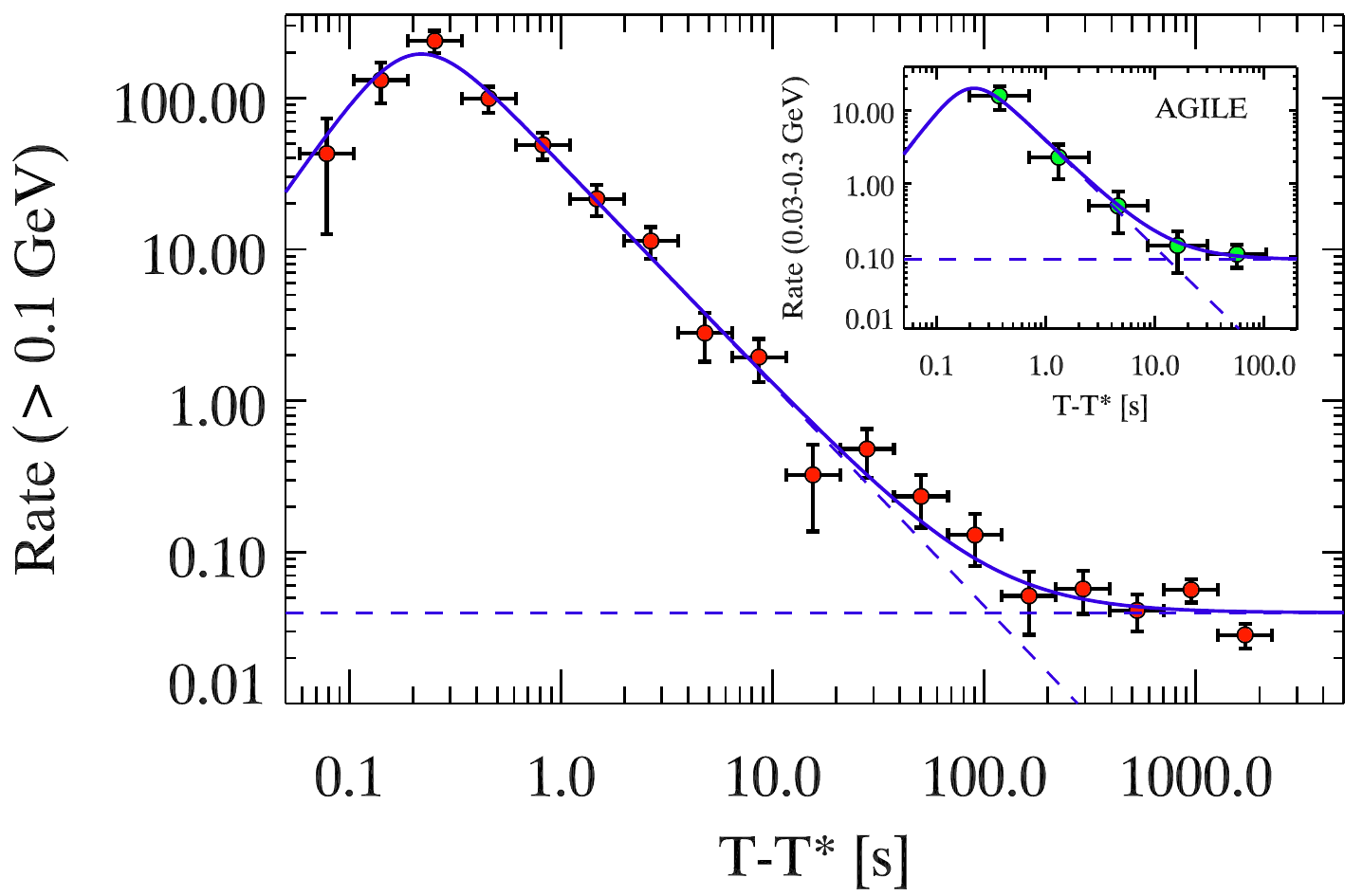}
\caption{{\it Fermi} and {\it AGILE} (in the inset) high-energy lightcurve of the short GRB~090510. The solid line is the best fit to the data, obtained as the sum of a smoothly broken power-law and a constant, background component. The first index of the smoothly broken PL has been fixed to $\beta_1=2$, while the index after the peak has best fit value $\beta_2=-1.46^{+0.06}_{-0.03}$. From Ref.~\protect\refcite{ghirlanda10}.
\label{fig:090510}}
\end{figure}

The behaviour of their HE component is very similar to the behaviour displayed by the HE emission of long GRBs.
Also short GRBs can display a temporally extended emission decaying in time as a PL\cite{ghirlanda10} (see the LAT and {\it AGILE} lightcurve of GRB~090510 in Fig.~\ref{fig:090510}).
Photons at tens of GeV have been detected also from short GRBs (see star symbols in Fig.~\ref{fig:HEph_arrivaltime}): in particular a 30.5$^{+5.8}_{-2.6}$\,GeV photon has been detected during the prompt phase of GRB~090510\cite{ackermann10}.
The only difference identified so far is in the ratio between the LAT and the GBM fluences, that in short GRBs is systematically larger as compared to long events, and is around 1 or more (Fig.~\ref{fig:fluence-fluence}).

\subsection{Energetics}\label{sec:energetics}
From the point of view of the energy radiated during the prompt emission, LAT GRBs are among the most energetic events.
Fig.~\ref{fig:fluence_eiso_redshift} shows, for the subsample of GBM bursts with measured redshift, the prompt fluence and \eiso\ versus redshift. 
LAT detected GRBs (denoted with star symbols) have systematically higher fluences and \eiso\ as compared to the whole GBM population.
Their sub-MeV prompt fluence and \eiso\ are reported in Table~\ref{tab:table}.

\begin{figure}[h]
{\includegraphics[scale=0.32]{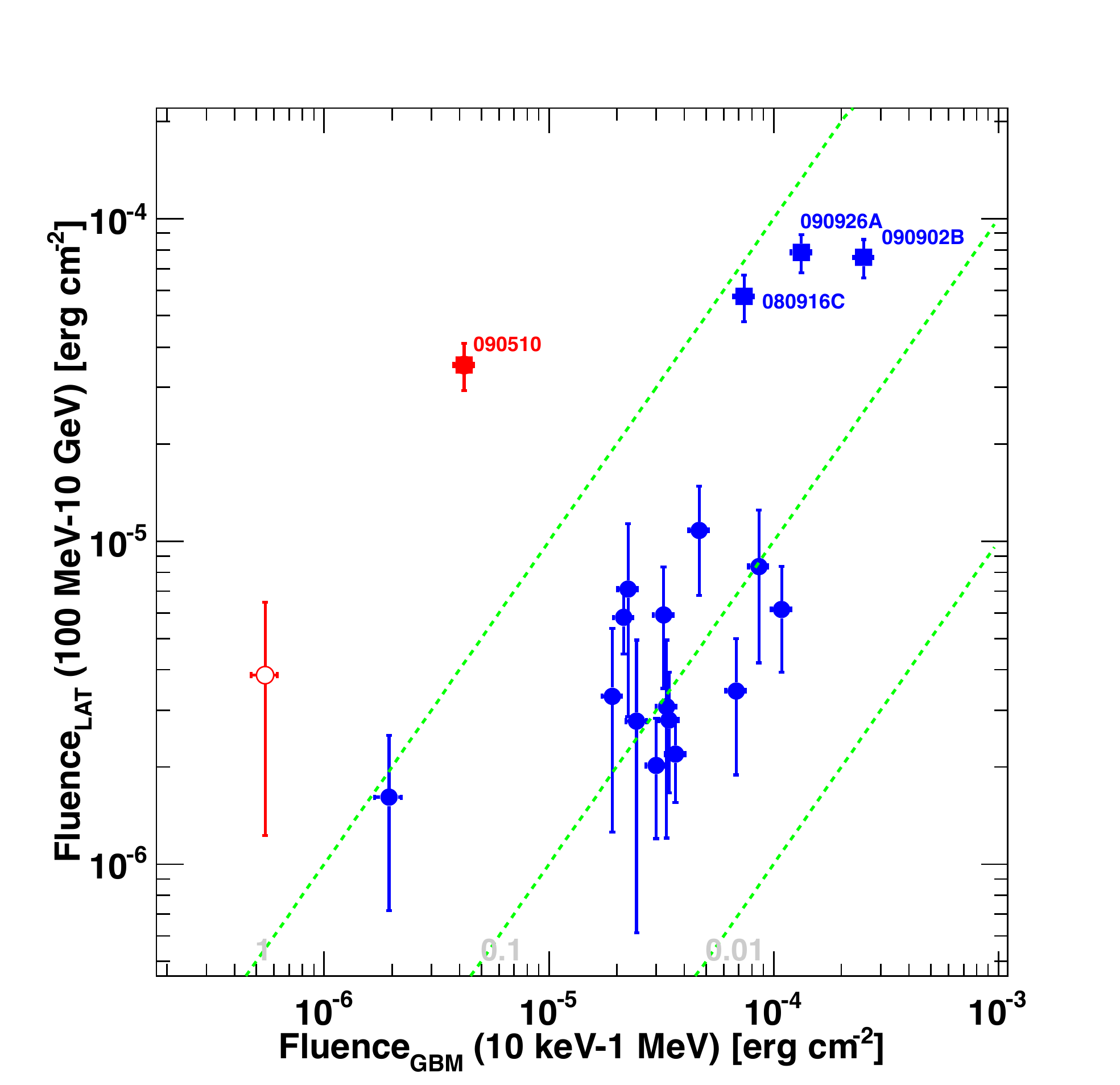}
\includegraphics[scale=0.4]{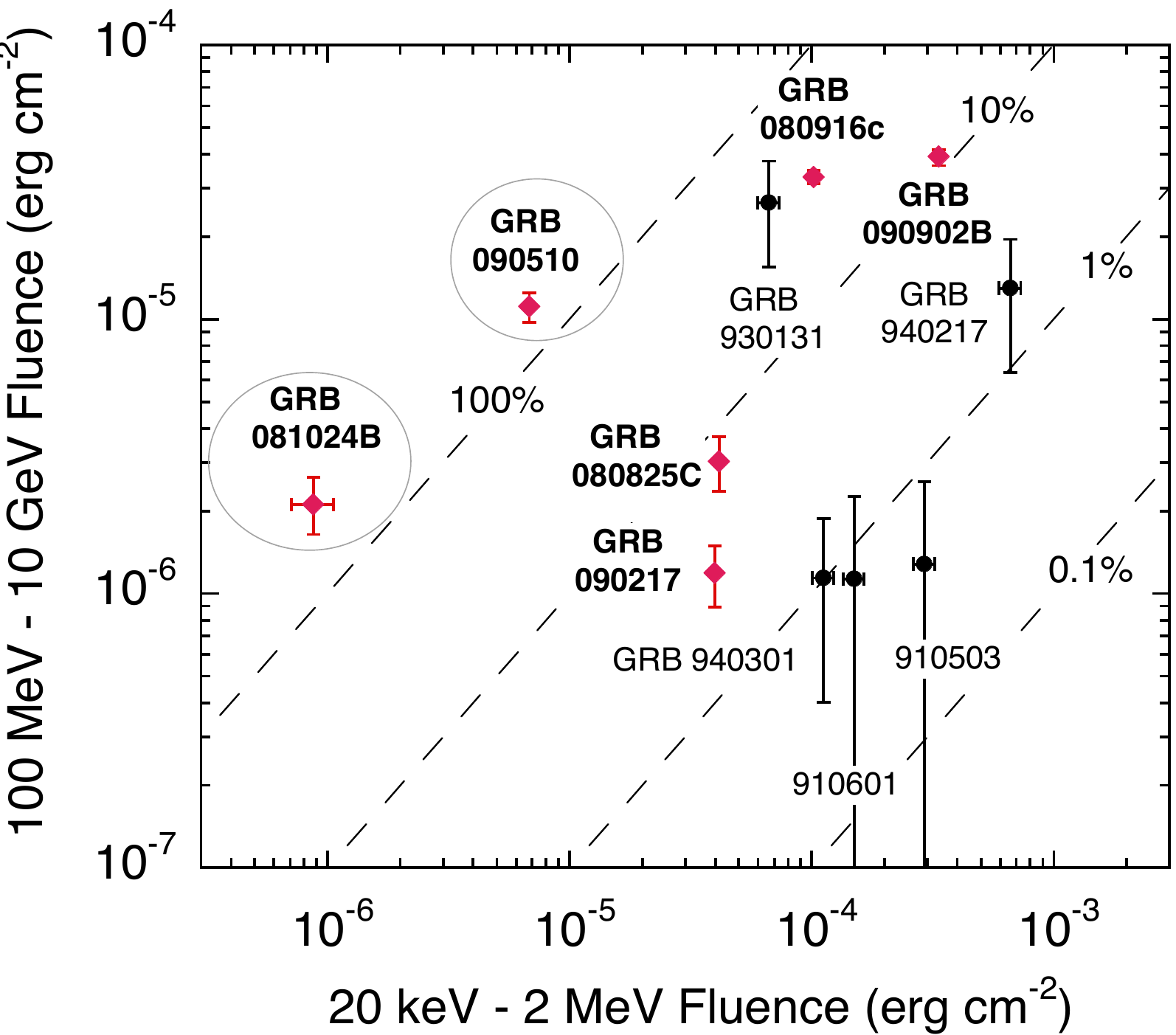}}
\caption{Left: fluence measured by the LAT in the energy range 100\,MeV-10\,GeV vs. fluence measured by the GBM in the energy range 10\,keV-1\,MeV. The LAT fluence is integrated in time over the whole LAT emission, and the GBM fluence is integrated over the duration of the prompt emission. Dashed lines correspond to Fluence$_{\rm LAT}=1,0.1,0.01 \times$Fluence$_{\rm GBM}$. Blue symbols denote long GRBs, while red symbols denote short GRBs (from Ref.~\protect\refcite{LATcatalog13}).
Right: a similar diagram including also EGRET GRBs (black filled circles, from Ref.~\protect\refcite{dermer10}).
\label{fig:fluence-fluence}}
\end{figure}

For the sample of LAT detected GRBs, one can compare the energy radiated in the keV-MeV range with the energy radiated in the LAT energy range. 
Examples of fluence-fluence diagrams are shown in Fig.~\ref{fig:fluence-fluence}.
The left-hand panel compares the LAT fluence integrated in time over the temporally extended emission, with the prompt fluence, for the GRBs included in the First {\it Fermi}-LAT GRB catalog\cite{LATcatalog13}. Red symbols refer to short GRBs, while blue symbols are used for long events. In general the ratio between LAT and GBM fluences is smaller than 1 for long GRBs (the typical value being around 0.1), and larger than 1 for short GRBs.
The right-hand panel shows a similar plot where also EGRET GRBs are included\cite{dermer10}.

\subsection{The X-ray and optical afterglow of LAT GRBs}
A systematic study published in Ref.~\refcite{racusin11} addresses the question whether the X-ray and optical afterglows of LAT bursts have properties typical of the general  population or they present some peculiarities.
To answer this question, the authors considered three different samples: GRBs detected by BAT only, BAT and GBM, GBM and LAT, searching for differences in their afterglow emission. 

\begin{figure}[t]
{\includegraphics[scale=0.242]{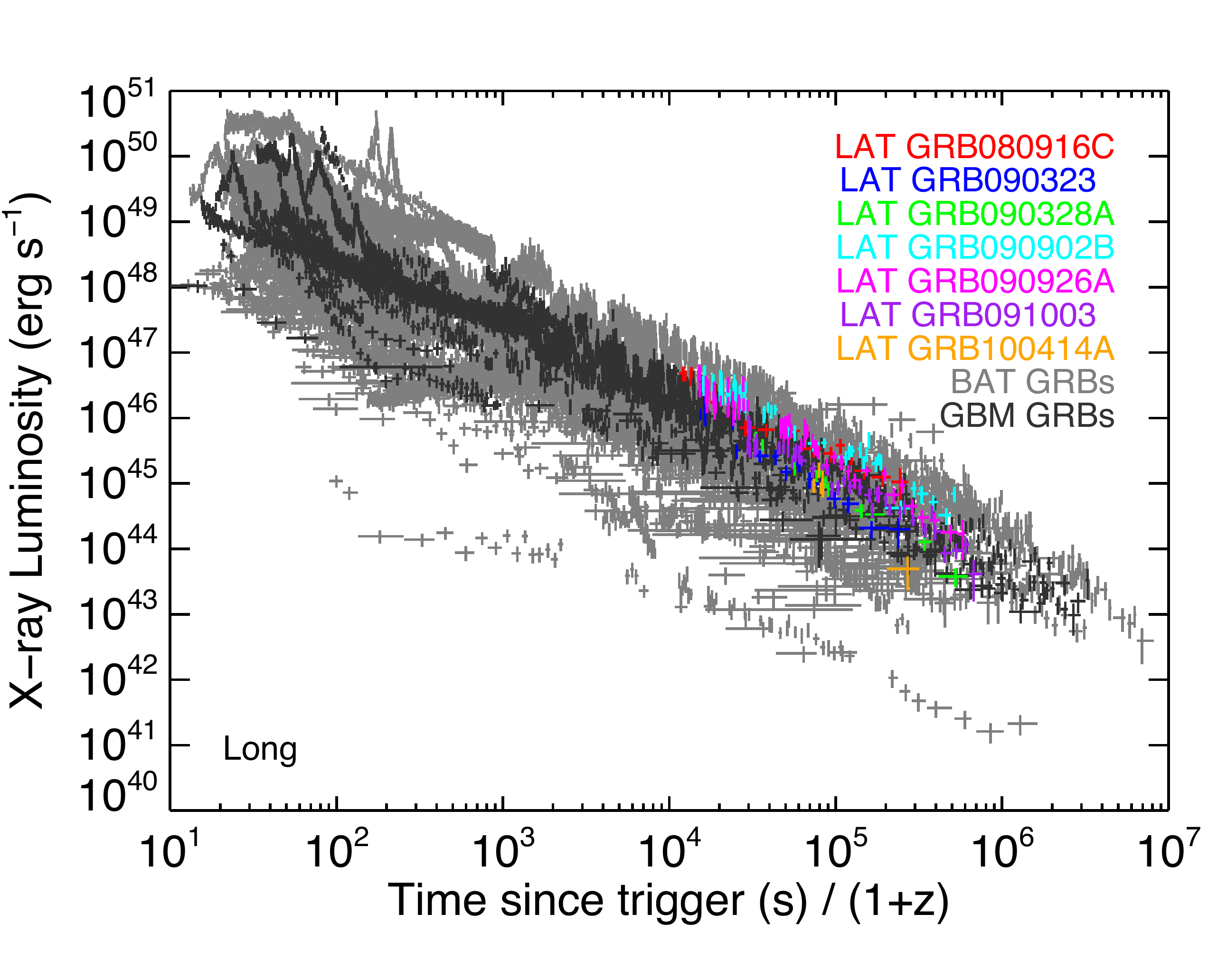}
\includegraphics[scale=0.242]{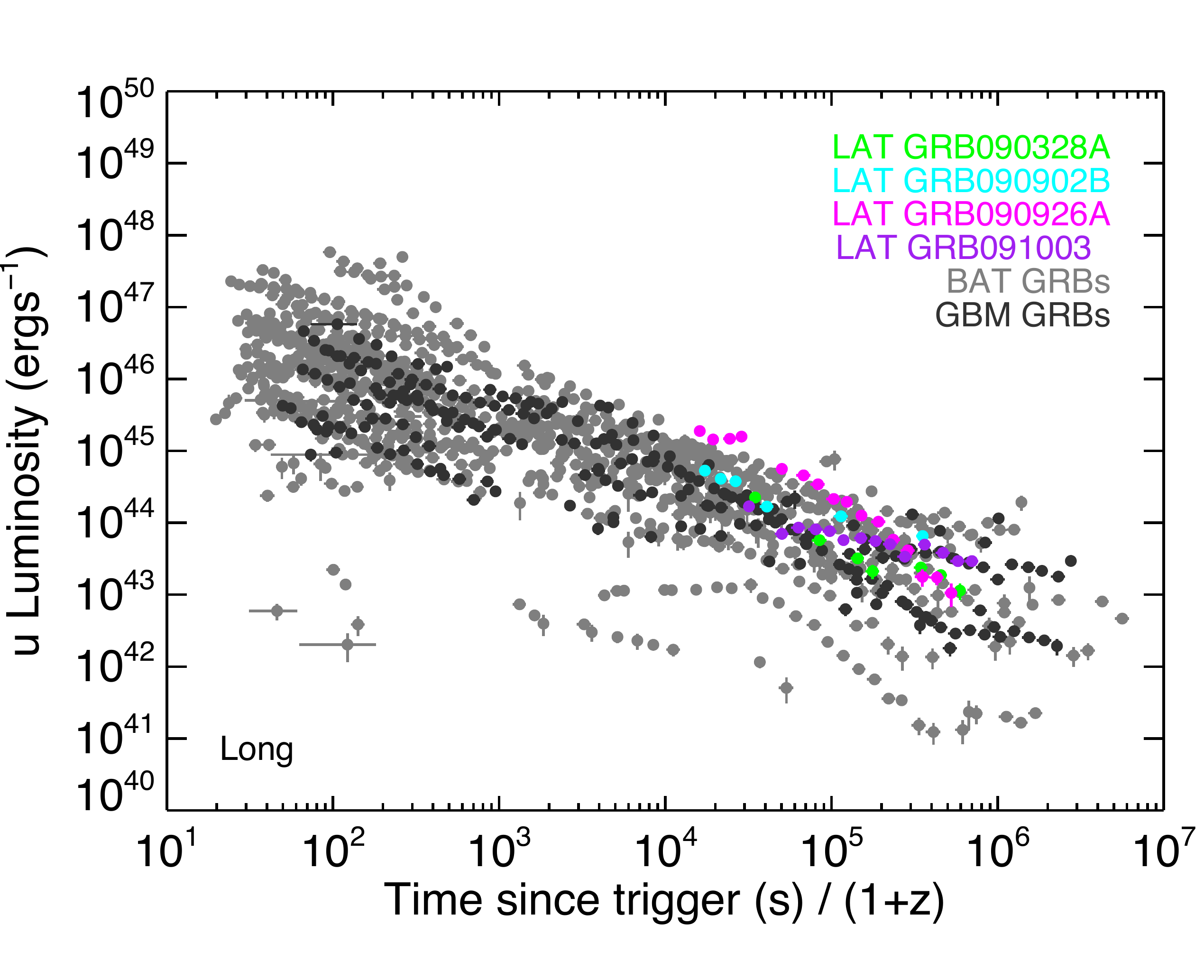}}
\caption{X-ray (left) and optical (right) afterglow lightcurves of long GRBs detected by the LAT (coloured symbols), GBM (dark grey) and BAT only (light grey). From Ref.~\protect\refcite{racusin11}.}
\label{fig:afterglows}
\end{figure}

For the sample of long GRBs, the X-ray and optical lightcurves of the three different samples are shown in Fig.~\ref{fig:afterglows} (left-hand and right-hand panels, respectively).
At the time of the study, early time afterglow observations were available only for the short GRB~090510.
No significant differences emerged from this investigation: the afterglow emission of LAT-detected GRBs seems to share the same properties of the more general population of GRBs, both in terms of temporal decay rate and spectral indices.

The most evident difference is in the ratio between the energy released during the prompt emission phase $E_{\rm \gamma,iso}$ and the luminosity of the X-ray afterglow: this ratio is larger for LAT bursts as compared to the BAT-only sample. This suggests that either the mechanism for prompt emission is more efficient in bursts with HE emission, or, conversely, their X-ray afterglows are suppressed. The origin of this suppression has been investigated in Ref.~\refcite{beniamini15}, and will be discussed in section~\ref{sec:origin_external}.

\section{Origin of the high-energy emission}\label{sec:origin}
Many different models have been invoked to explain the HE emission detected from GRBs, including synchrotron and SSC radiation from electrons accelerated within the jet and/or in the (forward or reverse) external shocks, and hadronic models.
While the initial, large variety of proposed models has been narrowed down by observations, it is difficult to argue in favour of one single origin able to explain all the observations, from early to late time, from 30\,MeV to 100\,GeV.
The coexistence of different components, one of internal origin (able to explain the early time observations and the correlation with the prompt emission) and one of external origin (explaining the long-lasting, smooth behaviour) is the most plausible scenario.

This section discusses separately the origin of the HE emission from processes occurring within the jet (giving rise to the prompt component) and the origin of the HE emission produced in external shocks (responsible for the long-lasting radiation).
The discussion focuses on synchrotron and SSC leptonic models, since there is a general consensus that they are able to explain all the most recurrent observed features.

\subsection{Prompt component: internal origin}
The temporal properties of the HE emission during the prompt phase indicate that this emission has an internal origin, or at least a dominant contribution from photons of internal origin.
Data analysis\cite{zhang11} and theoretical modeling\cite{maxham11,he11} suggest that, in general, the contribution from an afterglow component during the prompt is sub-dominant, and the external shock emission starts to dominate the HE radiation at the end of the prompt phase. 
In support of this picture, some GRBs show a steep-to-shallow decay in their GeV lightcurve, which has been interpreted has the transition from prompt to afterglow dominated emission\cite{LATcatalog13}. In this section, I review the mechanisms that can produce a HE radiation of internal origin. 

\subsubsection*{Single spectral component}
The emission detected above few tens of MeV during the prompt phase might simply be the high-energy continuation of the prompt spectrum.
This would naturally produce a HE lightcurve with temporal characteristics resembling those of the emission in the keV-MeV energy range.
The prompt spectrum can extend to the LAT energy range as a simple PL or can display a softening, depending on the position of the cutoff that is expected to arise because of photon-photon annichilation. 
The interest in these cases is provided by the possibility to infer the bulk Lorentz factor from the measure of the cutoff energy.

In Ref.~\refcite{vianello17}, a standard model (applicable for example to pair production affecting the synchrotron radiation in internal shocks) and a photospheric model in a highly magnetized outflow are considered and applied to two GRBs with a well defined exponential cutoff at 20-150\,MeV. Since the cutoff energy is clearly identified also in the time resolved analysis, the evolution of $\Gamma$ with time and a tentative correlation with the flux are discussed.
The estimated values of $\Gamma$ are in the range 100-400, for both GRBs and for both theoretical models.
These estimates are similar to the upper limits derived in Ref.~\refcite{LATcatalogUL12} for a sample of six GRBs with no detection in the LAT range of sensitivity (see section~\ref{sec:lack}).

In these models, a delay in the onset of the HE emission as compared to the onset of the keV radiation can be explained by different conditions at the emitting region, leading to a flux increase of the Band component, a hardening of the Band component (e.g., due to an increase of the peak energy $E_{\rm peak}$ and/or a harder high-energy spectral index, caused for example by variations in the injected particle spectrum), or an increase of the cutoff energy due to a lower opacity to pair production.

\subsubsection*{Extra-component}
The GRB prompt emission in the keV-MeV domain is likely related to synchrotron emission from electrons accelerated after energy dissipation within the jet, in internal shocks or magnetic reconnection events.
Even though a synchrotron interpretation faces some difficulties in explaining the properties of the observed spectra, recent progresses in the characterisation of prompt spectra strongly support synchrotron radiation as the dominant contribution to prompt emission \cite{oganesyan17a,oganesyan17b,ravasio17}.
In this context, a HE additional component during the prompt phase can arise from SSC radiation, i.e., inverse Compton scattered synchrotron photons. 

At least in one case (GRB~090926A) the HE spectral component is better described by a PL with a high-energy cutoff rather than a simple PL function\cite{ackermann11,yassine17}. 
In this case the spectral peak may correspond to the peak of the SSC emission (likely affected by the Klein-Nishina suppression of inverse Compton scatterings) or, alternatively, may be caused by $\gamma$-$\gamma$ annihilation.
In the SSC scenario, the temporal delay can be explained if inverse Compton scattering occurs in Klein-Nishina regime at early times, while at later times, conditions in the shocked region are such that the scatterings enter the Thomson regime\cite{bosnjak09,daigne11,daigne12}. 
It is not clear, however, if the predicted delay - which should be comparable with the duration of the MeV pulse - are too short to satisfactorily explain the observations\cite{asano11}.

\subsubsection*{Low-energy excess}
An SSC component of internal origin can not explain the flux excess which is sometimes observed below $\lesssim$\,50\,keV (GRB~090510, GRB~090902B and time resolved spectrum of GRB~090926A, see Fig.~\ref{fig:090926A}). 
From the point of view of empirical modelling, this flux excess is compatible with the extension of the high-energy extra-PL component to the lowest energies. 
A simple way to explain these two excesses as a single component is to invoke the contribution of synchrotron from the early afterglow. This might however be in contradiction with a possible temporal correlation between the flux of the extra-component and the flux of the prompt keV-MeV component.
Alternatively, in the synchrotron-SSC internal scenario, the assumption that the low and high-energy excesses have the same origin should be relaxed and a further component must be invoked in order to model the low-energy excess.
Late synchrotron emission from the cooled electrons has been proposed as a viable mechanism for the soft spectrum below 50\,keV\cite{asano11}.

In GRB~090902B, where the low-energy excess is most evident, the hardness and narrowness of the prompt spectrum point to a photospheric emission for the main prompt component. In this case the non-thermal PL component from keV to GeV energies may be synchrotron/SSC radiation from electrons accelerated at larger radii\cite{peer12}. 
In Ref.~\refcite{toma11}, the main emission from GRB~090902B is interpreted as photospheric emission, while Compton up-scattered photospheric emission is invoked to explain the $>10$\,MeV radiation, and synchrotron from internal shocks to explain the excess at low energies. In this interpretation, the high-energy component and the low-energy excess are two separated components, both produced at the same internal shock.

Hadronic models, (for example synchrotron radiation from protons\cite{razzaque09} or photohadronic interactions\cite{asano09}, can also explain a 
component producing a low energy excess via direct and/or cascade radiation (e.g., synchrotron from secondary pairs). 
The main disadvantage of hadronic models, however, is in their total energy budget, that is required to be extremely high. 
Energetic requirements can be relaxed by a very narrow jet with jet opening angle $<1^\circ$.

\subsection{Late time component: external origin}\label{sec:origin_external}
This section focuses on theoretical models for HE emission lasting much longer than the prompt phase.
This emission is likely produced as a consequence of the interactions between the fireball and the external medium.
In particular, the most viable model, that will be largely discussed in this section, invokes synchrotron radiation from electrons energized at the external shock, with a possible contribution from SSC.

\subsubsection{Synchrotron}
A first simple consistency check between late time HE observations and the afterglow model can be performed by comparing the typical lightcurve decay rates and spectral photon indices with theoretical predictions.
The synchrotron afterglow theory, applied at frequencies $\nu>[\nu_{\rm c},\nu_{\rm m}]$, predicts a flux proportional to $F_\nu\propto\nu^{-p/2}\,t^{-(3p-2)/4}$, where $p$ is the spectral index of the injected accelerated electrons. 
The typical decay rate observed in LAT lightcurves ($F_\nu\propto t^{-1.2}$) and the photon index (usually slightly softer than $-2$, see section~\ref{sec:observations}) agree with model expectations for values of $p<-2$\cite{kumar09,kumar10,ghisellini10,ghirlanda10,corsi10b}, compatible with predictions from particle acceleration models. Moreover, also the lack of strong spectral evolution is in agreement with this interpretation.

While these model predictions refer to the phase when the blastwave is decelerating, at earlier times (i.e. when the bulk Lorentz factor is constant) the flux should rise as $t^{2}$ for a medium with a constant density profile, and as $t^0$ if the medium has a density decreasing as $R^{-2}$.
In a few cases, an early time increase of the LAT flux has been identified and interpreted as early afterglow emission during the coasting phase.
An example is shown in Fig.~\ref{fig:090510}\cite{ghirlanda10}, where the HE lightcurve of GRB~090510 (as observed by the LAT and by the {\it AGILE}-GRID) is modeled with a smoothly broken PL with the first index fixed to +2 and the second one having best fit value $-1.46^{+0.06}_{-0.03}$. 
In this context, the peak of the lightcurve should roughly correspond to the deceleration time and can then be used to estimate $\Gamma_0$, similarly to what is usually done with optical data exhibiting a peak in their early time lightcurve\cite{ghirlanda12}. This method has been applied to LAT data by several authors and usually returns Lorentz factors of the order of 10$^3$.
The analysis presented in Ref.~\refcite{maxham11}, however, cautions against the use of this method: at early time the contribution from internal origin is likely dominating the emission and the identification of the afterglow peak is not straightforward.

The closure relations mentioned earlier for the spectral and temporal indices refer to the case of a blast-wave that decelerates adiabatically. 
A radiative blast-wave would result in a different (faster) temporal decay  $F_\nu\propto t^{(2-6p)/7}$. This regime was initially invoked in Ref.~\refcite{ghisellini10} to explain why LAT lightcurves have a steep decay ($F_\nu\propto t^{-1.5}$), steeper than what usually observed at later times in X-ray lightcurves, suggesting an initial radiative phase in the blast-wave evolution\cite{nava13}. Subsequent analyses of the temporal decay of LAT lightcurves\cite{LATcatalog13} recovered, for most GRBs, flatter temporal decay indices. Steep decays are yet found in GRB~080916C, GRB~090902B, and GRB~110731A.

\subsubsection*{Clustering of the HE lightcurves}
\begin{figure}[b]
{\includegraphics[scale=0.33]{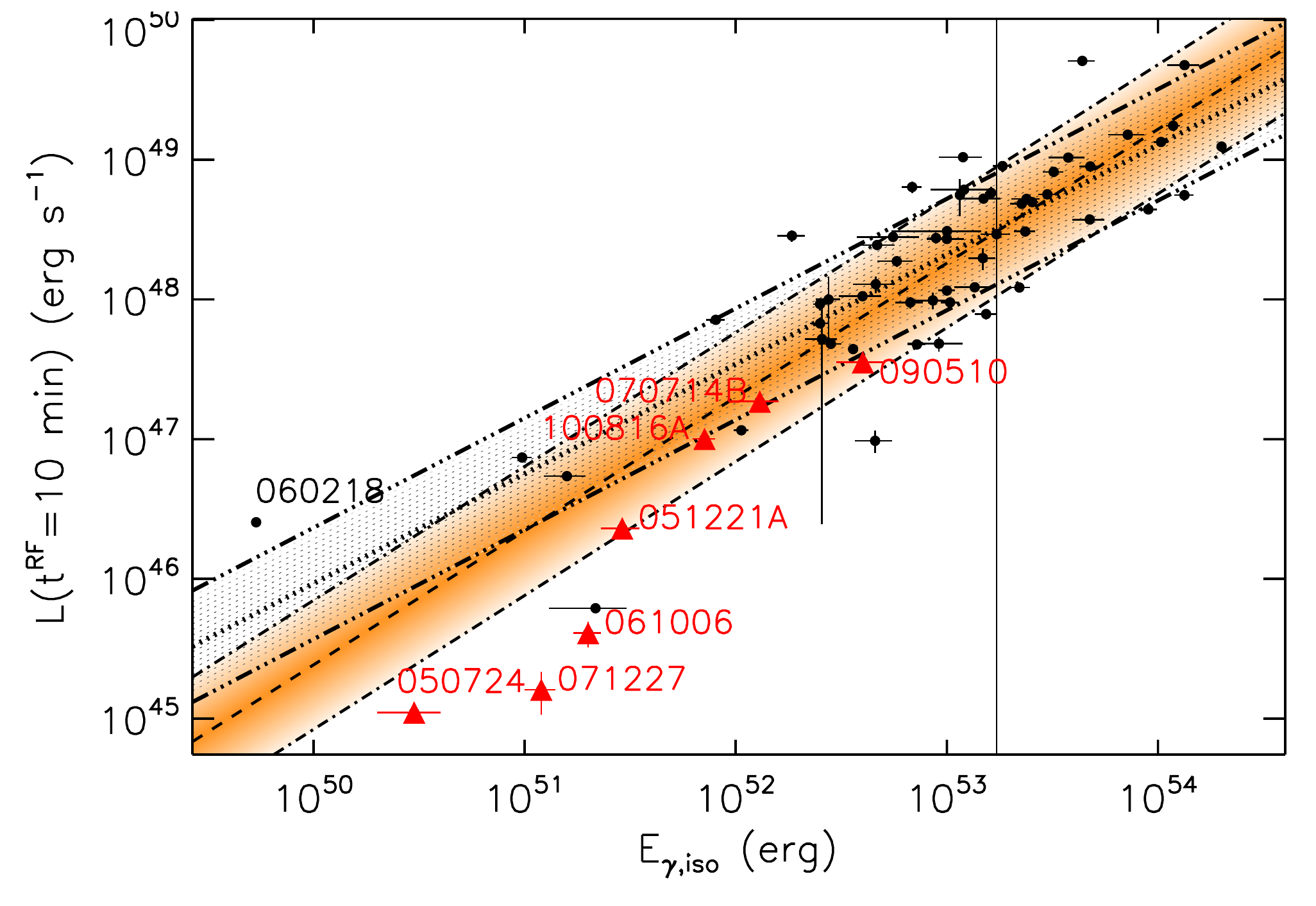}
\includegraphics[scale=0.32]{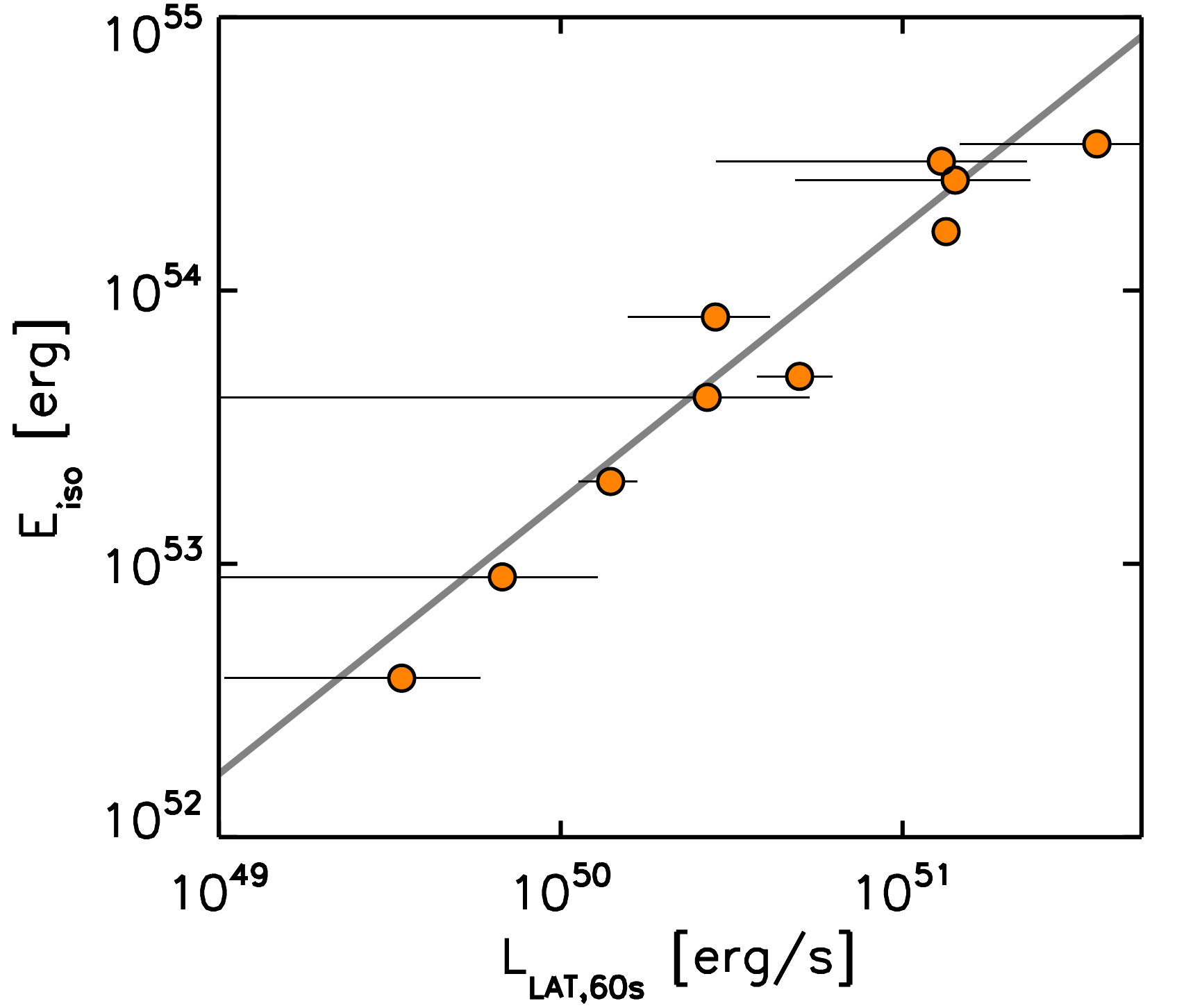}}
\caption{Left: X-ray (0.3-3-\,keV) afterglow luminosity measured at 10 minutes versus the prompt energy $E_{\rm \gamma,iso}$ for a sample of long (black symbols) and short (red symbols) GRBs (from Ref.~\protect\refcite{margutti13}). Right: $E_{\rm \gamma,iso}$ versus the LAT luminosity (0.1-10\,GeV) measured at 60 seconds for LAT GRBs with temporally extended emission and known redshift (from Ref.~\protect\refcite{nava14}). 
\label{fig:x_clustering}}
\end{figure}
Another prediction of the afterglow synchrotron theory is the existence of a linear correlation between the afterglow luminosity of the high-energy part of the synchrotron spectrum (i.e. above both $\nu_{\rm c}$ and $\nu_{\rm m}$) and the energy emitted during the prompt phase \eiso.
The reason for this relation can be understood as follows.
The afterglow luminosity at high-energy is produced by electrons in fast cooling regime, and then it 
depends only on two parameters\cite{kumar00}: the fraction \ee\ of dissipated energy that goes into the non-thermal electrons, and the energy content of the fireball $E_{\rm k,aft}$. This last quantity is related to the energy emitted during the prompt $E_{\rm \gamma,iso}$ and to the efficiency of the prompt emission mechanism $\eta_{\gamma}$ through the equation $E_{\rm k,aft} = E_{\rm \gamma,iso} (1-\eta_\gamma)/\eta_\gamma$. 
This means that the high-energy luminosity during the deceleration phase is determined by two unknown parameters (\ee\ and $\eta_\gamma$) and is proportional to $E_{\rm \gamma,iso}$:
\begin{equation}
L^{\rm aft}_{\nu>[\nu_{\rm c}, \nu_{\rm m}]}\propto \epsilon_{\rm e}\,E_{\rm k,aft}\,\nu^{-p/2}\,t^{-(3p-2)/4} \propto \epsilon_{\rm e}\,E_{\rm \gamma,iso} \frac{1-\eta_\gamma}{\eta_\gamma}\,\nu^{-p/2}\,t^{-(3p-2)/4}
\label{eq:L_aft}
\end{equation}

The proportionality between the high-energy afterglow luminosity and $E_{\rm \gamma,iso}$ has been indeed found using X-ray data of both short and long GRBs\cite{berger07,kaneko07,davanzo12,margutti13,berger14}. These studies found that the correlation between $L^{\rm aft}_{\rm X}$ (estimated at a fixed time $t$, usually 10\,hr-1\,day) and $E_{\rm \gamma,iso}$ is approximately linear, as predicted by the afterglow model. 
According to the model, its scatter must be related to the variation of the parameters \ee\ and $\eta_\gamma$ among different GRBs (see equation~\ref{eq:L_aft}). 
An example of this correlation is shown in Fig.~\ref{fig:x_clustering} (left-hand panel) for a sample of long and short GRBs\cite{margutti13}.

If late time GeV emission is the high-energy part of the synchrotron spectrum, this correlation must be recovered also when the LAT luminosity is considered.
This test has been first performed on a early sample of four GRBs in Ref.~\refcite{ghisellini10}, and then repeated on a larger sample of 10 GRBs with measured redshift in Ref.~\refcite{nava14}. 
The results are shown in Fig.~\ref{fig:x_clustering} (right-hand panel). A strong correlation between the LAT luminosity at 60\,s and \eiso\ is indeed found. The solid line (that is not a fit to the points) shows a linear relation.
\begin{figure}[b]
\centering
\includegraphics[scale=0.63]{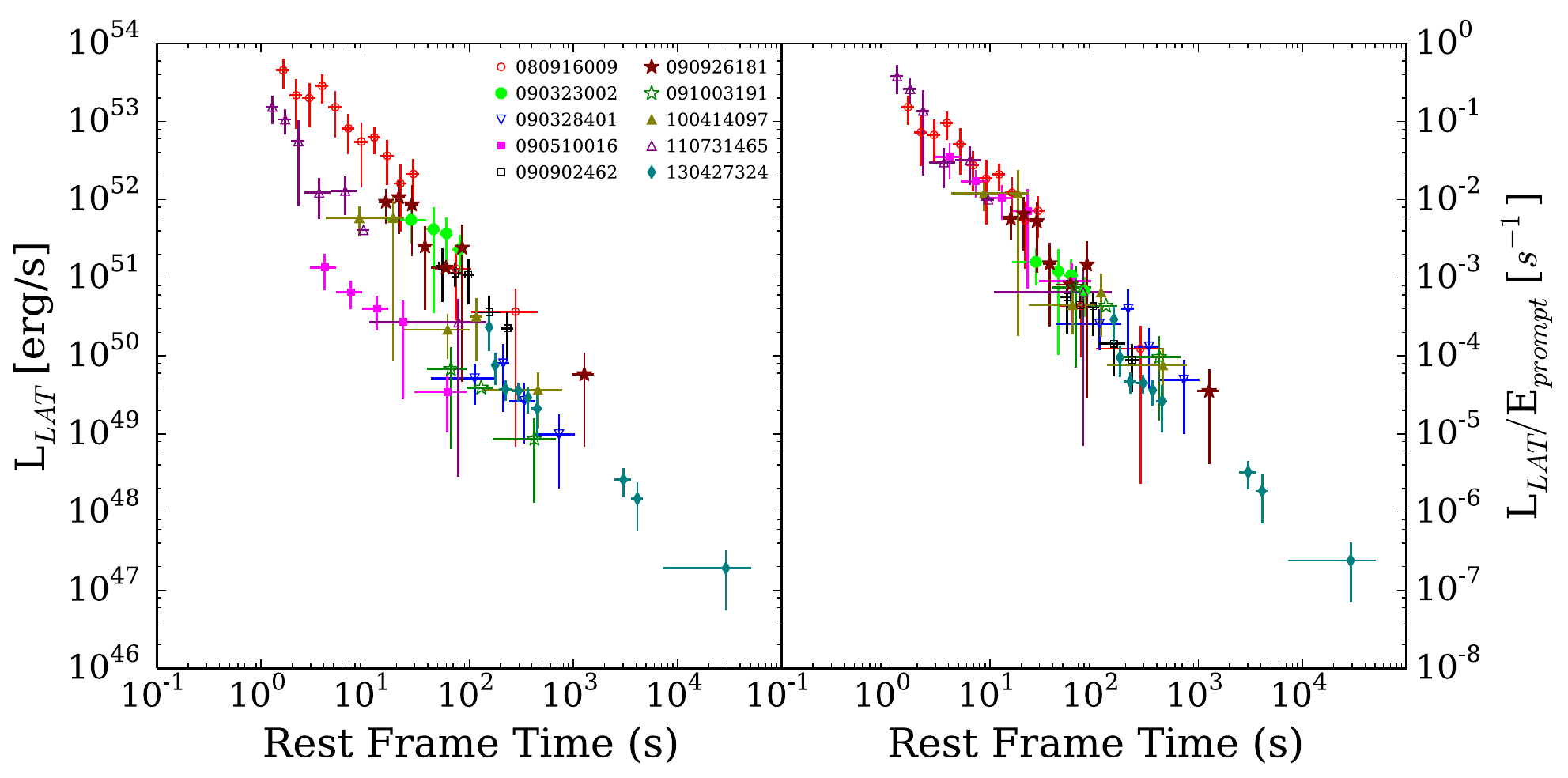}
\caption{Left: LAT lightcurves (above 0.1\,GeV) of 10 GRBs with measured redshift. Right: in this panel, each lightcurve has been normalised to the energy \eiso\ emitted during the prompt. From Ref.~\protect\refcite{nava14}.
\label{fig:clustering}}
\end{figure}

A different way of displaying this relation is to plot the afterglow lightcurves normalised to $E_{\rm \gamma,iso}$, instead of showing the luminosity at a fixed time as a function of $E_{\rm \gamma,iso}$.
The results for the sample of 10 LAT GRBs are reported in Fig.~\ref{fig:clustering}. The panel on the left shows the luminosity lightcurves as a function of the rest frame time. The panel on the right shows the overlap (clustering) obtained when each lightcurve is normalised to the prompt energy $E_{\rm \gamma,iso}$ (called simply $E_{\rm prompt}$ in the plot label). 
The fact that the LAT lightcurves overlap when the time-evolving luminosity is divided by $E_{\rm \gamma,iso}$ not only supports the afterglow scenario, but is also an indication that \ee\ and $\eta_\gamma$ must have a small dispersion in this sample of GRBs (the standard deviation of the gaussian distributions of their logarithmic values are $\sigma_{\rm Log\epsilon_{\rm e}}< 0.19$ and $\sigma_{\rm Log_{\eta_\gamma}}<0.23$)\cite{nava14}. 
The normalisation of the LAT lightcurves is consistent with typical values \ee\,=\,0.1 and $\eta_\gamma\sim$\,0.2. Whether this result on the small variation of \ee\ and $\eta_\gamma$ can be extended to the general GRB population is still an open question. Note however that very similar results on the typical value and narrow distribution of \ee\ have been found with a completely different method and sample, from the modeling of radio emission\cite{beniamini17}.

\subsubsection*{Broadband modeling}
The most direct way to test the forward shock interpretation is to perform broadband temporal and spectral modeling of all the available afterglow data (from radio to GeV) and investigate if it is possible to model the broadband data in the context of the afterglow scenario.

Broadband modeling of three LAT GRBs (GRB~080916C, GRB~090510, and GRB~090902B) was performed in Refs.~\refcite{kumar09,kumar10}. Besides showing that the synchrotron external shock model is able to account for GeV data in a self-consistent scenario, the authors pointed out that the addition of GeV data constraints the magnetic field to be much smaller ($\epsilon_{\rm B}\sim10^{-6}-10^{-4}$) than the value traditionally assumed  in afterglow studies ($\epsilon_{\rm B}\sim10^{-2}-10^{-1}$). 
They argue that the magnetic field is consistent with a shock-compressed circum-stellar medium characterised by pre-shock values of the order of a few tens of $\mu$G. 
These results were confirmed on a larger number of LAT GRBs, by different authors\cite{feng11,lemoine13a,lemoine13b,beniamini15}.
In this context, a different interpretation for the small values of $\epsilon_{\rm B}$ was proposed in Refs.~\refcite{lemoine13a,lemoine13b,wang13}. 
These values may be suggestive of a decayed magnetic field: the authors introduced a model where the micro-turbulence generated at the shock decreases as a function of distance from the shock front, and showed that the GeV to optical data can be consistently modeled within the synchrotron external shock scenario if $\epsilon_{\rm B,+}\sim$\,0.01 close to the shock, and $\epsilon_{\rm B,-}$ from $10^{-6}$ to $10^{-4}$ at the rear of the blast, where radio, optical and X-ray photons are produced.

\begin{figure}
\centering
{\includegraphics[scale=0.31]{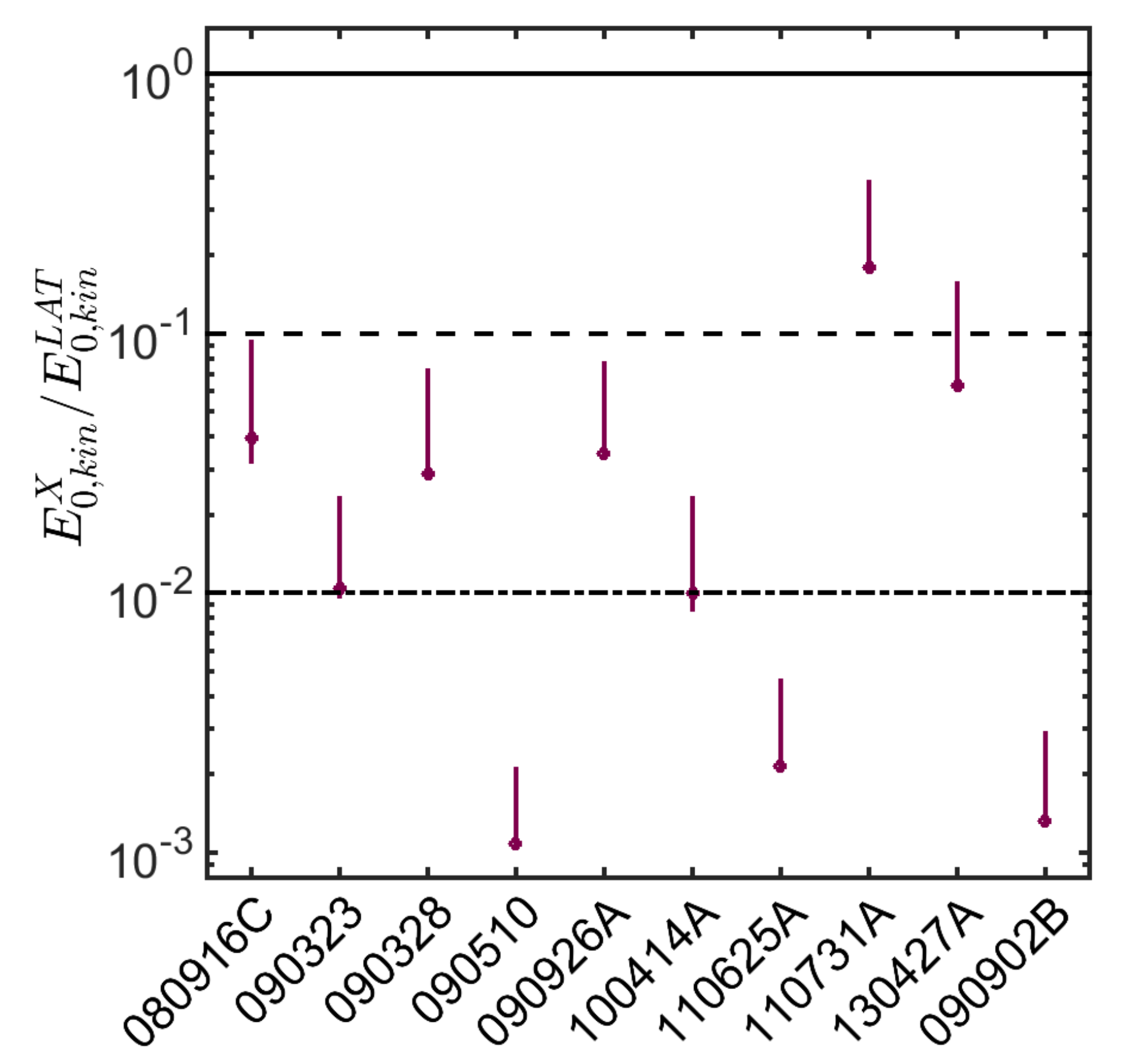}
\includegraphics[scale=0.31]{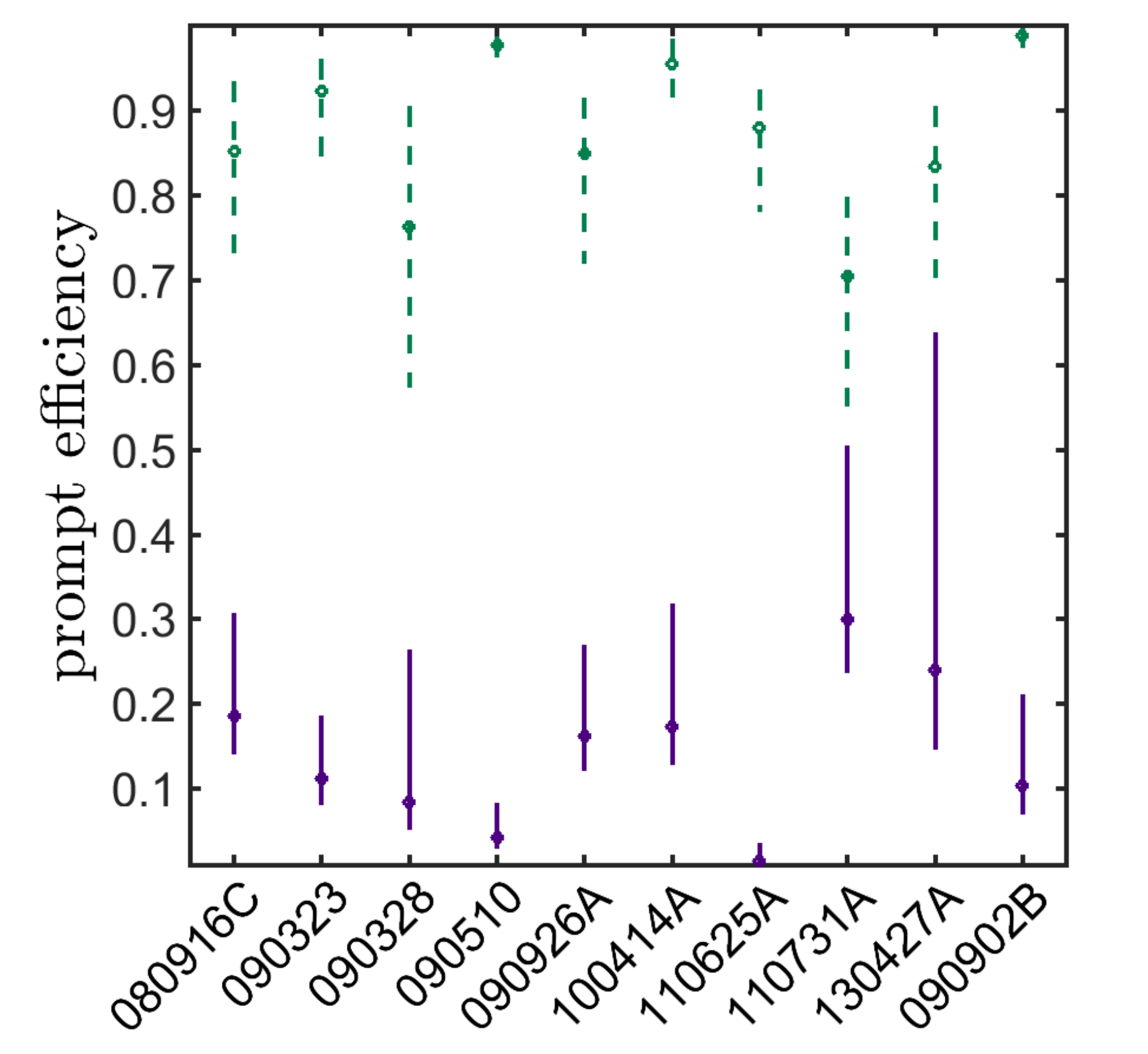}}
\caption{Left: ratio between the blast-wave energy $E_{\rm k,aft}$ estimated from X-rays observations and from GeV observations, inferred assuming that both the X-ray and the GeV band are above $\nu_{\rm c}$ at the time of observations. The error bar is due to different assumptions on $p$ (between 2.1 and 2.8).
Filled circles correspond to $p=2.5$. Right: efficiency of the mechanism producing prompt radiation, as inferred from X-ray (green) and GeV (purple), data using equation~\ref{eq:L_aft}. 
Also in this case the uncertainty on $p$ has been taken into account. From Ref.~\protect\refcite{beniamini15}.
\label{fig:paz_paper1}}
\end{figure}

An effort of performing broadband afterglow modeling of a larger sample of LAT GRBs was performed in Ref.~\refcite{beniamini15}. 
The authors collected all LAT events for which X-ray afterglow observations are also available. If available, optical data have also been included in the analysis. 
Before performing broadband modeling, the authors outlined a striking result coming from the simple comparison of GeV and X-ray observations. Assuming that both GeV and X-ray late time ($\sim$1\,d) data lie above the synchrotron break frequencies $\nu_{\rm c}$ and $\nu_{\rm m}$, the two sets of data can be used independently to estimate the prompt efficiency $\eta_\gamma$ and the blastwave energy content during the afterglow $E_{\rm k,aft}$ (see equation~\ref{eq:L_aft}). 
The results are shown in Fig.~\ref{fig:paz_paper1}. The left-hand panel displays, for each GRB of the sample, the ratio between $E_{\rm k,aft}$ (called $E_{\rm 0,kin}$ in the figure) estimated from X-ray data and from GeV data. The names of the GRBs are reported on the $x$-axis. The estimates of the energetics performed using X-ray data are systematically larger by a factor 10 to 10$^2$, or even more. Note that the ratio is independent from the assumed values of $\epsilon_{\rm e}$, provided that its value is constant in time. The uncertainty on $p$ is taken into account in the error bars, but it can not explain the difference between the two independent estimates of $E_{\rm k,aft}$.

These estimates imply also different inferred values of the prompt efficiency, as shown in the right-hand panel of the same figure. Prompt efficiencies $\eta_\gamma$ inferred from X-ray data (green dashed) are in the range 0.7-1, in agreement with results from studies adopting the similar method and exploiting X-ray data\cite{ioka06,nousek06,zhang07}. The values of $\eta_\gamma$ inferred from GeV data (purple) are systematically lower, between 0.05 and 0.3, allowing for a less efficient mechanism.
The two different estimates are inconsistent with each other, raising the question of which one of the initial assumptions (i.e., both X-ray and GeV are synchrotron radiation from the forward shock and they lie on the high-energy segment of the synchrotron spectrum) is not correct. 

Two possible explanations have been considered in Ref.~\refcite{beniamini15} for the apparent inconsistency. One simple explanation is that the X-ray band in these GRBs falls below the cooling frequency: $\nu_{\rm X}<\nu_{\rm c}$. In this case, equation~\ref{eq:L_aft} can not be applied to X-ray data, but it can still be applied to GeV data, since the GeV band is more safely above $\nu_{\rm c}$. The second possibility is that the X-ray band does lie above $\nu_{\rm c}$, but the flux in this energy band is partially suppressed by SSC scattering. The suppression factor to be applied to equation~\ref{eq:L_aft} is (1+$Y$), where $Y$ is the Compton parameter, and will in general depend on the unknown parameters \eb\ and the density $n$. Since the GeV energy band is in Klein-Nishina regime, it is not affected by this effect. Also in this case, the method to infer $E_{\rm k,aft}$ and $\eta_\gamma$ can be safely applied to GeV radiation, but not to X-ray radiation, i.e., the X-ray afterglow luminosity is no longer a robust proxy for the blastwave energetics. 

\begin{figure}
\centering
{\includegraphics[scale=0.44]{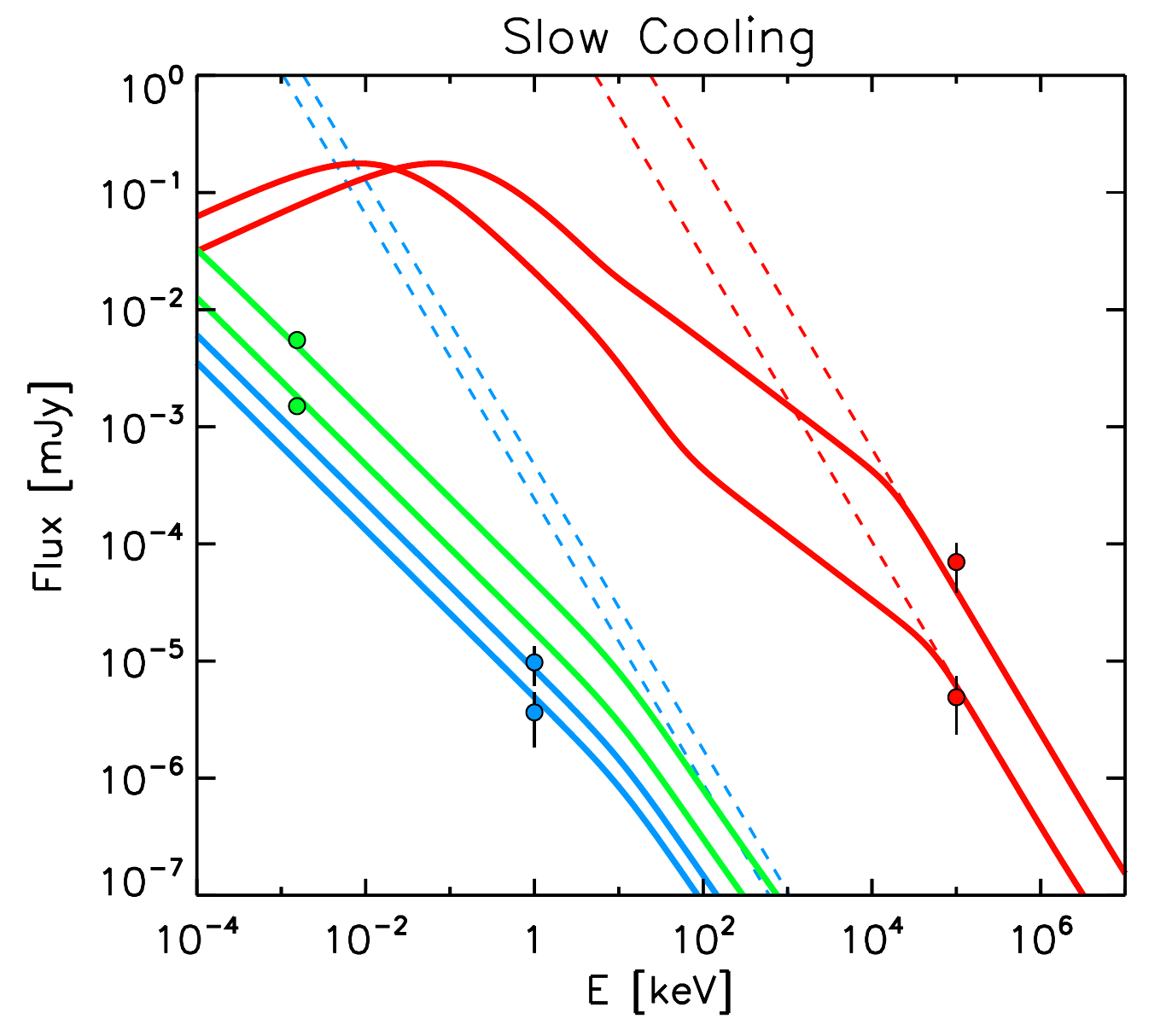}
\includegraphics[scale=0.44]{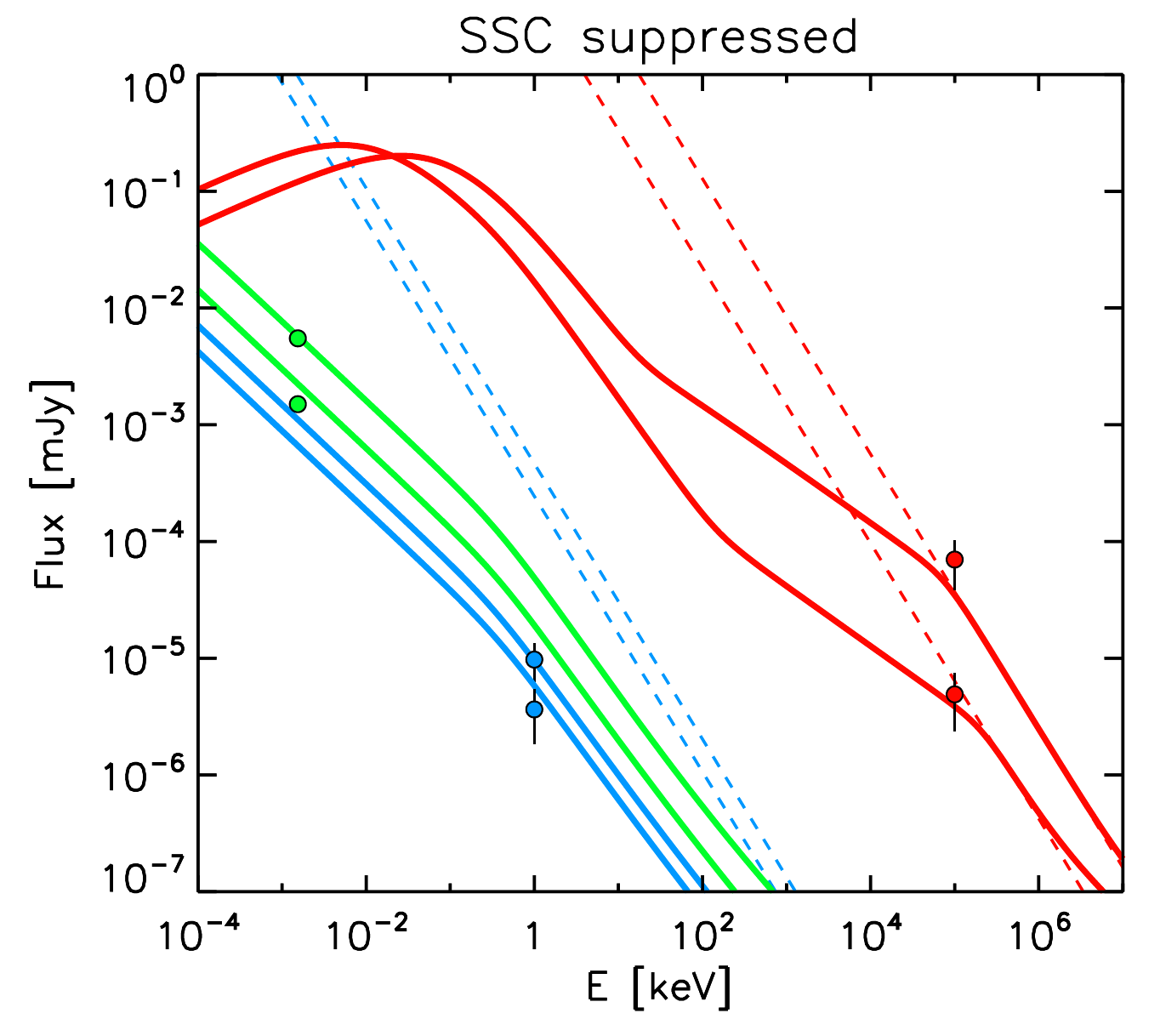}}
\caption{LAT (red), X-ray (blue), and optical observations (green) of GRB~080916C at different times. 
Left: solid lines show an example of modelling where the X-ray band is located below the cooling frequency, while the GeV band is located above. 
Right: an example of modelling where X-rays are produced by SSC-suppressed fast-cooling synchrotron emission, while the GeV radiation is produced by non-suppressed fast-cooling synchrotron emission. The solid curves are the results of numerical afterglow modeling. The red and blue dashed lines are the expected flux in case of no SSC suppression and for frequencies above $\nu_{\rm c}$. It is evident that, according to this modelling, this assumption is correct for the GeV data, while it is not correct for X-ray observations, either because the X-ray observations lie below $\nu_{\rm c}$ (left-hand panel) or because at the X-ray frequency the flux is strongly suppressed by SSC cooling (right-hand panel). From Ref.~\protect\refcite{beniamini15}.
\label{fig:sc_SSCsuppressed}}
\end{figure}

To test the viability of one (or both) explanations, modeling of GeV, X-ray and optical data has been performed by the authors.
The first result of this study is that for all GRBs in the sample, external shocks are a viable model to explain the long lasting GeV radiation. Second, in most cases the allowed solutions correspond to the case $\nu_{\rm X}<\nu_{\rm c}$ (see one example in the left-hand panel in Fig.~\ref{fig:sc_SSCsuppressed}), and part of the parameter space correspond to cases where $\nu_{\rm X}>\nu_{\rm c}$ but the X-ray flux is suppressed by SSC scatterings, while the GeV flux is in KN regime (see one example in the right-hand panel in Fig.~\ref{fig:sc_SSCsuppressed}).
These findings imply that GeV data are a more robust proxy for the blastwave energy $E_{\rm k,aft}$ and give more reliable estimates of the prompt efficiency. Whether this result is peculiar of this sample or can be extended to the general GRB population is not clear. 

The third important result of this study is that in all 10 cases, the constraints derived on \eb\ show that its value is always much smaller than the one traditionally assumed in afterglow studies, consistently with findings from similar analysis ($\epsilon_{\rm B}\sim10^{-6}-10^{-4}$).

\subsubsection*{Limit to the maximum energy of synchrotron photons}\label{sec:synchro}
A simple way to derive the theoretical limit for synchrotron photons is to equate the electron acceleration timescale and the synchrotron energy loss timescale
\cite{dejager92,piran10,kumar12}.
While the physics of particle acceleration, especially in relativistic plasmas, is uncertain, one can derive a conservative limit by assuming a very efficient acceleration (i.e., assuming the Bohm diffusion limit, in which the scattering mean free path is equal to the particle Larmor radius) and equating its timescale to the synchrotron losses timescale after which the particle loses half of its energy:
\begin{equation}
\frac{R_{\rm L}}{c} = \frac{\gamma_{\rm e,max}\,m_{\rm e}\,c}{q\,B} \simeq \frac{3\,\pi\,m_{\rm e}\,c}{\sigma_{\rm T}\,B^2\,\gamma_{\rm e,max}}\,,
\label{eq:max_lf}
\end{equation}
where $B$ is strength of the magnetic field in the comoving frame.
Under the simple scenario of homogeneous magnetic field, the maximum particle energy $\gamma_{\rm e,max}$ is then inversely proportional to the square root of the magnetic field. This implies that the maximum synchrotron photon energy does not depend on $B$ and is given by (comoving frame):
\begin{equation}
E^\prime_{\rm syn,max} = \frac{q\,h}{2\,\pi\,m_{\rm e}\,c} \gamma_{\rm e,max}^2\,B \simeq \frac{9\,h\,m_{\rm e}\,c^3}{16\,\pi\,q^2}\,.
\label{eq:max_ene}
\end{equation}
This estimate returns a maximum synchrotron photon energy of $\simeq$\,50\,MeV in the comoving frame. Different choices on numerical factors may yield to estimates that differ by a factor of a few from the simple estimate reported here.
The maximum photon energy in the observer frame is then:
\begin{equation}
E^{\rm obs}_{\rm syn,max}(t) \sim\,\frac{50}{1+z}\,\Gamma(t)\,{\rm MeV}. 
\label{eq:obs_max_ene}
\end{equation}

In Fig.~\ref{fig:HEph_arrivaltime} we calculated this limit as a function of time (orange and blu solid lines for wind and homogeneous medium, respectively) and compared it with the most energetic photon detected by the LAT in different GRBs. 
The limiting curves have been estimated for typical values of the involved quantities (energetic, density and redshift). However, different choices of these parameters do not strongly impact the estimate of $\Gamma(t)$, since in a homogeneous medium we have $\Gamma(t)\propto [E(1+z)^3/n]^{1/8}t^{-3/8}$ and in a wind-like medium $\Gamma(t)\propto [E(1+z)/A_\star]^{1/4}t^{-1/4}$.
According to the results shown in Fig.~\ref{fig:HEph_arrivaltime}, the limit is violated by several LAT-detected photons.

One possibility to explain the tension between observations and predictions is to invoke a revision of the present understanding of particle acceleration in relativistic shocks\cite{kumar12}.
An alternative explanation consists in abandoning the assumption that these are synchrotron photons, and considering the contribution of an inverse Compton component, either of external origin, or on the same synchrotron photons (SSC).

\newpage
\subsubsection{Synchrotron Self Compton}\label{sec:ssc}
The first effect of SSC cooling is a modification of the high-energy part of the synchrotron spectrum. The transition from Thomson to KN regime will indeed result in an enhancement of the high-energy synchrotron flux as compared to the synchrotron flux at lower frequencies, suppressed by the SSC cooling (see an example in Fig.~\ref{fig:sc_SSCsuppressed}). This modification of the synchrotron spectrum has been invoked in Ref.~\refcite{beniamini15} to explain the apparent inconsistency between X-ray and GeV fluxes, with GeV fluxes being higher than what expected from a simple spectral extrapolation of the X-ray flux.
A suppression factor $(1+Y_{\rm X})$ up to several tens is required in order to account for the apparent discrepancy.

Beside modifying the synchrotron spectrum, IC scatterings produce a separated spectral component at higher energies.
The contribution of this component to the LAT flux during the afterglow phase has been invoked by several authors, either to explain the full LAT late-time emission, or to explain at least those photons exceeding the synchrotron maximum energy\cite{wang10,wang13,liu13,tang14,beniamini15}.

\begin{figure}
\centering
\includegraphics[scale=0.53]{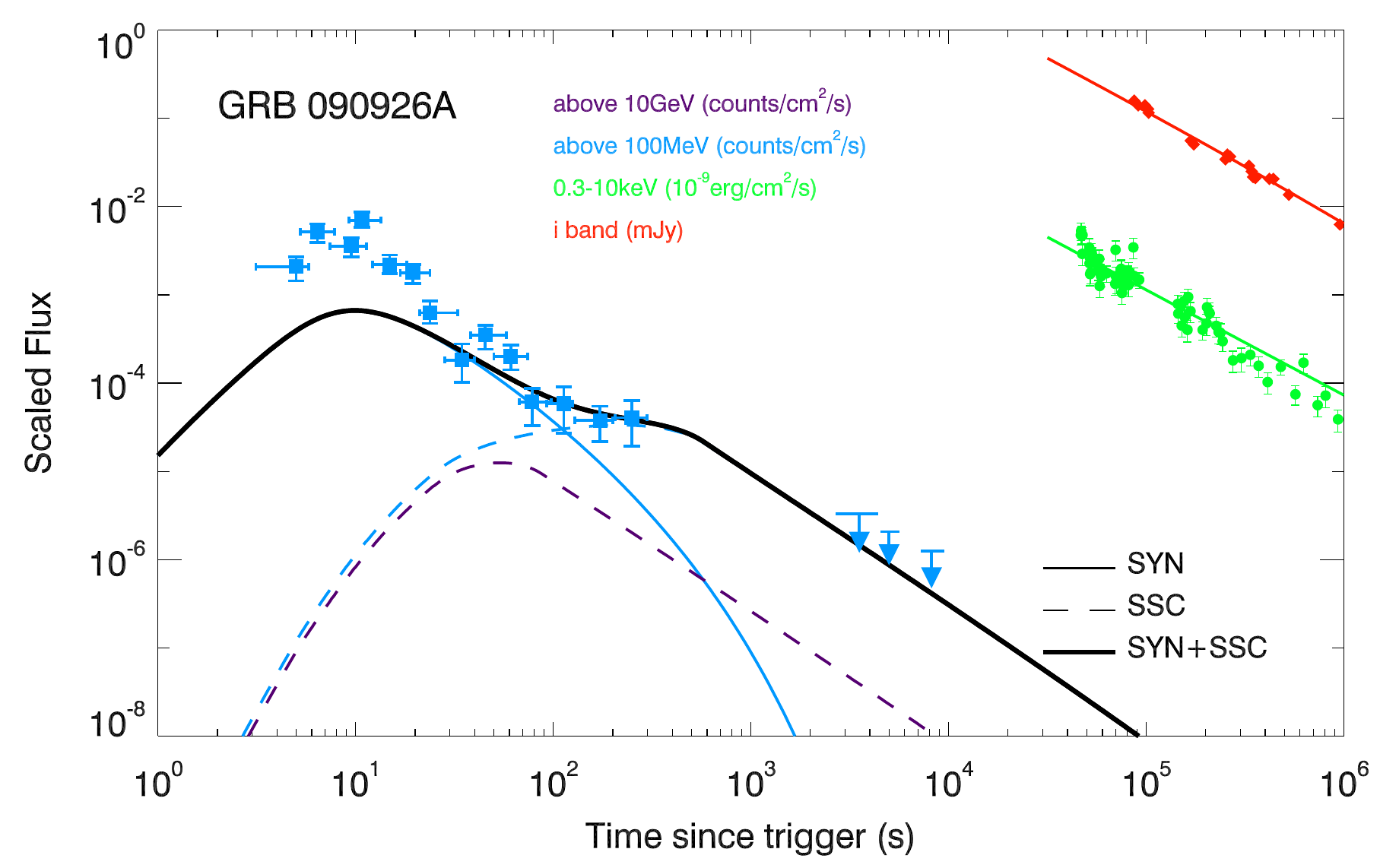}
\caption{Left: broad band modeling of GRB~090926A proposed in Ref.~\protect\refcite{wang13}. The parameter values are $E=2\times10^{55}$\,erg, $n$=1.2\,cm$^{-3}$, $\epsilon_{\rm e} = 0.1$, $\epsilon_{\rm B,+} = 0.005$, $\epsilon_{\rm B,-}=6\times10^{-6}$, $\Gamma_0 =600$, and $p=2.5$. According to this modeling, after $\sim10^2$\,s the LAT emission is dominated by the SSC component.
\label{fig:wang_SSC}}
\end{figure}

In Ref.~\refcite{beniamini15} authors find that for three GRBs (090926A, 090902B, and 090510) the GeV emission is dominated by SSC radiation, whereas the X-ray flux is dominated by synchrotron from fast-cooling electrons.
A similar conclusion on GRB~090926A has been reached in Ref.~\refcite{wang13}, where it is shown that at late time (after $10^2$\,s), when the Lorentz factor is decreasing and the synchrotron component at high-energies is suppressed by the maximum photon limit, the emission is dominated by the SSC (Fig.~\ref{fig:wang_SSC}). 

\section{Non-detections: GRBs lacking high-energy emission}\label{sec:lack}
Most GRBs observed at MeV-GeV energies show no hint of HE emission.
In these cases it is possible to estimate upper limits to the average source flux in a given time window.
Flux upper limits have been estimated both from {\it AGILE}-GRID\cite{longo12} and  {\it Fermi}-LAT\cite{LATcatalogUL12} observations.
They can be used to constrain the unknown parameters of all those theoretical models predicting emission in the high-energy band.

Upper limits have been used in literature mainly to constrain two emission components: the high-energy part of the prompt spectrum and the high-energy part of the afterglow synchrotron spectrum. In the first case, the upper limit (estimated on the prompt duration timescale) is compared to the high-energy PL extrapolation of the sub-MeV prompt spectrum in order to infer the possible presence of a spectral cutoff and place constraints on the bulk Lorentz factor. 
In the second case, limits derived on longer timescales have been used to constrain the brightness of the high-energy part of the afterglow synchrotron component.
As it will be discussed in the following, also in this case it is possible to infer information on the bulk Lorentz factor.

\subsubsection*{Lack of prompt high-energy radiation}
The lack of HE detection during the prompt phase may be simply caused by the softness of the Band spectrum above its spectral peak and/or the faintness of the emission. 
To understand if non-detections imply instead the presence of a high-energy cutoff in the prompt spectrum, one simple method consists in extrapolating the best fit spectral model of the emission detected by the GBM into the HE range and compare the predicted flux with the upper limits imposed by the HE non-detection.
This analysis has been performed in Ref.~\refcite{LATcatalogUL12} on LAT observations. 
The authors collected a sample of 288 GRBs detected by the GBM and observed by the LAT (i.e., falling within the LAT field of view), but with no hint of LAT emission. 
For each burst, the upper limits to the average 0.1-10\,GeV flux have been estimated on the timescales corresponding to the prompt emission duration. 

A statistically significant presence of a spectral cutoff has been found in six GRBs. 
Assuming that the cutoff is caused by $\gamma$-$\gamma$ attenuation within the source, the bulk Lorentz factor has been estimated from the location of the cutoff. Since in these six cases there is no direct measurement of the cutoff energy, only an upper limit $\Gamma_{\rm max}$ can be estimated, conservatively assuming that the cutoff energy is below 100\,MeV.
For one GRB with measured redshift, this method led to infer $\Gamma_{\rm max}\sim155$, while for the other 5 GRBs the upper limits were derived as a function of the redshift. 
For redshifts in the range 1 to 5, the upper limits varies from $\sim$\,200 to $\sim$\,600.

A similar analysis has been performed on 64 {\it AGILE} GRBs observed by the GRID\cite{longo12}. For the 28 events with known prompt spectral properties, the GRID flux upper limits (estimated in the energy range 30\,MeV-3\,GeV) revealed no strong evidence for the presence of high-energy spectral cutoffs.

\subsubsection*{Lack of late time high-energy radiation}
\begin{figure}[t]
{\includegraphics[scale=0.32]{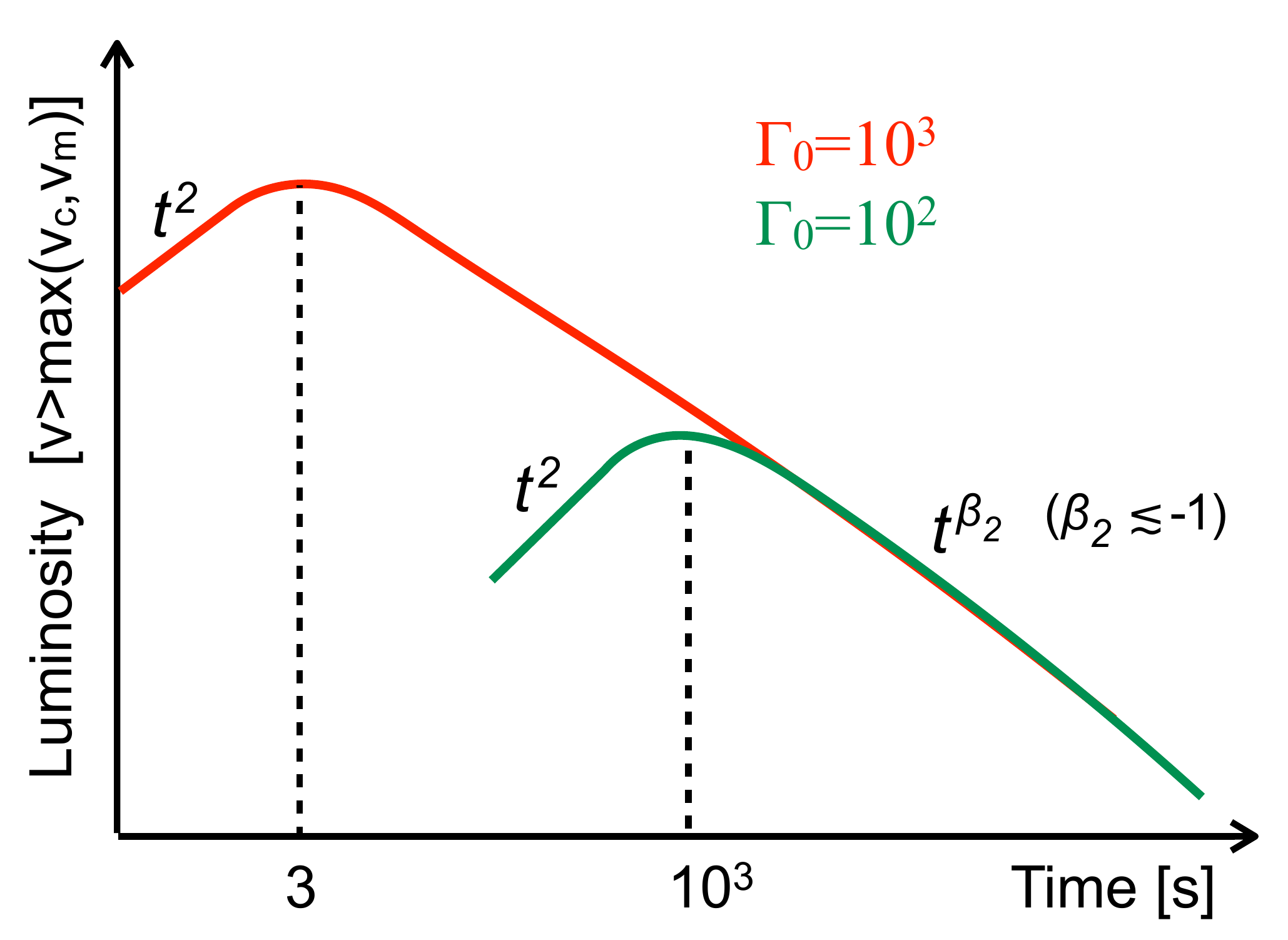}
\vskip -5.35truecm
\hskip 6.9truecm
\includegraphics[scale=0.31]{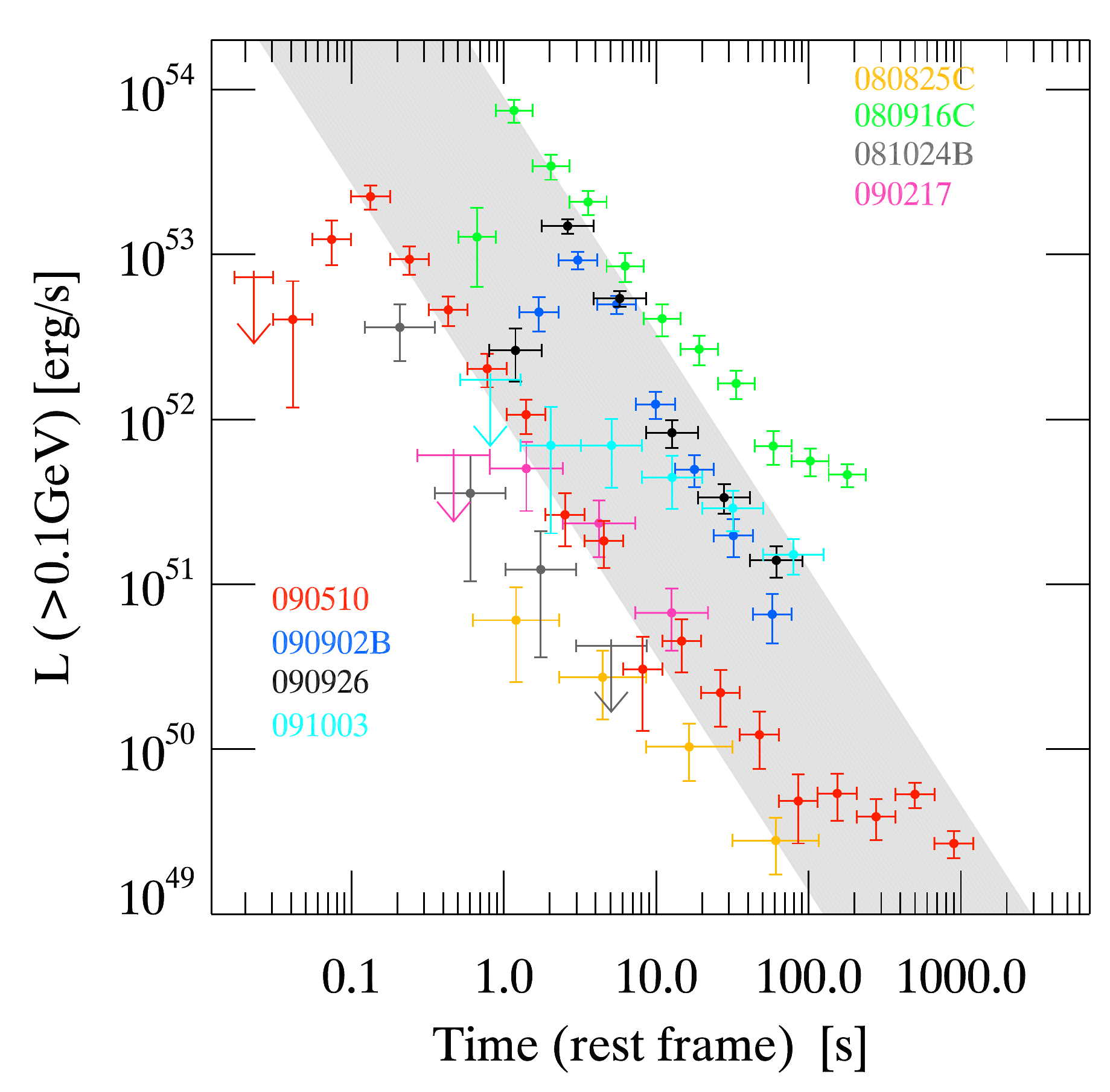}}
\vskip 0.5truecm
\caption{
Left: simulation of two afterglow lightcurves observed at a frequency $\nu>\nu_{\rm c}$ for two GRBs with similar energetics but different initial bulk Lorentz factors $\Gamma_0$ (from Ref.~\protect\refcite{nava17}).
Right: luminosity lightcurves of 8 LAT GRBs, in the energy range 0.1-100\,GeV (from Ref.~\protect\refcite{ghisellini10}). 
\label{fig:LAT_lc}}
\end{figure}
One of the spectral components that might contribute to the $\sim$\,GeV flux on time scales much longer than the prompt emission duration is synchrotron radiation from electrons energised at the external shock (i.e., the high-energy part of the synchrotron afterglow spectrum). 
It is then interesting to determine the constraints on the afterglow model parameters inferred from the absence of a detectable afterglow emission at high energies.
 
It is reasonable to assume that the \gr\ energy band is located above the cooling frequency of the synchrotron afterglow spectrum (i.e., $\nu_{\rm LAT}>\nu_{\rm c}$). 
In this frequency regime, the afterglow flux increases in time during the coasting phase, reaches a maximum around the deceleration time and then decays, following a power-law behaviour in time (Fig.~\ref{fig:LAT_lc}, left-hand panel). 
This behaviour is consistent with observations of the brightest LAT GRBs (see Fig.~\ref{fig:LAT_lc}, right-hand panel).
In order to model the expected HE afterglow emission and infer constraints based on the comparison with HE flux upper limits it is necessary to determine the position of the peak of the HE lightcurve and the overall lightcurve normalisation.

As already discussed in section~\ref{sec:origin}, during the power-law decay,
the afterglow luminosity at a frequencies $\nu>\nu_{\rm c}$ depends only on the fraction $\epsilon_{\rm e}$ of energy gained by the shocked electrons and on the energy $E_{\rm k,aft}$ of the blastwave\cite{kumar00}. 
This last quantity is related to the energy emitted during the prompt $E_{\rm \gamma,iso}$ and to the efficiency of the prompt emission mechanism $\eta_{\gamma}$ through the equation $E_{\rm k,aft} = E_{\rm \gamma,iso} (1-\eta_\gamma)/\eta_\gamma$. 
The afterglow luminosity at $\nu>\nu_{\rm c}$ during the deceleration can then be predicted from $E_{\rm \gamma,iso}$, assuming some values for $\epsilon_{\rm e}$ and $\eta_\gamma$. 
As previously discussed, there is evidence that these two parameters do not vary too much among different GRBs\cite{nava14}.
If this is the case, the HE afterglow luminosity is determined mainly by $E_{\rm \gamma,iso}$.
$E_{\rm \gamma,iso}$ determines then the normalisation of the power-law decay of the high-energy afterglow flux, while the power-law index is given by $t^{-(3p-4)/2}$. 
\begin{figure}[t]
{\includegraphics[scale=0.407]{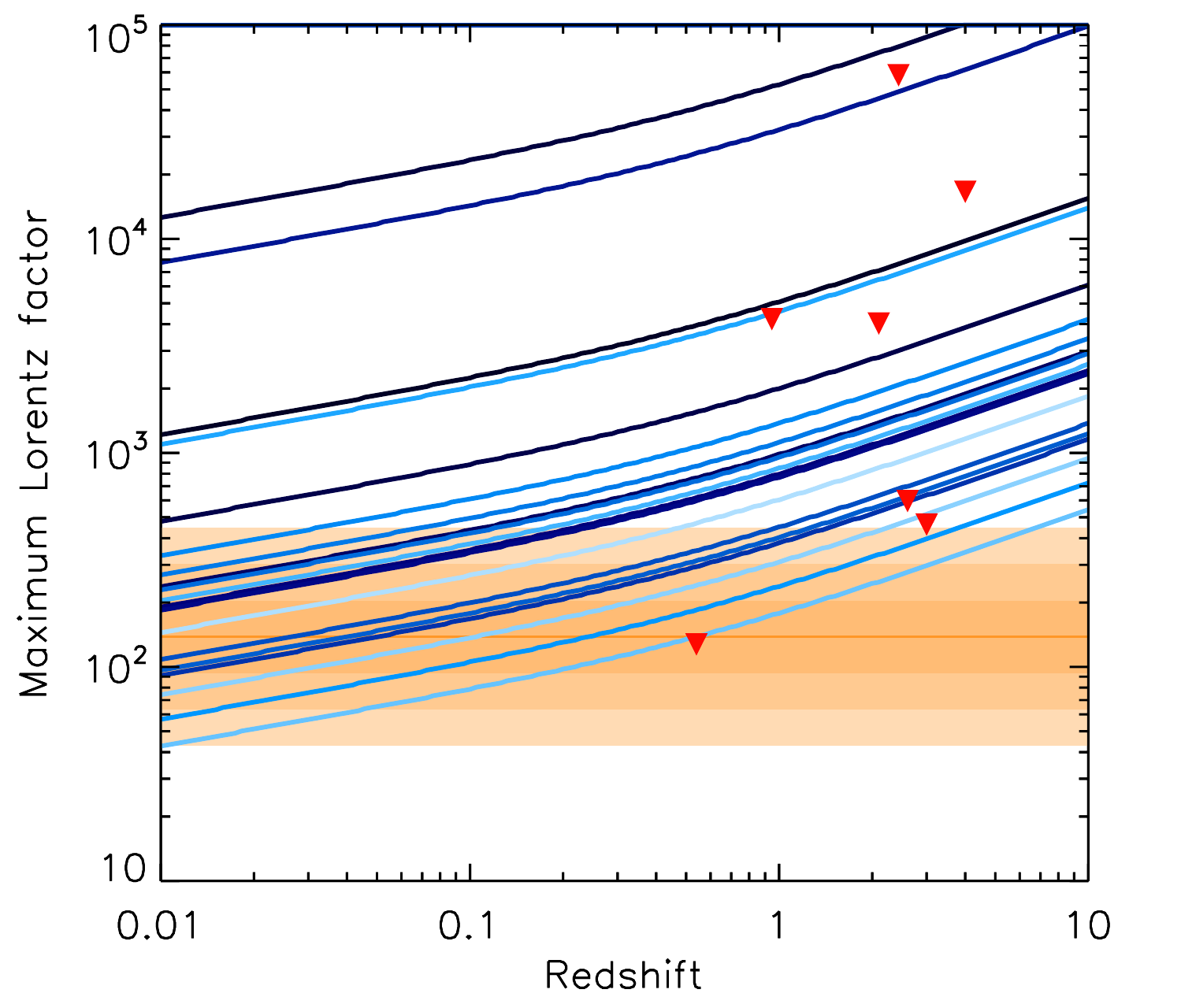}
\includegraphics[scale=0.496]{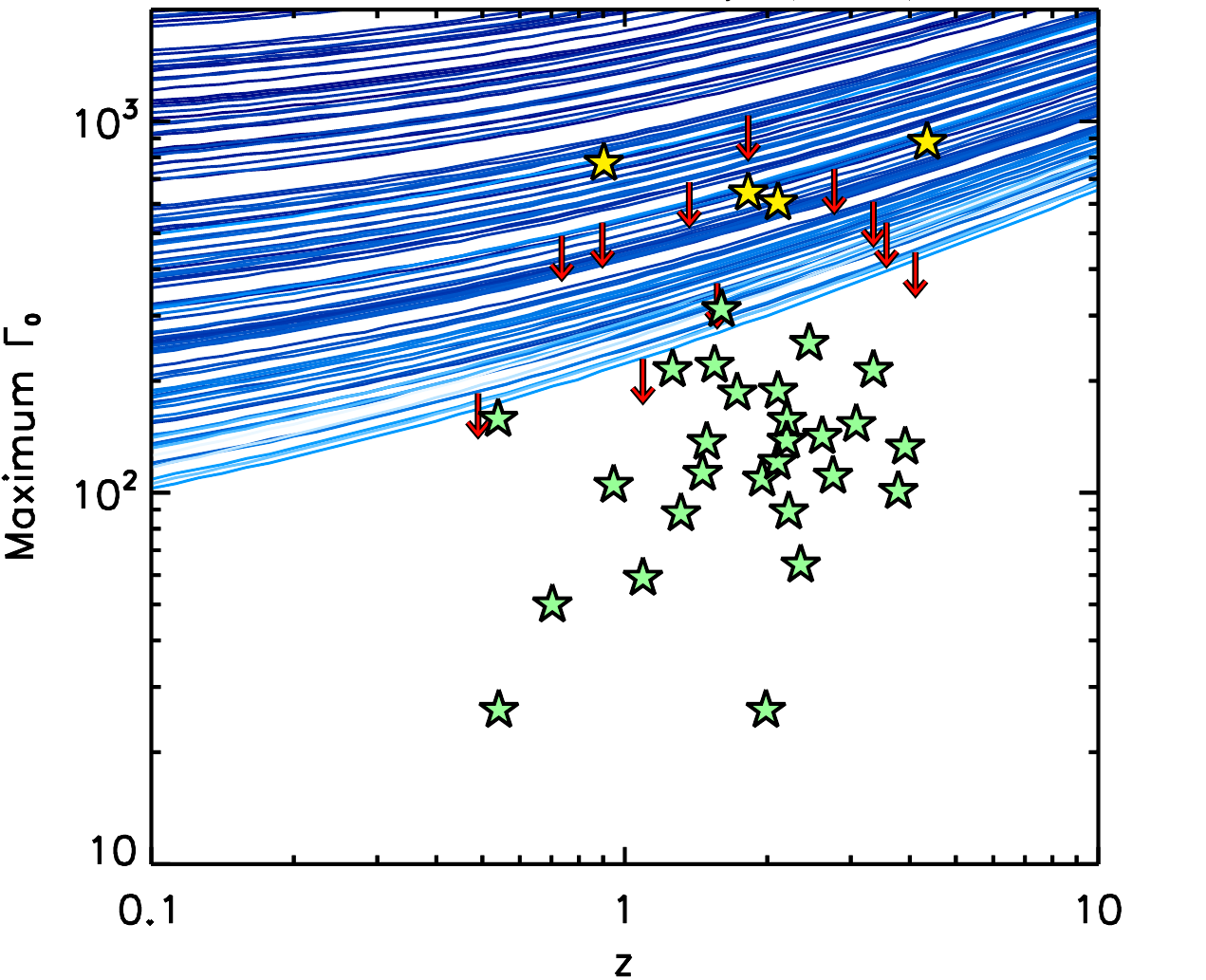}}
\caption{The blue curves show the upper limits on the bulk Lorentz factor as a function of redshift inferred from the non detection of the high-energy part of the synchrotron afterglow spectrum. Higher prompt fluences are marked with lighter colours. Left: results from {\it AGILE}-GRID flux upper limits (from Ref.~\protect\refcite{longo12}). The red triangles denote upper limits for GRBs with measured redshift. The orange horizontal line show the mean value of distribution of bulk Lorentz factors inferred by \cite{ghirlanda12} mainly from the measure of the peak of the optical lightcurves. Shaded stripes show the 1-2-3$\sigma$ of the distribution. Right: results inferred from LAT flux upper limits (from Ref.~\protect\refcite{nava17}). Red arrows show the upper limits for GRBs with measured redshift. Star symbols denote GRBs for which $\Gamma$ has been estimated from the peak of the early optical lightcurve (green stars) and GeV lightcurve (yellow stars)\cite{ghirlanda12}.}
\label{fig:ul}
\end{figure}

The peak time of the lightcurve is set by the initial bulk Lorentz factor $\Gamma_0$: for large $\Gamma_0$, the fireball decelerates at early time, the afterglow lightcurve peaks earlier, and is initially brighter. For smaller $\Gamma_0$, the afterglow lightcurve peaks at later times and the early time luminosity is dimmer as compared to the large-$\Gamma_0$ case.
This is evident in Fig.~\ref{fig:LAT_lc} (left-hand panel)
where two simulated high-energy lightcurves are reproduced, corresponding to two events with same $E_{\rm \gamma,iso}$ (and then same behaviour after the deceleration time) but different initial Lorentz factors $\Gamma_0$. 

As a consequence, for a given \eiso, that is known from prompt observations, the HE emission detectability strongly depends on the value of $\Gamma_0$.
This implies that 
from the non-detection of the afterglow component at high-energies and from the estimate of the energy $E_{\rm \gamma,iso}$ radiated during the prompt emission, it is possible to set an upper limit to the value of $\Gamma_0$.

This method has been applied both to {\it AGILE}-GRID upper limits\cite{longo12} and to LAT upper limits\cite{nava17}, and the results are shown in Fig.~\ref{fig:ul} (left and right panel, respectively). In both panels, a homogeneous density medium has been assumed. A wind medium results in stronger constraints (a factor of 3-4) on the Lorentz factor. When the redshift is unknown, the upper limits are represented by curves as a function of $z$.

\section{GRB observations at very high energies}\label{sec:vhe}
The detection of several photons with energies between few tens of GeV and 0.1\,TeV at relatively late times (see Fig.~\ref{fig:HEph_arrivaltime}) encourages searches for GRB emission at very-high energies (VHE, $>$\,100\,GeV). 
Thanks to the several tens of thousands times larger effective area as compared to telescopes in space, ground-based TeV astronomy allows for counterpart searches with orders of magnitudes better sensitivity.

Presently, it is not clear whether GRBs are TeV emitters, the only hint for $\sim$\,TeV emission coming from the Milagrito experiment\cite{atkins00a}, a TeV air-shower array, prototype of the Milagro detector.
Milagrito was a water Cherenkov telescope that operated between 1997 and 1998, and observed 54 GRBs detected by BATSE and within its field of view.
Evidence of TeV emission was found in one case (GRB~970417), with a probability of 1.5$\times$10$^{-3}$ of being a background fluctuation\cite{atkins00b}. 
An excess of events was observed from the direction of the burst, during the emission observed by BATSE.
Despite the difficulty in obtaining information on the energy of individual events, the excess observed by Milagrito, provided its association with GRB~970417, is likely caused by photons of at least 650\,GeV\cite{atkins03}. No GRB detections were found by the later Milagro experiment.

Except for this case, all the GRB follow-up observations conducted so far by experiments sensitive at VHE resulted only in upper limits.
The reasons are manyfold. First, it is still not clear whether GRBs radiate a significant amount of photons above 0.1\,TeV, and in any case, being cosmological sources with mean redshift $\sim2$, they suffer from absorption by the Extragalactic Background Light (EBL). The most promising energy range for ground detections is then 10-100\,GeV, both because less affected by the EBL and because photons with such energies have already been detected, though they are rare.
A low energy threshold of ground-based $\gamma$-ray detectors, as low as a few tens of GeV, is then fundamental to detect GRBs.

Another characteristic that strongly affects the probability of detecting photons from a GRB is the possibility to observe the source in the early phase of the emission. According to LAT observations, the high-energy flux fades indeed in time as $t^{-1.2}$ or even faster. 
Current observations of GRBs at VHE have led so far to upper limits on time scales ranging from a few tens of seconds to days. 

Apart from the faintness of most GRBs at very high energies, the reasons for this lack of success include the relatively high-energy threshold ($\sim100-200$\,GeV) of past  imaging atmospheric Cherenkov telescopes (IACT), a low duty cycle ($\sim10$\%), as well as the IACT narrow field of view (a few degrees) combined with the poor localization capabilities of the space detectors such as the {\it CGRO}/BATSE and the {\it Fermi}/GBM: for these detectors, the error box can be quite limiting when performing follow-up observations, since it is of the order of the size of a typical IACT field of view.
Extensive air shower (EAS) arrays overcome some of these limitations, but suffer from higher energy thresholds and lower sensitivities than IACTs, caused by a smaller effective area and a more difficult background rejection.

GRB follow-up programs are currently active for
all IACTs,
such as H.E.S.S.\footnote{\label{fn:HESS}https://www.mpi-hd.mpg.de/hfm/HESS}, 
MAGIC\footnote{\label{fn:magic}https://magic.mpp.mpg.de}, 
VERITAS\footnote{\label{fn:veritas}http://veritas.sao.arizona.edu}, and also for the extensive air shower arrays HAWC\footnote{\label{fn:hawc}https://www.hawc-observatory.org}.
I summarize in the following the present status of observations with the major, currently operative IACT and EAS detectors. The last part of this section will be instead dedicated to future facilities for observations at VHE.

\subsection{Imaging atmospheric Cherenkov telescopes}

\subsubsection*{H.E.S.S.}
The High Energy Stereoscopic System (H.E.S.S.\footref{fn:HESS}) is a system of IACTs operating in Namibia since 2004.
During phase-I, H.E.S.S. was featuring four 12\,m diameter mirrors (called CT1-4), characterised by a collecting area of 108\,m$^2$ each, a low energy threshold of $\sim$\,100\,GeV at zenith, and a $\sim5^\circ$ field of view.
H.E.S.S. was upgraded in 2012 with a 28\,m diameter telescope (CT5), characterised by a faster repointing (a full rotation - 180$^\circ$ in azimuth - in $\sim$\,110 seconds), larger light collection area (600\,m$^2$), improved sensitivity to low-energy $\gamma$-rays, and lower energy threshold ($\sim$\,50\,GeV). 
The introduction of CT5 marked the beginning of the second phase of H.E.S.S. operations and the start of the H.E.S.S.-II GRB program\cite{parsons17}. 

H.E.S.S.-I GRB observations from 2003 and 2007 are summarised in Ref.~\refcite{aharonian09a}. 
The upper limits inferred for two particularly interesting bursts, GRB~060602B and GRB~100621A are presented in Refs.~\refcite{aharonian09b} and \refcite{lennarz14}, respectively.

Since 2012, H.E.S.S.-II has followed more than 30 GRBs\cite{hoischen17}, and most of them were observed with CT5. 
Also with the upgraded facility, neither individual GRB observations nor stacked analyses provided the detection of a VHE signal.

\begin{figure}
\centering
\includegraphics[scale=0.63]{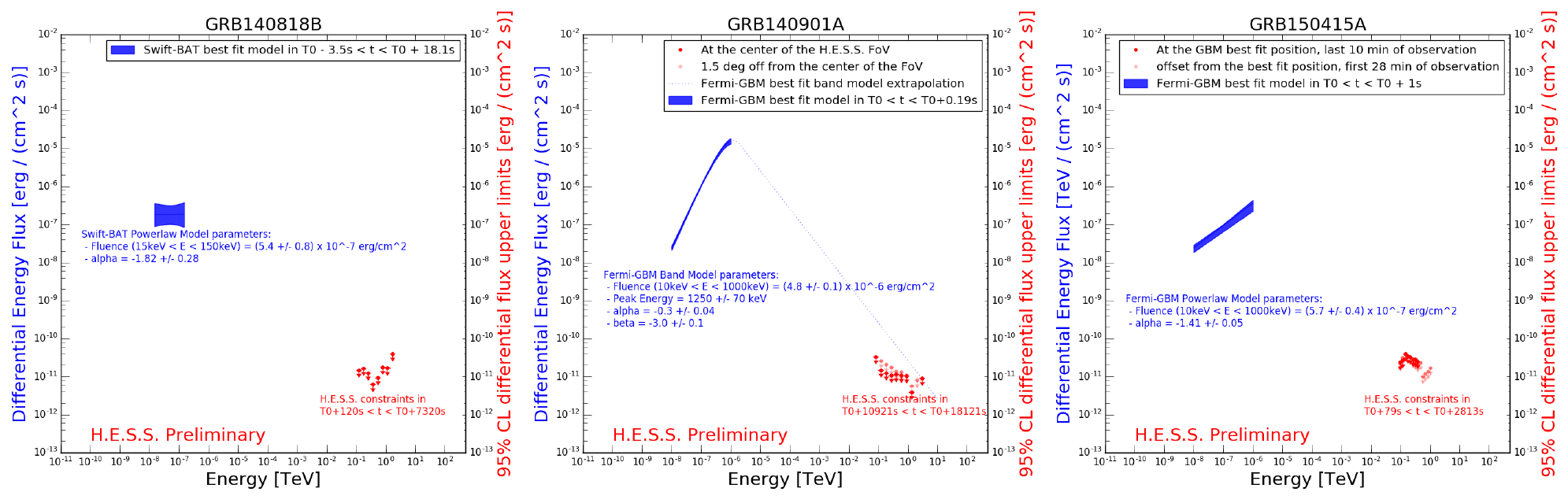}
\caption{Flux upper limits inferred from H.E.S.S. observations (red symbols) for three different GRBs. The best fit for the prompt spectra are shown in blue. From Ref.~\protect\refcite{hoischen17}.
\label{fig:hess_UL_3grbs}}
\end{figure}

The flux upper limits for 3 GRBs observed by H.E.S.S.-II are shown in Fig.~\ref{fig:hess_UL_3grbs}.

\subsubsection*{VERITAS}
The Very Energetic Radiation Imaging Telescope Array System (VERITAS\footref{fn:veritas}), located in southern Arizona, is an array of four 12\,m IACTs, mostly sensitive from $\sim$\,100\,GeV to $\sim$\,10\,TeV, with a field of view of $\sim$\,3.5$^\circ$.
The cameras were upgraded in 2012, to reach an improved sensitivity and a lower energy threshold.

The VERITAS Collaboration has activated a GRB observing program since the beginning of full array operations in 2007. Data taken over the first three years from 16 GRBs triggered by the {\it Swift}-BAT are presented in Ref.~\refcite{acciari11}:
among the 9 bursts with measured redshifts, 3 could be constrained to have VHE afterglows less energetic than the prompt measured by {\it Swift}-BAT in the 15-350\,keV energy range.

\begin{figure}
\centering
\includegraphics[scale=0.33]{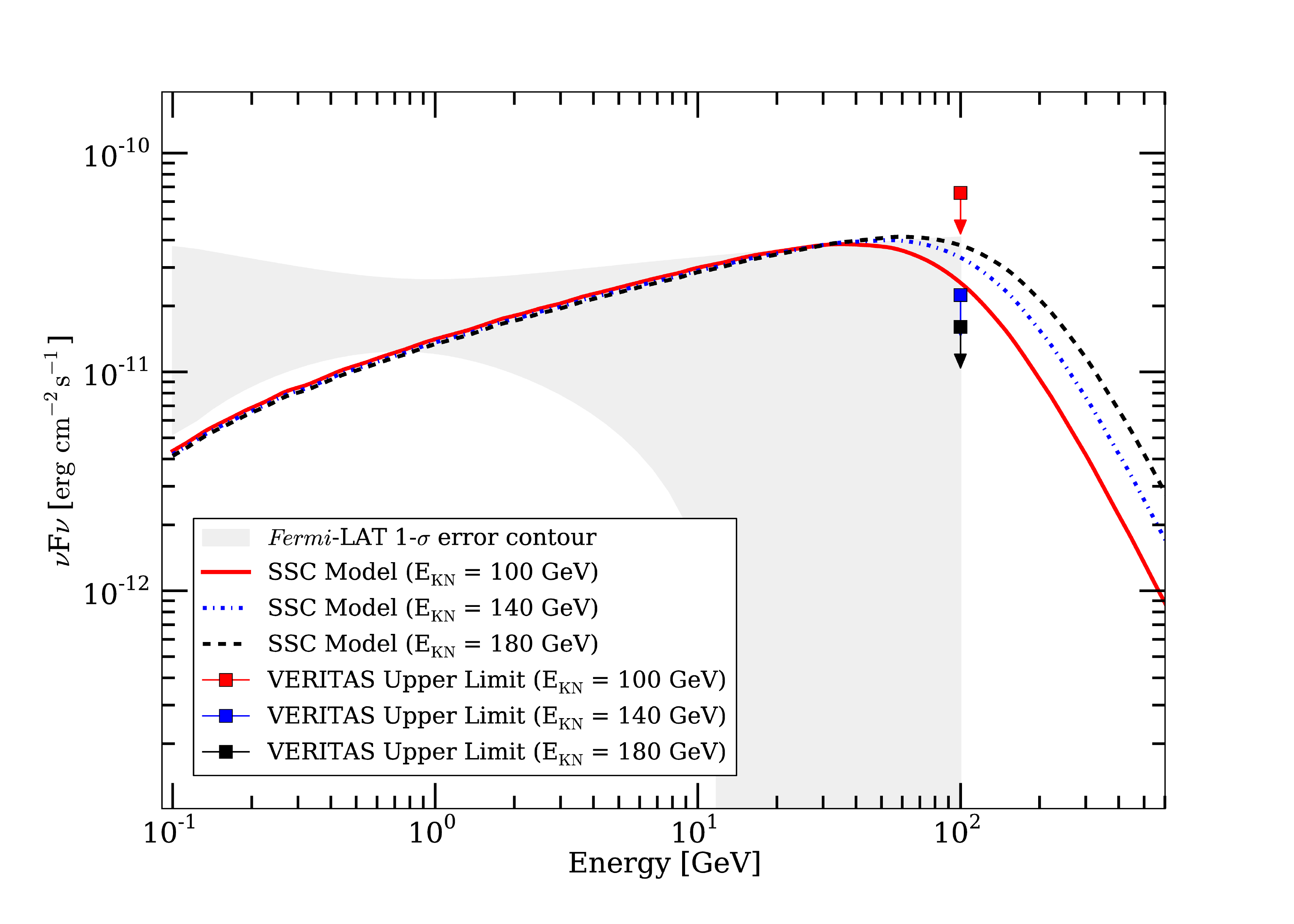}
\caption{Fermi-LAT spectrum (shaded blue area, including 1-$\sigma$ uncertainties) and VERITAS upper limits (square symbols) compared with theoretical models for GRB~130427A. From Ref.~\protect\refcite{aliu14}. 
\label{fig:veritas_130427A}}
\end{figure}

VERITAS performed follow-up observations of GRB~130427A from 71 to 75\,ks ($\sim$\,20\,hours after the onset of the burst)\cite{aliu14}. 
At that time, the LAT was still detecting radiation from the source. 
The joint LAT-VERITAS spectrum in the observing time window of VERITAS is shown in Fig.~\ref{fig:veritas_130427A} and compared with modeling of afterglow radiation.
The SSC models are estimated from Ref.~\refcite{sari01}, that is limited to the case of Thomson regime. The tension between model predictions and VERITAS upper limits might then be a hint for the Klein-Nishina effect on the SSC spectrum (see the dashed, dotted, and solid line in Fig.~\ref{fig:veritas_130427A}).

\subsubsection*{MAGIC}
The Major Atmospheric Gamma Imaging Cherenkov (MAGIC\footref{fn:magic}), located in the Canary Island of La Palma, is a system of two 17\,m IACTs, with a 236\,m$^2$ reflective surface each, and a 3.5$^\circ$ field of view.
Observations started in 2004 with one single telescope. The second telescope was added in 2009, improving energy and angular resolution, and significantly increasing the sensitivity. A major hardware upgrade in 2013 allowed further performance improvements. 

\begin{figure}
\centering
{\includegraphics[scale=0.31]{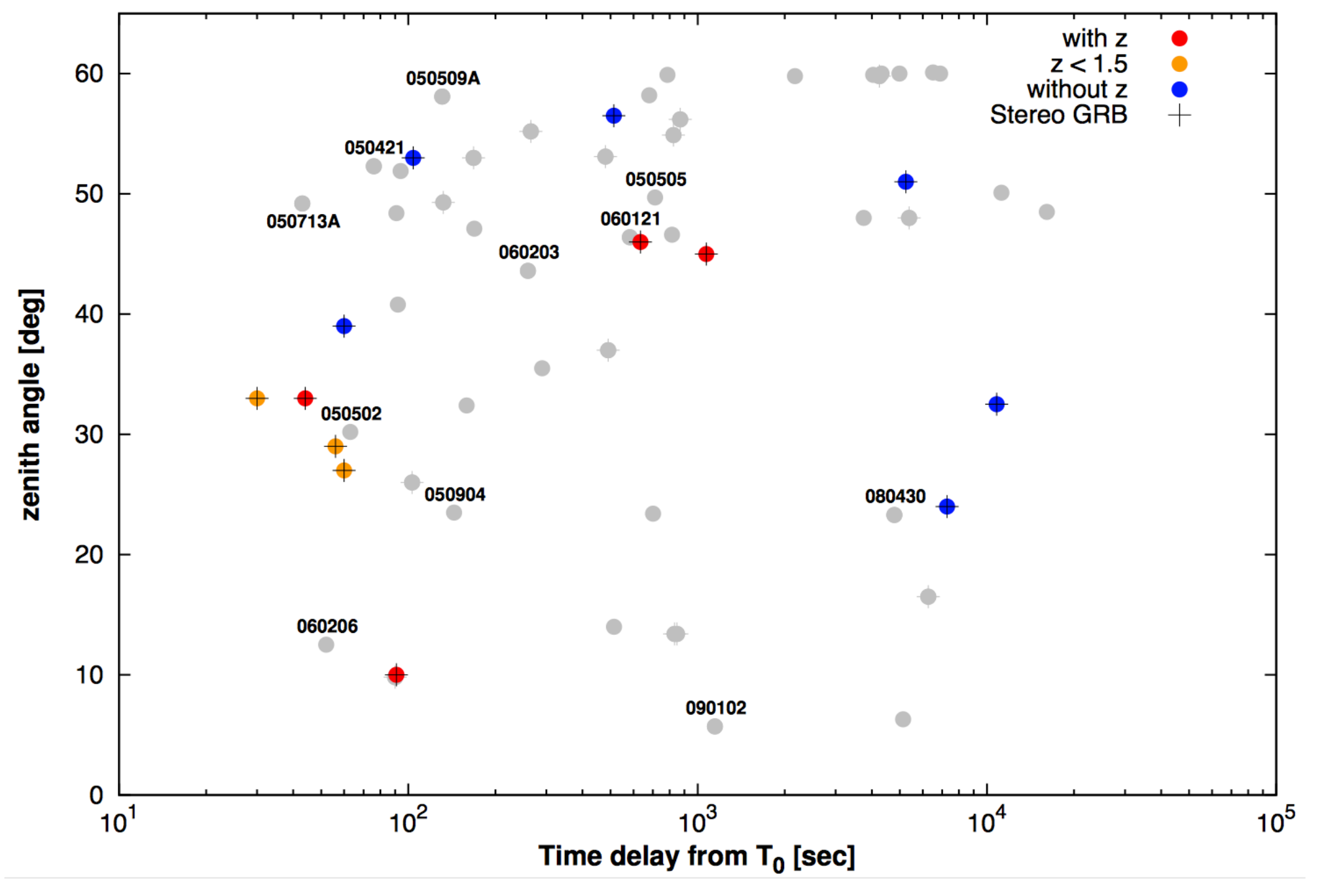}
\includegraphics[scale=0.31]{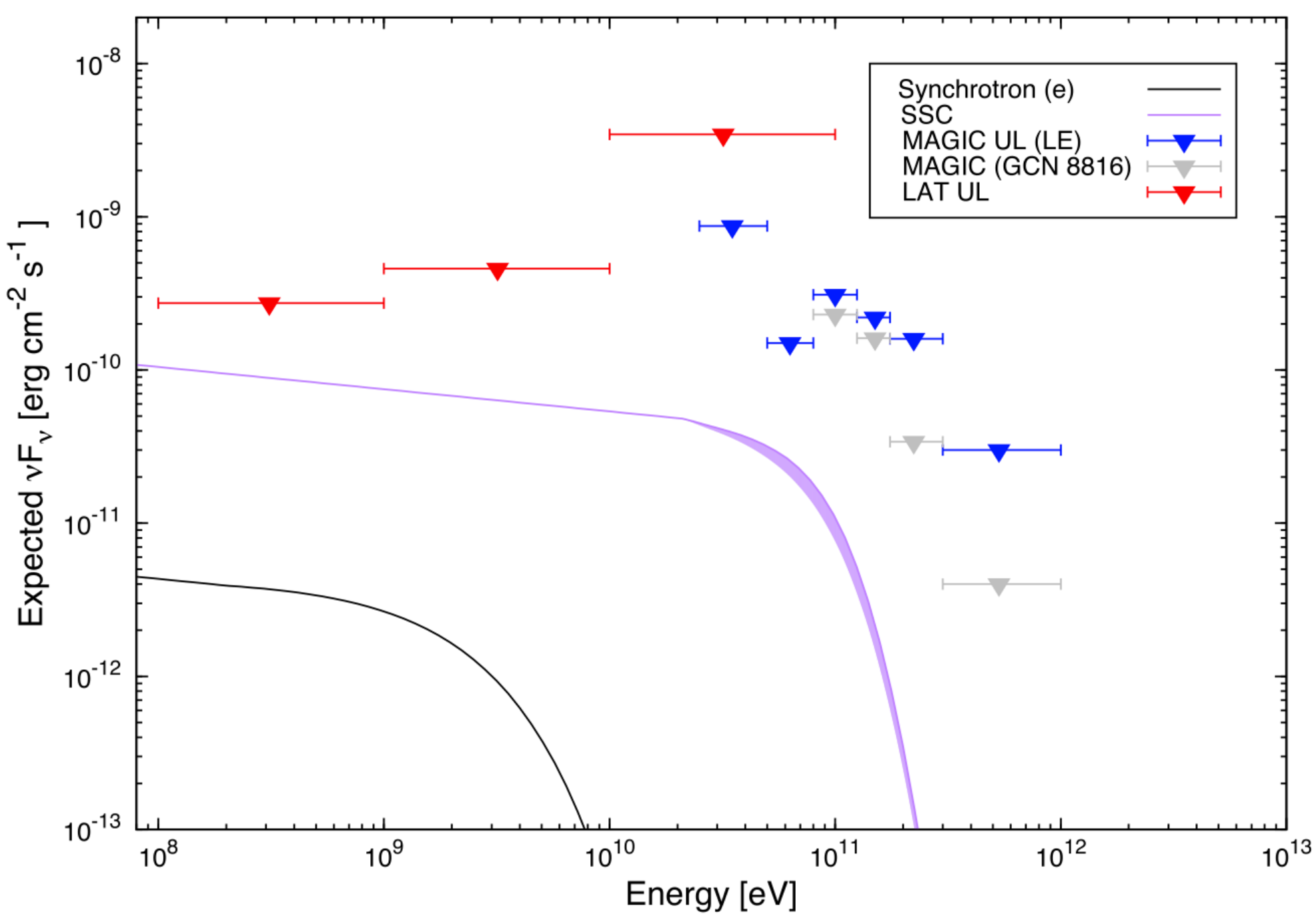}}
\caption{Left: zenith angle versus time delay of observations for GRBs observed by MAGIC before and after 2013 (gray and coloured circles, respectively). Crosses are used to mark GRBs for which stereoscopic observations have been performed (from Ref.~\protect\refcite{carosi15}).
Right: observations of GRB~090102 by MAGIC and LAT, compared to afterglow modeling. MAGIC upper limits are shown by blue triangles, while LAT upper limits are shown in red. The purple and black curve show the predicted SSC and synchrotron emission, for standard parameters $\epsilon_{\rm e}=0.1$ and $\epsilon_{\rm B}=0.01$ (from Ref.~\protect\refcite{aleksic14}).
\label{fig:magic}}
\end{figure}

MAGIC is particularly suited for the study of GRBs, thanks to its low energy threshold ($\lesssim$\,50\,GeV), combined to a very fast repositioning speed ($\sim$\,7$^\circ$ per second), achieved by minimising the device weight.
As of 2016, MAGIC has observed 89 GRBs in good observing conditions\cite{berti17}. 
The zenith angle versus delay time of MAGIC observations are shown in Fig.~\ref{fig:magic} (left-hand panel) for a sample of GRBs.
The recent upgrade of the MAGIC system and an improved GRB observation procedure have increased the rate of follow-ups starting within 100\,s after the GRB trigger (coloured filled circles), increasing the probabilities of detection.

Among the follow-up observations deserving mentioning, the observations of GRB~090102 (performed when MAGIC was operating as a single telescope) is particularly interesting\cite{aleksic14}. Observations started 1161\,s after the {\it Swift}-BAT trigger, when both {\it Fermi} and {\it Swfit} observations were ongoing.
Simultaneous observations by {\it Fermi}-LAT and MAGIC allowed to place upper limits on the high-energy SSC component. The upper limits derived by MAGIC and LAT for GRB~090102 are shown in Fig.~\ref{fig:magic} (right-hand panel) and compared with modeling of synchrotron and SSC afterglow emission. 
In the overlapping energy range of sensitivity, MAGIC upper limits are more stringent than LAT upper limits, by an order of magnitude.

\subsection{Extensive air shower arrays}
IACTs are pointed instruments with  limited field of view that need to slew to the GRB position.
For this reason, in general they miss the brightest emission phase. 
The small field of view also implies that IACTs need GRBs to have a small localization error.
Another disadvantage affecting the performances of IACTs is that they operate only in good weather conditions, at night, with no or limited moon light. 
The duty cycle of IACTs is then reduced to 10-15\%.  

EAS arrays, on the contrary, have the advantage of very high duty cycle (over 95\%) and very large instantaneous field of view (nominally 2 sr) so they are able to catch GRBs even with a localization error of several degrees. This comes at the price of a lower sensitivity as compared to IACTs, due to their smaller power in rejecting background events induced by charged cosmic-rays. 

With reference to future planned facilities, The Large High Altitude Air Shower Observatory (LHAASO) is expected to study a wide energy range (100\,GeV - 1\,PeV) with unprecedented sensitivity (as compared to other EAS arrays), thanks to a multi-component approach\cite{disciascio16}. 
LHAASO will be located at an altitude of 4410\,m, in China. Completion of the installation is expected at the end of 2021.

In the next section, I will review GRB observations performed by the currently operative observatory HAWC.

\subsubsection*{HAWC}
The High Altitude Water Cherenkov (HAWC\footref{fn:hawc}) observatory is an air shower array exploiting the water Cherenkov technique, operating in central Mexico at an altitude of 4100\,m.
It consists of 300 tanks, totaling an instrumented area of 22000\,m$^{2}$.
Science operations of the HAWC array started with a partially built array in August 2013. Full operation started at the end of 2014.
The stage of the detector between August 2, 2013 and July 8, 2014 is called HAWC-111. Observations performed during this time interval are presented in Ref.~\refcite{lennarz15}: none of the observed GRBs has significant signal to claim for a detection.
Presently, HAWC has a large instantaneous field of view ($\sim$\,2\,sr, corresponding to $\sim$\,16\% of the sky), $>$\,95\% duty cycle, and low energy threshold as small as $\sim$\,50\,GeV.
The lack of temporal delays makes HAWC an ideal detector for studying transient sources like GRBs.

A recent analysis of GRB observations during the first 18 months of HAWC full operation has been published in Ref.~\refcite{alfaro17}.
HAWC successfully observed 64 GRBs detected by {\it Swift} and/or {\it Fermi} (including 3 GRBs detected also by the LAT), and reported no statistically significant excess of events.
Of particular interest is GRB~170206A, a bright short GRB for which the fluence in the HAWC energy range implied by the HAWC upper limits is below the fluence detected in the {\it Fermi}-GBM energy range. From simultaneous GBM-LAT-HAWC observations, the authors infer that, provided that an SSC prompt component is present and for a reasonable small redshift, VHE upper limits constrain the high-energy cutoff to be smaller than 100\,GeV.

\subsection{Future prospects for VHE observations}

\subsubsection*{CTA}
The Cherenkov Telescope Array (CTA\footnote{http://www.cta-observatory.org}) is the next generation of ground-based observatory for $\gamma$-ray astronomy at very-high energies, sensitive in the range from $\sim$\,20\,GeV to 300\,TeV.
To cover such a large energy range with good sensitivity, CTA will feature three classes of telescope types. 
For its core energy range (100\,GeV-10\,TeV), CTA is planning 40 Medium-Sized Telescopes (MSTs). 
Eight Large-Sized Telescopes (LSTs) and 70 Small-Sized Telescopes (SSTs) are planned to extend the energy range below 100\,GeV and above 10\,TeV, respectively. 
The telescopes will be located at two different sites (La Palma and Paranal), covering the northern and southern hemispheres.
The MSTs and LSTs will be installed on both sites, while the SSTs will only be installed on the southern hemisphere site.
First telescopes are scheduled to be on site in 2019, operations should begin in 2021, and the array construction is expected to end in 2024.

\begin{figure}
\centering
\includegraphics[scale=0.5]{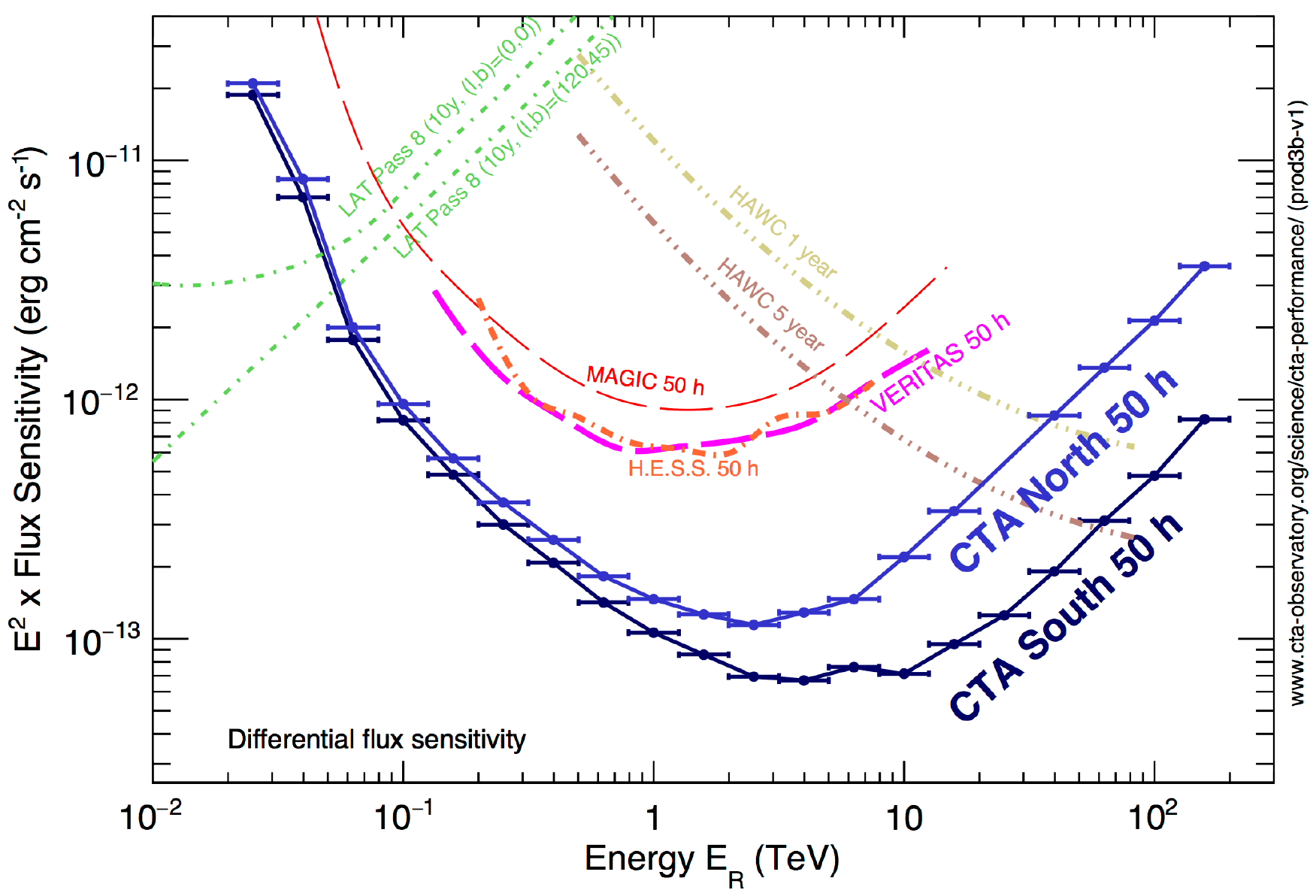}
\caption{Sensitivity curves of different instruments. The comparison is approximate, as the specific methods applied to infer the sensitivity curves for different instruments are different.
\label{fig:CTA_sensitivity}}
\end{figure}

The LSTs, sensitive in the energy range $\sim$\,20-200\,GeV, are the most interesting ones for GRB studies. They have a 23\,m diameter, 400\,m$^2$ collecting surface, and a $\sim4.5^\circ$ field of view. The repositioning time is expected to be around 20\,s.
An LST prototype telescope is under construction in La Palma. Its installation is expected before the end of 2018 with commissioning phase until mid-2019. 
With its low energy threshold, large effective area, and rapid slewing capabilities, CTA will be able to measure the spectra and variability of GRBs at multi-GeV energies with unprecedented photon statistics (see Fig.~\ref{fig:CTA_sensitivity} for a comparison with other Cherenkov telescope arrays\footnote{https://www.cta-observatory.org/science/cta-performance}).

Estimates  of GRB detection rates performed by different authors roughly amount to $\approx0.1-1$\,GRB per year\cite{gilmore13,inoue13}.
However, these estimates suffer from important uncertainties, because they depends on the final CTA configuration and performances, on the availability of triggers and localisation precision, and on the still uncertain nature and properties of the VHE emission in GRBs.

In Ref.~\refcite{inoue13}, the authors have explored a phenomenological approach to perform simulations of GRB lightcurves and spectra as seen by CTA. 
They considered two among the brightest GRBs detected by LAT and extrapolated their spectra up to higher energies using PL functions with photon index equal to the photon index measured by LAT. 
The temporal decay of the flux is also taken from the temporal index measured by LAT. To assess the role of different redshifts and different intrinsic luminosities, the template source has been also moved at different redshifts (from 1 to 6.5) and dimmed by a factor 0.1 to account for more typical luminosities. Different models of EBL have been considered. 

Encouraging results are derived from these simulations. For GRB~090902B ($z=1.822$), CTA would have detected 1000-2000 photons with a 50\,s exposure time, starting observations 50\,s after the burst trigger. A number of events ten times larger would have been detected from a GRB with the properties of GRB~090902B but located at $z=1$. In this case, photons up to 400\,GeV might be potentially detected. For larger redshifts, the capability of detecting HE photons rapidly decreases, although a GRB similar to 080916C but located at $z=6.5$ would result in about 200 photons detected (all below 100\,GeV).

The detection rate, however, is quite modest, mainly because such bright events are rare, and because of the small expected duty cycle of CTA ($\sim$\,10\%).
To increase the detection chances, a rapid follow-up and an energy threshold as low as possible are crucial and have a strong impact on the expected detection rate.

\section{Conclusions}\label{sec:conclusions}
The HE radiation detected from GRBs carries a wealth of information and a strong potential to shed new light on the physics of processes occurring both within the jet and in interactions between the jet and the external medium.
The temporal and spectral properties of this emission are indeed strongly connected to the nature of some of the least understood mechanisms at work in GRBs.
As illustrated in this review paper, the HE radiation is indeed strictly related to the nature of prompt emission mechanism, to the efficiency of the acceleration mechanism, to the configuration and origin of magnetic fields, to the properties of the emitting region, and finally to the density of the external medium.

The exploitation of HE observations as a tool to constrain models and learn about the mentioned open issues requires, however, a good understanding of the origin of this emission.
The first obstacle is represented by the difficulty in disentangling different components: while it seems clear that photons arriving well after the end of the prompt phase have an external origin, during the prompt emission a (possibly dominant) contribution from radiation produced within the jet is required. 
Disentangle the two components is however a hard task.  
Additional spectral components have been clearly identified during the prompt phase, but it is not clear if they originate from inverse Compton scatterings of the keV-MeV spectrum or if we are observing the early afterglow. 
In a few cases the HE additional component seem to extend also to lower energies, dominating the emission below 20-50\,keV. This behaviour is at odd with an interpretation of prompt emission in a simple synchrotron--SSC scenario. In general, the presence of an inverse Compton component associated to prompt emission (that was one of the main expectations from {\it Fermi}-LAT observations) has not been clearly identified yet.

The HE radiation detected during the afterglow phase is better understood.
Usually, it can be satisfactorily modeled as synchrotron radiation from the external shock. 
At odd with this interpretation is the detection of photons with energies in excess of the maximum limit expected for synchrotron photons. However, these photons might be explained by the presence of an SSC component dominating at least the high-energy part of the LAT energy range.

The presence of an afterglow SSC component at GeV-TeV energies represents an appealing possibility for GRB detections with the CTA\cite{lemoine15,vurm17}.
Past and present follow-up observations from Cherenkov telescopes have not detected so far any excess of events coming from GRBs.
However, for the brightest GRBs (e.g., GRB~130427A) the derived flux upper limits lie close to theoretical expectations. 
The improved sensitivity expected to be achieved by the CTA gives reasonable hope for future GRB detections: a few but invaluable GRB detections are expected at 20-100\,GeV and possibly even beyond.

\section*{Acknowledgments}
The author acknowledges funding from the European Union's Horizon 2020 research and innovation programme under the Marie Sk\l odowska-Curie grant agreement n. 664931.
The author is deeply thankful to all the colleagues who, over the years, have been collaborating with her on the topic of this review paper, sharing ideas, and raising fruitful discussions. 
The author is particularly grateful to G.~Ghisellini, G.~Ghirlanda, A.~Celotti, F.~Longo, P.~Beniamini, T.~Piran, R.~Barniol~Duran, P.~Kumar, N.~Omodei, G.~Vianello, R.~Desiante.

\end{document}